Artificial Intelligence Augmented Medical Imaging Reconstruction in Radiation Therapy

by
DI XU

DISSERTATION
Submitted in partial satisfaction of the requirements for degree of
DOCTOR OF PHILOSOPHY

in

Bioengineering

in the

GRADUATE DIVISION

of the

UNIVERSITY OF CALIFORNIA, SAN FRANCISCO
AND
UNIVERSITY OF CALIFORNIA, BERKELEY

Approved:

Ke Sheng
_________________________________________
Chair

Martina Descovich
_________________________________________

Yang Yang
_________________________________________

Chunlei Liu
_________________________________________
_________________________________________
Committee Members

Copyright 2025

by

Di Xu

# Dedication and Acknowledgement

First and foremost, my greatest and most sincere thank goes to my supervisor Dr. Ke Sheng. He is the best mentor I could ever ask for. He provided the strongest support through all the ups and downs in my PhD journey. For the starter, he gave me the most valuable opportunity to switch my career path from computer science to medicine, where my deepest passion always lies in since I was little. Additionally, my start year happened to fall in a very rare and chaotic time when COVID widely spread and international travel ban was issued. He provided all the possible support he could to help my peers and me get through the hard time. In the beginning of my PhD studies, he guided me to understand the field and formulate my own vision. In the later stage of my PhD studies, he let me navigate through our own research interest, lead my own projects, and learn from my rights and wrongs. He always encourages us to explore deeply and enlightens us to become independent and well-round researchers. Meanwhile, he is extremely patient, kind, generous, considerate, and understanding. He gave me full trust even when I am in deep self-doubt.

I would also like to thank Dr. Yang Yang and Dr. Dan Ruan for their constant guidance and strong support throughout my graduate journey. Talking to them is always fun and inspiring.

I would also like to thank Dr. Martina Descovich and Dr. Chunlei Liu on my dissertation committee. I benefited greatly from discussing my projects with them.

I would also like to thank Dr. Duan Xu, Ms Victoria Starrett and Ms Rocio A. Sanchez for advising my PhD study in the Bioengineering program. UCSF-UCB Bioengineering is such a wonderful diverse, and unique program. I am more than honored to be part of it.

My thanks would also go to my other collaborators, Dr. Wensha Yang, Dr. Fabien Scalzo and Dr. Xin Miao, for their valuable data and precious insights.

My thanks would also go to my wonderful labmates, Yi, Hengjie, Lu, Jiayi, Qifan, Xinyi, Alan, Pav, Holly, Shusen, Menghan, Wan-jin, Jingjie, Phoebe and Josh. It has been a great fortune to be a part of the team and witness many exciting research happening. Beyond that, their friendship and companion filled the journal with full of joy. In particular, I want to thank Yi for being my both my best colleague and best friend. It's such a pleasure to work and hang out with you.

My thanks should also go to my CAMPEP-Medical Physics program director Dr. Adam M. Cunha, instructor Dr. Katelyn Hass, Dr. Emily Hirata, Dr. Jose Ramos-Mendez, Dr. Jessica E. Scholey, Dr. Alon Witztum, and Dr. Benjamin Paul Zimer. Thank them for guiding me understand the field and laying the fundamental ground for my career.

I would also like to personally thank my parents Zihui Xu and Yaxin Wang and my in-laws Chunguang Zhou and Hongmei Yin for their unconditional love and support.

Finally, I want to thank my husband Yinsheng Zhou, who is not only my partner but also my best friend. He provides his deepest support and love throughout my PhD studies. He is the person who understands me the most, knows me the most and I can share with and trust the most. Thank you for supporting all my decisions and blue-sky thinkings. Thank you for growing up with me together, You made me who I am today.

*"Research is what I'm doing when I don't know what I'm doing."*

*-Wernher von Braun*

# Abstract

**Background and Purpose:** Efficiently acquired and precisely reconstructed imaging are crucial to the success of modern radiation therapy (RT). Computed tomography (CT) and magnetic resonance imaging (MRI) are two common modalities for providing RT treatment planning and delivery guidance/monitoring. In recent decades, artificial intelligence (AI) has emerged as a powerful and widely adopted technique across various fields, valued for its efficiency and convenience enabled by implicit function definition and data-driven feature representation learning. Here, we present a series of AI-driven medical imaging reconstruction frameworks for enhanced radiotherapy, designed to improve CT image reconstruction quality and speed, refine dual-energy CT (DECT) multi-material decomposition (MMD), and significantly accelerate 4D MRI acquisition.

**Method**: Our CT reconstruction framework focuses on ultra-sparse view reconstruction using Neural Radiance Field (NeRF), a powerful technique for reconstructing and rendering 3D scenes from limited views. We developed a novel framework, TomoGRAF, which integrates X-ray transport physics to reconstruct high-quality 3D volumes from ultra-sparse projections without requiring priors. TomoGRAF models CT imaging geometry, simulates the X-ray casting and tracing process, and penalizes the difference between simulated and ground-truth CT sub-volumes during training. Next, we introduce a sinogram-domain material decomposition framework designed for breast tissue differentiation. This non-recursive approach operates directly on raw projection data. Our rFast-MMDNet employs a two-stage algorithm: (1) SinoNet, which performs dual-energy projection decomposition on tissue sinograms, and (2) FBP-DenoiseNet, which facilitates domain adaptation and image post-processing.

Finally, we developed three distinct 4D MRI acceleration pipelines leveraging Transformers, generative adversarial networks (GANs), and diffusion networks. For the Transformer-based approach, we propose Reconstruction Swin Transformer (RST), which builds upon the Video Swin Transformer architecture with a novel reconstruction head to restore pixel-wise intensity. To enhance efficiency, we introduce SADXNet, a convolutional network that rapidly initializes 2D MR frames before RST learning, significantly reducing model complexity, GPU demand, and training time. For the GAN-based pipeline, we present Reconstruct Paired Conditional Generative Adversarial Network (Re-Con-GAN), which accelerates 4D MRI reconstruction while preserving image quality. Re-Con-GAN is trained using paired input-output data, exploring ResNet9, UNet, and RST as potential generators, with PatchGAN as the discriminator. For diffusion-based reconstruction, we introduce the Chained Iterative Reconstruction Network (CIRNet), designed for efficient sparse-sampling reconstruction while maintaining clinically deployable quality. CIRNet employs a denoising diffusion probabilistic framework, conditioning image reconstruction through a stochastic iterative denoising process. During training, a forward Markovian diffusion process progressively adds Gaussian noise to the densely sampled ground truth, while CIRNet learns to reverse this process iteratively. At inference, CIRNet runs the reverse process to recover signals from noise, conditioned on the undersampled input.

**Result**: In the task of TomoGRAF, we evaluated the performance of TomoGRAF on an unseen dataset of distinct imaging characteristics from the training data and demonstrated a vast leap in performance compared with state-of-the-art deep learning and NeRF methods. TomoGRAF provides the first generalizable solution for image-guided radiotherapy and interventional radiology applications, where only one or a few X-ray views are available, but 3D volumetric information is desired. For rFast-MMDNet, MMD for breast fibroglandular, adipose tissues, and calcification was performed using the 2022 DL-Spectral-Challenge dataset. Our results show that models trained from the sinogram domain can yield high-fidelity decomposition where rFast-MMDNet is particularly dominant. In our 4D MRI acceleration project series, RST was evaluated on a lung respiratory dataset, demonstrating promising reconstruction performance. Re-Con-GAN and the diffusion-based CIRNet were tested on 4D liver MRI, where Re-Con-GAN achieved sub-second inference while maintaining high-quality reconstruction at 10× acceleration. Meanwhile, CIRNet, though requiring slightly longer inference time (~10 seconds), pushed acceleration beyond 30× while preserving high reconstruction quality.

**Conclusion**: The proposed AI frameworks achieve fast image reconstruction with favorable quality in diverse imaging modality including CT, DECT and MRI.

Table of Contents

# List of Figures

# List of Tables

# List of Abbreviations

2D: two dimensional

3D: three dimensional

4D: four dimensional

AAPM: American Association of Physicists in Medicine

AI: artificial intelligence

AP: anterior posterior

ASD-POCS: adaptive steepest descent projection onto convex sets

ASDL: artifact suppressed dictionary learning

AwTV-POCS: adaptive-weighted TV POCS

BN: batch normalization

CBCT: cone beam computed tomography

CIRNet: chained iterative refinement network

CS: compressed sensing

CSLD: Coronal and Sagittal Lung Data

CT: computed tomography

CTOA: CT on rails

DECT: dual energy CT

DL: deep learning

DRRs: digitally reconstructed radiographs

DSA: digitally subtracted angiography

DDPM: Denoising Diffusion Probabilistic Model

FBP: filtered back-projection

FDK: Fekdkamp-Davis-Kress

FOV: field-of-view

FFT: fast Fourier transform

GAN: generative adversarial network

GRAF: generative radiance field

GPUs: graphics processing units

GT: ground truth

HCC: hepatocellular carcinoma

HMVCT: helical megavoltage CT

IAEA: International Agency for Research on Cancer

IGRT: image-guided radiation therapy

IMRT: intensity-modulated radiation therapy

IRB: Institutional Review Board

ITV: internal target volume

KNN: k nearest neighbors

KVCBCT: kilovoltage CBCT

LDCT: low dose CT

LINAC: linear accelerator

LPIPS: Learned Perceptual Image Patch Similarity

LSTM: long short-term memory

MACs: multiply–accumulate operations

MC: motion compensation

ME: motion estimation

ML: machine learning

MLP: multi-layer perceptron

MRI: magnetic resonance imaging

MRgRT: MRI-guided radiation therapy

MS-SSIM: multi-scale SSIM

MSA: multi-head self-attention layer

MVCBCT: Megavoltage CBCT

MV: megavoltage / mega-voltage

NLM: patch-based nonlocal means

NL-Means: non-local means

NeRF: neural radiance field

NLL: non-negative log likelihood

NNs: neural networks

OARs: organ at risks

PET: positron emitted tomography

PICCS: prior image-constrained CS

PSNR: peak signal to noise ratio

ReLU: rectified linear unit

ResNet: residual network

RMSE: rooted mean squared error

RNNs: recurrent neural networks

RST: Reconstruction Swin Transformer

RST-S: RST small

RST-T: RST tiny

RT: radiation therapy

SDCT: standard dose CT

SECT: single energy CT

SOTA: state-of-the-art

SR3: Super Resolution via Repeated Refinement

SSIM: structure similarity indexed measurement

S-DFR: sinogram-discriminative feature representation

SW-MSA: shifted window based multi-head self-attention

TV: total variation

TVS-POCS: TV-stokes-projection onto convex sets

US: ultrasound

VMAT: volumetric-modulated arc therapy

VMIs: virtual monochromatic images

VST: Video Swin Transformer

# List of Symbols

$t$: time step

$h_t$: the hidden state (or internal memory) of the RNN at time step $t$

$x_t$: the input at time step $t$

$W_{hh}$: the weight matrix that govern the transformation of the previous state $h_{t-1}$

$W_{xh}$: the weight matrix that and connecting the hidden state to the output $y_t$.

$W_{yh}$: the weight matrix that connecting the hidden state to the output $y_t$.

$b_h$: the bias vector

$b_y$: the bias vector

$f$ are an activation function

$f$ are an activation function

$f_t$,: forget gate

$i_t$: input gate

$o_t$: output gate

$\tilde{C}_t$: candidate cell state

$C_t$: cell state

$X$: input image

$H \times W \times D$: height×width×depth

$K$: filters

$W_k$: weight matrix of resulting of $K/$

$b_k$: bias term

$Z_k$: feature map

$p$: pool size

$A_{fc}$: the activations from the pooling layer

$W_{fc}$: weights in fully connected layers

$b_{fc}$: bias of fully connected layers

$Y$: output

$F$: function

$Q$: queries

$K$: keys

$V$: values

$d_k$: dimension of the keys

$\delta$: scale factor

$G$: generator

$\hat{Y}$: prediction

$Y$: ground truth

$\boldsymbol{\xi}$: direction (pose)

$V'$: sub-volume

$\boldsymbol{K}$: X-ray source setup matrix

$P'$: projection patch

$r$: rendered rays

$\boldsymbol{v}, \boldsymbol{u}, s$: image location parameters

$z_{sh}, z_a, \boldsymbol{M_{sh}}, \boldsymbol{M_a}, \gamma(\cdot)$ : network parameter encoding

$h_\vartheta, d_\vartheta$: network components

$D_{1,\emptyset}, D_{2,\emptyset}$: discriminators

$\emptyset_i(\cdot)$: output from previous networks

$\boldsymbol{f_v}, \boldsymbol{f_p}$: feature maps

$\mathcal{G}, \mathcal{T}$: pre-processing methods

$L$: loss

$\lambda$: hyperparameters

$\mu$: means

$c, k$: constant

$I_w$: Transmission

$S$: spectral sensitivities

$\mu_m$: linear attenuation

$\mathcal{P}$: forward projection.

# 1. Introduction

The introduction section is organized as outlined in **Figure 1-1**, beginning with a broad overview of **Radiation Therapy**, followed by **Medical Imaging in Radiation Therapy**. It then explores **Artificial Intelligence**, covering its general overview and applications in radiation oncology. Next, **Computed Tomography (CT) Imaging** is discussed, detailing its applications in radiation therapy, reconstruction techniques, and multi-material decomposition methods. This is followed by **Magnetic Resonance Imaging (MRI)**, which highlights its role in radiation therapy and reconstruction approaches. The section concludes with an **Overview of the Projects**, which are further detailed in **Sections 2–4**.

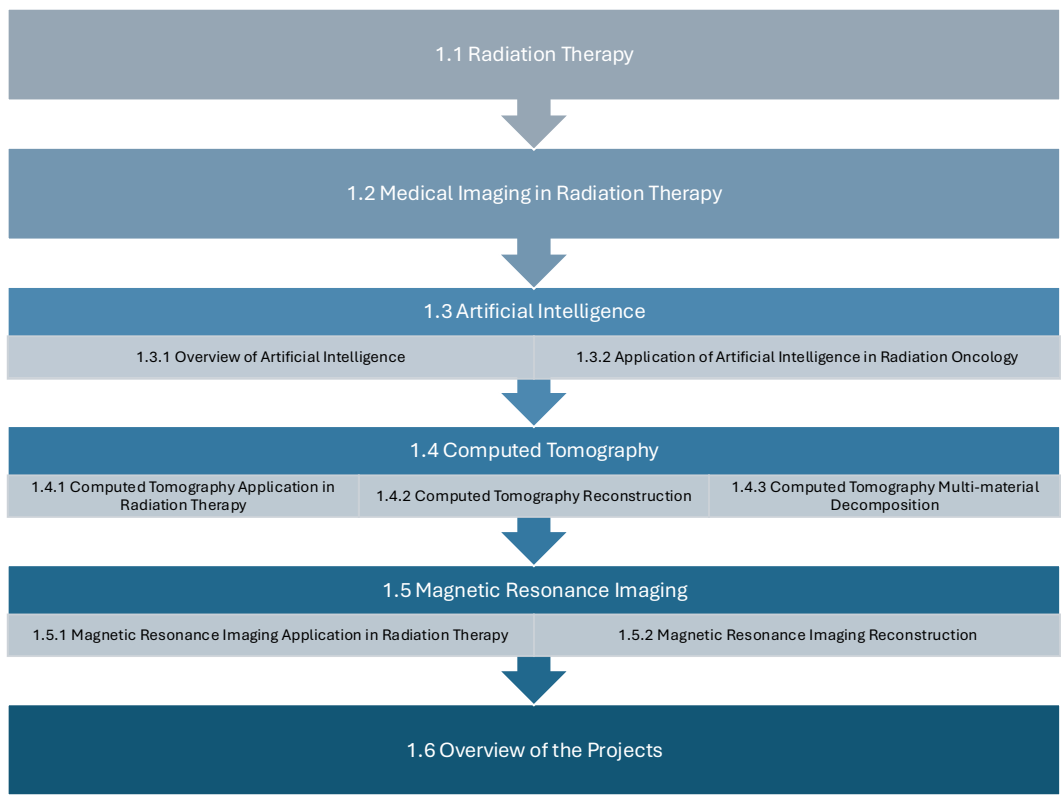

*Figure 1-1 The organization of the Section 1-Introduction*

## 1.1 Radiation Therapy

Cancer remains to be one of the leading causes of death worldwide. Cancer is a multicellular and multigenic disease that can occur in every cell type or organ with a muti-factorial etiology[1]. Identified cancer cell types can be distinguished into six distinct hallmarks, including cells with unlimited proliferative potential, environmental independence for growth, evasion of apoptosis, angiogenesis, and invasion, as well as metastasis to different parts of body. Failure in localized cell growth or metastatic spread control will lead to patient death[2]. According to the recent statistics estimated by the International Agency for Research on Cancer (IARC), 7.6 million deaths were owing to cancers, where 12.7 million incremental new cases were reported annually worldwide[3].

Over the past few decades, vast progress has been made toward understanding the response of different hallmarks for effective treatment. Surgery, chemotherapy, radiotherapy, immunotherapy, cell therapy, gene therapy and etc. are alternatives used to treat cancer[4]. Retrospective statistics have shown that radiotherapy, on average, accounts for only around 5% of the total cancer care cost[5]. On the other hand, radiotherapy contributes around 40% towards curative treatment response[6]. Rapid progress in radiotherapy continues to be boosted by advanced computerized treatment planning systems, improved treatment delivery, a better understanding of radiobiology, and enhanced imaging techniques[7].

Radiation therapy utilizes external high-energy radiation beams or internal radioactive emitters to cause irrepairable DNA double-strand breaks (Figure 1-2) and suppress cancer cell growth, division, and proliferation. Radiotherapy can be further classified into two general sub-categories – brachytherapy and teletherapy. Brachytherapy is a treatment operated by placing the radioactive source near or in contact with the treatment target. Teletherapy, also known as external beam therapy, performs treatment with the radioactive source at a distance from the treatment target. The voltage applied to produce X-ray photons can be seen in Table 1-1, where low-energy beams are mainly used for superficial tumor treatment, and high-energy beams are applied for deep-seated tumor treatment.

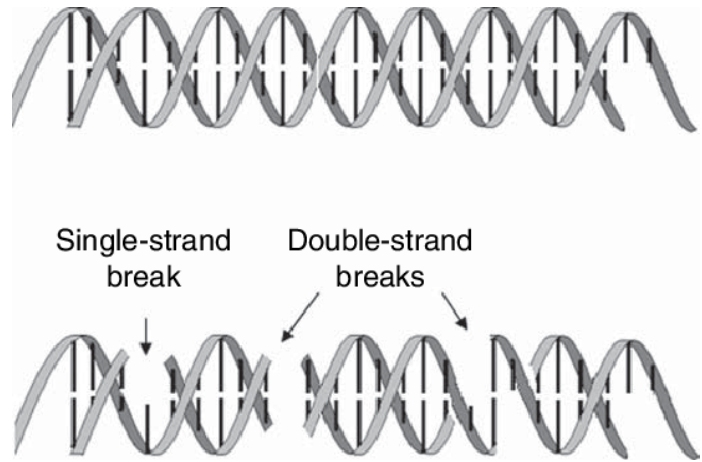

*Figure 1-2: Schematic representation of DNA with none-, single- and double-strand breaks.*

*Table 1-1 X-ray voltage and beam energy classification in teletherapy.*

| Voltage in Accelerator (kV) | Mean Photon Energy (KeV) | Energy Category |
|:---:|:---:|:---:|
| 10-150 | 3-50 | Superficial |
| 150-500 | 50-166 | Orthovoltage |
| 500-1000 | 166-333 | Supervoltage |
| >1000 | >333 | Megavoltage |

## 1.2     Medical Imaging in Radiation Therapy

Conventionally, non-invasive medical imaging, such as computed tomography (CT), enables the possibility for three-dimensional (3D) dose calculation and patient positioning (Figure 1-3).

Specifically, CT accurately describes 3D anatomy, which is essential for modern radiotherapy. The electron density quantified in CT is used for dose calculation. Organ-at-risk (OAR) and tumor structures delineated on CT guide treatment planning so a high prescription dose covers the tumor while the dose to surrounding critical organs is minimized. CT also helps achieve accurate patient positioning before treatment. Additional CT images, typically in the form of cone beam CT, are acquired and matched with the planning CT to reproduce the planning position. Besides 3D images, 2D radiographs are also used to match with digitally reconstructed radiographs (DRRs), virtual projection images calculated based on the planning CT, for patient positioning.

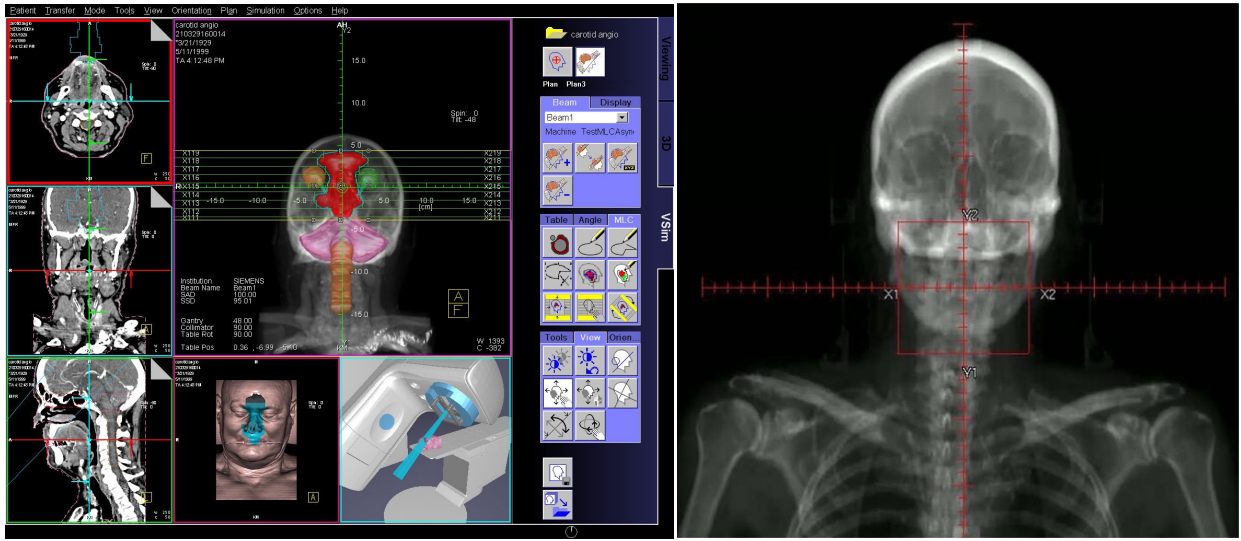

(a) A treatment planning example using CT    (b) An example for patient alignment using DRR

*Figure 1-3 Examples for treatment planning and patient alignment in radiotherapy with CT.*
Modern radiation therapy is fully computerized with advanced techniques for precise dose delivery, such as intensity-modulated radiation therapy (IMRT), volumetric-modulated arc therapy (VMAT) and image-guided radiation therapy (IGRT). Combining these technical advances helps modern radiotherapy achieve ~1 millimeter[9] targeting accuracy, high dose conformity, and steep dose gradient outside the target. While modern radiotherapy based on the CT ecosystem has achieved remarkable success, there are limitations intrinsic to CT and X-ray physics. CT has suboptimal soft tissue contrast, lacks functional information, is not sensitive to the presence of small metastatic tumors, and is not a real-time imaging method due to the mechanical gantry rotation needed to collect projections.

Other imaging modalities, such as magnetic resonance imaging (MRI), are often incorporated in the radiotherapy workflow for improved target localization, OAR delineation, and response monitoring[10–12]. Specifically, MRI is superior to CT in the soft tissue contrast (Figure 1-4), which can be critical to identify tumors embedded in tissues of similar electron density[13,14]. MR also differs from CT in its rich sequences that are designed to probe physiological and pathological processes beyond anatomy. For example, diffusion-weighted images can be used to measure cellularity highly correlated to tumor proliferation and response to therapies[15].

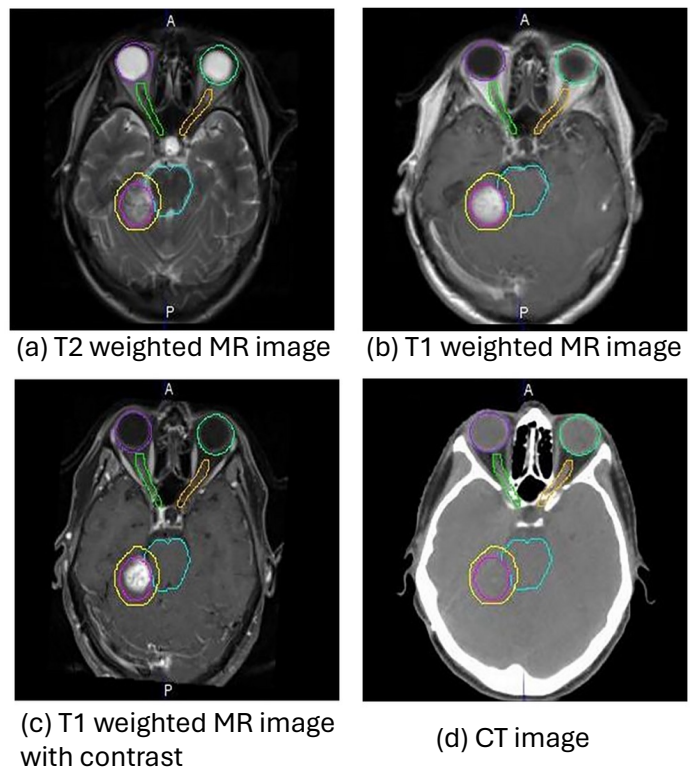

*Figure 1-4 Comparison in the soft tissue contrast from four imaging modalities of the same patient.*

Compared with diagnostic imaging and chemotherapy, a unique requirement in radiotherapy is absolute geometrical accuracy of dose delivery to the tumor target. The inter- and intra-fractional organ motion can significantly influence geometric accuracy and treatment efficacy. X-ray-based 2D and 3D imaging modalities are widely present in IGRT. Stereoscopic 2D radiographs are employed for the localization of bony anatomies, tumors with high X-ray contrast, such as lung tumors, and implanted fiducials[16–18]. Cone beam CTs (CBCT) are widely available on C-arm linacs for 3D imaging of patients, which enhances the visualization of soft tissues and enables internal organ-based registration and localization in comparison to 2D radiographs[19]. Similar to the diagnostic imaging domain, MR are also introduced in the treatment room to provide additional dimensions of information for better localization and delineation of the tumor and surrounding normal tissues[20–22].

The aforementioned advanced imaging techniques also enable adaptive radiotherapy (ART), which allows for treatment plans to be dynamically adjusted based on real-time patient anatomy and organ motion. By leveraging frequent imaging updates, ART enhances targeting precision, reduces healthy tissue toxicity, and improves treatment efficacy[23]. For instance, the Elekta Unity MRI Linear Accelerator (LINAC) combines a 1.5T scanner with a LINAC, where the beam-on MR is used for precise tumor tracking enabled time-resolved dose accumulation and real-time replanning[24]. Additionally, Accuray CyberKnife leverages X-ray imaging and infrared tracking (Xsight®) algorithm for real-time beam-on tumor position tracking and adaptive beam adjustment to account tumor motions[25].

## 1.3     Artificial Intelligence

### *1.3.1     Overview of Artificial Intelligence*

Artificial Intelligence (AI) is a sub-field of computer science that aims to simulate human intelligence in machines programmed to perform tasks like perception, speech recognition, decision-making, and language translation. AI technologies, such as machine learning (ML) and deep learning (DL), enable these systems to improve over time through data and experience.

Earlier AI algorithms were condition-based reasoning realized with "if-else" setup defined by human experts[26,27]. Some of these frameworks achieved varying degrees of clinical utilities[28]. However, they required explicit labels for training, which had limited availability. They also suffer from low generalizability in unseen scenarios and edge cases. Lately, the rapid development of graphics processing Units (GPUs) vastly empowers the applicability of deep neural network (NN). These algorithms include a layered data representation structure, where the high-level features can be extracted from the last layers of the networks while the low-level features are extracted from the lower layers. The design of DL was originally inspired by the classification function of human brain, which automatically extract data representation from various scenes with input received from human vision and output of the classified objects. Different type of NNs were previously proposed to tackle problems from various domains. The proposed NNs can be classified into three general categories, including recurrent neural networks (RNN)[29], Transformers[30], and convolutional neural networks (CNNs). Details regarding the structure of each network architecture are elaborated as follows.

### 1.3.1.1 Recurrent Neural Network

RNNs are a specialized type of artificial neural network designed to process sequential data by maintaining an internal memory. Unlike traditional feedforward NNs, RNNs can capture temporal dependencies and patterns in data, making them particularly powerful for tasks involving sequences such as speech recognition, natural language processing, and time series prediction. The key feature of RNNs is their ability to persist information over time through recurrent connections, allowing them to learn from previous inputs to influence subsequent predictions.

Figure 1-5 (a) illustrates a typical unfolded RNN diagram. At each time step $t$, the RNN takes an input $x_t$ and updates its internal state $h_t$ based on the current input and the previous state $h_{t-1}$. This updating process is mathematically represented as:

$$h_t = f(W_{hh}h_{t-1} + W_{xh}x_t + b_h) \tag{1-1}$$

The output of the RNN at each time step $t$, denoted as $y_t$, depends on the current state $h_t$:

$$y_t = g(W_{yh}h_t + b_y) \tag{1-2}$$

Where $h_t$ is the hidden state (or internal memory) of the RNN at time step $t$, $x_t$ is the input at time step $t$, $W_{hh}$, $W_{xh}$ and $W_{yh}$ are weight matrices that govern the transformation of the previous state $h_{t-1}$ and current input, $x_t$ into the current state $h_t$, and connecting the hidden state to the output $y_t$. $b_h$ and $b_y$ are the bias vector, $f$ and $g$ are an activation function applied elementwise, such as the sigmoid, tanh, or ReLU function.

Long Short-Term Memory (LSTM)[31] networks are an improvement of traditional RNN architecture, which fails in handling long-term dependencies. As shown in Figure 1-5 (b), LSTMs incorporate memory cells and gating mechanisms to control the flow of information. Each LSTM cell contains three gates: the input gate, the forget gate, and the output gate, which decide which data to add to the cell state, which to discard, and which to output. Mathematically, this is represented by:

$$f_t = \sigma(W_f \cdot [h_{t-1}, x_t] + b_f) \tag{1-3}$$

$$i_t = \sigma(W_i \cdot [h_{t-1}, x_t] + b_i) \tag{1-4}$$

$$o_t = \sigma(W_o \cdot [h_{t-1}, x_t] + b_o) \tag{1-5}$$

$$\tilde{C}_t = tanh(W_C \cdot [h_{t-1}, x_t] + b_C) \tag{1-6}$$

$$C_t = f_t * C_{t-1} + i_t * \tilde{C}_t \tag{1-7}$$

$$h_t = o_t * \tanh(C_t) \tag{1-8}$$

Here, $f_t, i_t$ and $o_t$ are forget, input, and output gates, respectively; $\tilde{C}_t$ is the candidate cell state; $C_t$ is the cell state; and $h_t$ is the hidden state. The gating mechanisms allow LSTMs to maintain and update cell states over long sequences, effectively capturing long-term dependencies without suffering from the vanishing gradient problem. LSTMs have been widely successful in applications such as language modeling, machine translation, and time series prediction, where understanding the context and retaining information over long periods is crucial.

### 1.3.1.2  Convolutional Neural Network

CNNs are also a class of popular DL algorithms that have revolutionized the field of computer vision[32], natural language processing[33], and speech recognition[34]. Inspired by the visual cortex of animals, CNNs are designed to automatically and adaptively learn spatial hierarchies of features from input images. They consist of multiple layers, primarily convolutional layers, pooling layers, and fully connected layers. Convolutional layers apply filters to the input image to extract essential features, such as edges, textures, and shapes. Pooling layers reduce the spatial dimensions of the feature maps, retaining the most relevant information while reducing computational complexity. Fully connected layers at the end of the network integrate these features to perform classification or regression tasks. CNNs have demonstrated remarkable performance in various applications, including image recognition, object detection, and even tasks beyond vision, due to their ability to capture and interpret complex patterns in data.

Figure 1-5 (c) illustrates a typical CNN structure. As briefly mentioned, the key components of a CNN include convolutional layers, pooling layers, and fully connected layers. In a convolutional layer, the input image $X$ of dimensions $H \times W \times D$ (height, width, and depth) is convolved with a set of filters $K$ (or kernels), $W_k$, each of the dimensions $f \times f \times D$ (filter height, filter width, and depth), resulting in $K$ feature maps. The convolution operation is defined as:

$$Z_k = X * W_k + b_k \tag{1-9}$$

where $*$ denotes the convolution operation, and $b_k$ is the bias term for the k-th filter. Each feature map $Z_k$ is then passed through an activation function, typically the ReLU (Rectified Linear Unit), to introduce non-linearity:

$$A_k = ReLU(Z_k) \tag{1-10}$$

Pooling layers, such as max pooling or average pooling, follow convolutional layers to reduce the spatial dimensions (height and width) of the feature maps while retaining important features. For example, max pooling with a pool size of $p \times p$ is given by:

$$P_k = maxpool(A_k) \tag{1-11}$$

Finally, fully connected layers (dense layers) take the flattened feature maps from the final pooling layer and perform linear transformations to classify the input. If $A_{fc}$ is the activations from the last pooling layer and $W_{fc}$ and $b_{fc}$ are the weights and bias for the fully connected layer, the output $Y$ is given by:

$$Y = softmax(W_{fc}A_{fc} + b_{fc}) \tag{1-12}$$

where the softmax function converts the final linear activations into probabilities for classification tasks. This layered structure allows CNNs to effectively learn and interpret hierarchical patterns and features in the data.

Residual Network (ResNet)[35], or residual network block, is designed to tackle the vanishing gradient problem in very deep neural networks (Figure 1-5 (d)). A ResNet block consists of a series of convolutional layers, batch normalization, and ReLU activations, but it uniquely incorporates shortcut connections, also known as skip connections. These connections allow the input of the block to bypass one or more convolutional layers and be added directly to the block's output. Mathematically, this is represented as:

$$y = F(x, \{W_i\}) + x \tag{1-13}$$

where $x$ is the input, $F(x, \{W_i\})$ is the function representing the convolutional layers, and $W_i$ are the weights. This structure enables the network to learn residual functions instead of direct mappings, making it easier to train very deep networks by allowing gradients to flow more effectively during backpropagation.

Compared to ResNet, standard CNNs do not have these shortcut connections and rely solely on stacked layers of convolutions, pooling, and nonlinear activations. While standard CNNs can perform well with a moderate number of layers, they often struggle with training deep networks due to issues like vanishing gradients, where the gradients become very small and learning effectively halts.

1.3.1.3 Transformers

Transformers, introduced by Vaswani et al.[30], is a type of NN primarily designed for handling sequential data. Transformers have revolutionized sequential processing by enabling the model to capture dependencies across an entire sequence, regardless of its length. As shown in Figure 1-5 (e), the core component of transformers is the self-attention mechanism, which computes the relevance of each element in a sequence to every other element, allowing the model to weigh the importance of different words dynamically. Mathematically, for an input sequence $X$, the self-attention mechanism produces an output sequence by computing attention scores using queries $Q$, keys $K$, and values $V$ derived from $X$:

$$Attention(Q, K, V) = softmax(\frac{QK^T}{\sqrt{d_k}})V \qquad (1\text{-}14)$$

where $d_k$ is the dimension of the keys. Transformers also use positional encoding to retain the order of the sequence, as the self-attention mechanism is inherently order-agnostic. This architecture enables transformers to handle long-range dependencies and parallelize training processes more effectively than traditional RNNs. As a result, transformers have become the foundation for SOTA NLP models like BERT and GPT, excelling in tasks such as translation, text generation, and question-answering.

More recently, transformers have also been applied in computer vision, with architectures like Vision Transformers (ViTs)[36] demonstrating impressive performance on image classification tasks by treating image patches as sequences, thereby leveraging the same powerful attention mechanisms. Different from CNNs, which process images by applying convolutional filters to local regions, ViTs treat images as sequences of fixed-size patches, embedding each patch into a vector and feeding these vectors into a standard transformer model (Figure 1-5 (f)). The self-attention mechanism within transformers allows ViTs to capture long-range dependencies and global context more effectively than CNNs, which are typically limited to local receptive fields. This global context modeling is one of the primary advantages of ViTs, enabling them to achieve SOTA performance on various image classification benchmarks. Additionally, ViTs benefit from the scalability and parallelization advantages inherent in transformer architectures, allowing for efficient training on large datasets.

However, ViTs also come with some disadvantages compared to CNNs. ViTs generally require a larger amount of training data to achieve comparable performance, as they do not have the built-in inductive biases of locality and translation invariance present in CNNs. This makes ViTs more data-hungry and potentially less effective on smaller datasets. Furthermore, the computational and memory requirements for ViTs can be significantly higher, especially for high-resolution images, due to the quadratic complexity of the self-attention mechanism. Despite these challenges, ViTs represent a promising direction in computer vision research, leveraging the strengths of transformer architectures to push the boundaries of image understanding.

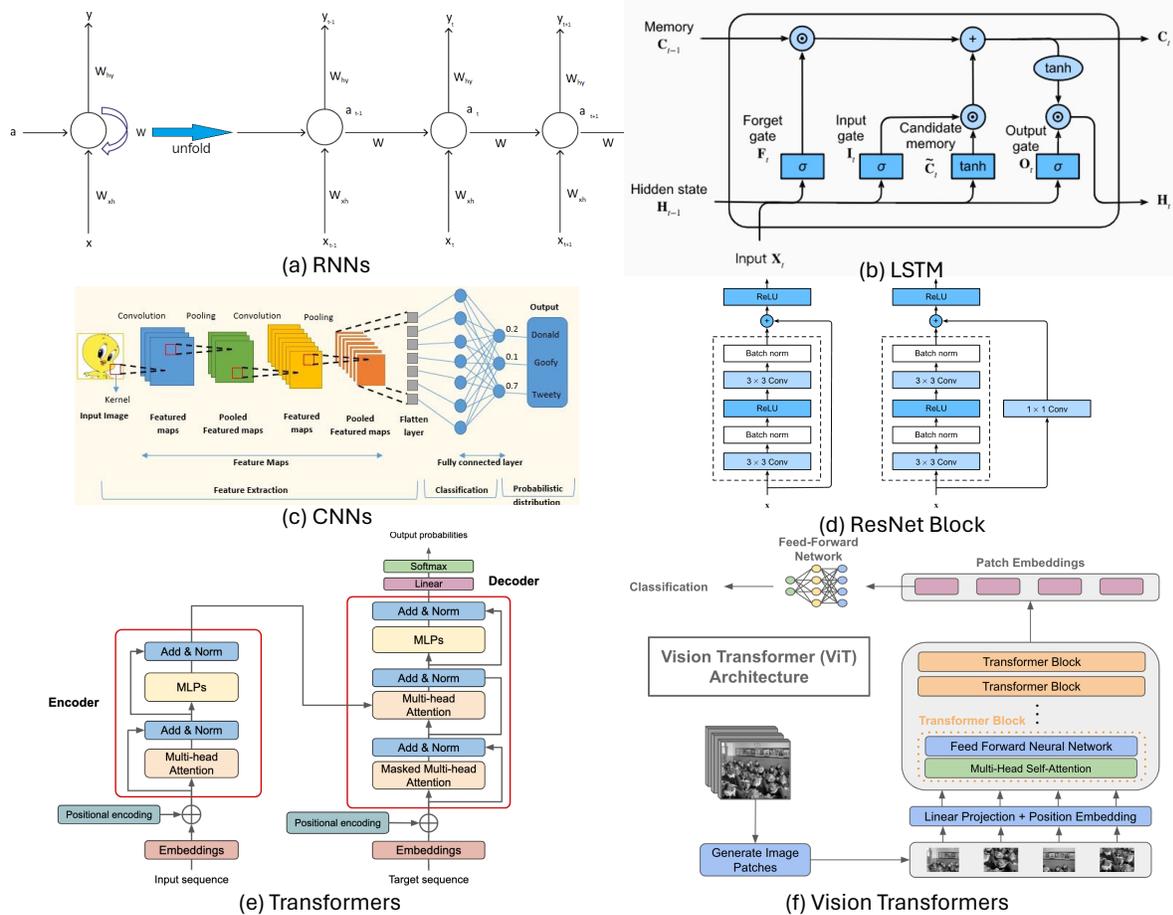

*Figure 1-5 Diagrams of six types of general NN architecture*

### 1.3.2 Application of Artificial Intelligence in Radiation Oncology

As shown in Figure 1-6, the workflow of radiotherapy can be divided into six steps: 1) treatment decision-making; 2) imaging and simulation; 3) treatment preparation and planning; 4) plan approval quality assurance; 5) treatment delivery; 6) follow-up care.

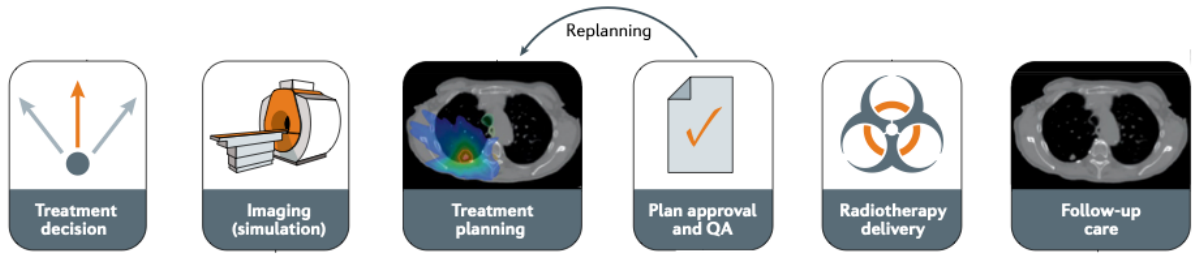

*Figure 1-6 Radiotherapy workflow*

Overall, AI applications in the radiotherapy workflow have penetrated each step. Firstly, AI can assist in decision-making (diagnosis and prescribing dose) by processing the patient's multi-modality data (various imaging and electronic medical records) at the treatment decision step[37,38]. Secondly, AI can facilitate the automatic tumor and organs-at-risk (OARs) segmentation[39–41] and treatment plan beam optimization[42]. Thirdly, AI can be applied in plan check and quality assurance processes to automate and augment the manual processes that are tedious and inconsistent[43]. Fourthly, AI algorithms can help analyze longitudinal imaging features for tumor response prediction and treatment adaptation[44,45]. Lastly, AI algorithms can enhance the treatment setup and delivery process (setup/delivery imaging reconstruction[16]). For instance, AI algorithm can be applied in adaptive, image-guided, and motion-controlled treatments by accelerating the image acquisition speed[46], improving the image quality to provide robust guidance[47], and automatically detecting treatment target on the in-treatment images to guide the adjustment of beam angles[40,48,49].

## 1.4 Computed Tomography

### *1.4.1 Computed Tomography Application in Radiation Therapy*

CT is an advanced medical imaging technique that utilizes X-ray equipment to create detailed cross-sectional images of the body's internal structures. This method involves rotating an X-ray source and detectors around the patient, capturing densely sampled projections covering a large angle that meets the data sufficient condition for reconstruction. The projections are then concatenated into a continuous matrix, termed sinogram, for reconstruction using filtered backprojection or constrained optimization methods[50]. CT scans offer superior resolution and improved tissue contrast compared to conventional X-rays.

As mentioned, CT is foundational to modern radiotherapy. Besides its function for tumor diagnosis and target definition, CT supports accurate 3D dose calculation, optimization, and quantitative plan evaluation[51].

For scanning radiotherapy planning CT, several restrictions needed to enforce to ensure effective dose calculation. 1) The setup of the patient during CT exam must match the setup of the following radiotherapy. 2) The body of the patient should be in treatment configuration. 3) The CT field of view must include the entire patient anatomy. 4) For radiotherapy options not accounting for motion, a time-sorted CT acquisition is needed to account for the patient's breathing movement during the treatment cycle.

Figure 1-7 illustrates the details of CT-aided radiotherapy treatment planning workflow. For the latter, various CT modalities, such as CT on rails (CTOA), helical megavoltage CT (HMVCT), and portal imaging - mega- and kilovoltage CBCT (MVCBCT and KVCBCT) - can be employed to provide in-room (i.e., inside radiotherapy suite) imaging guidance (Figure 1-8). The scanned volumetric images can provide online verification of tumor location. Specifically, CTOA is implemented by installing a conventional CT scanner in the treatment rooms, while helical MVCT integrates a LINAC and CT scanner with a setup similar to conventional CT. A pre-treatment CT scan is conducted in both imaging modalities once the patient is at the treatment position. Then, the pre-treatment CT images are registered with the planning CT for treatment verification[52]. The typical image and registration time for CTOA is 7-8 minutes[53]. The lengthy time is considered an impediment to CTOA being used more broadly in radiotherapy. On-board CBCT allows the images to be taken at the same isocenter position and eliminates the time to transfer the patients back and forth between the treatment and imaging isocenters. There are two types of CBCTs. MVCBCT uses the same MV source for treatment and the electronic portal imaging device, while KVCBCT adds an additional imaging axis with kV X-ray tube and flat panel detector. The pre-treatment CBCT images are then fused with planning CT for treatment alignments. The time needed for CBCT imaging read-out plus fusion is 1-3 minutes[52]. CBCT is faster than CTOA but has degraded image quality and Honsfield Unit (HU) value fidelity, which has been partially mitigated with advanced reconstruction and image processing[54–58]. CBCT does not require the additional space for a separate imaging gantry and is easily integrated into the existing radiotherapy workflow. As a result, CBCT has become the most popular IGRT solution in modern radiotherapy. Nevertheless, the speed of CBCT is insufficient to capture dynamic processes for real-time treatment adaptation[59].

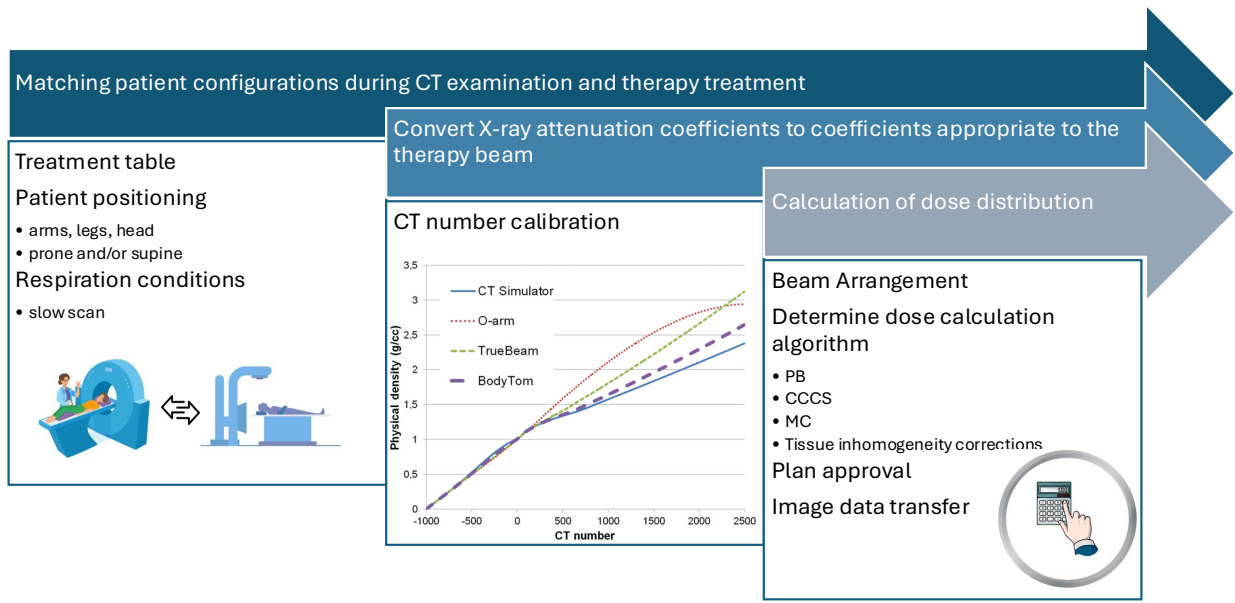

*Figure 1-7 CT-aided radiotherapy planning pipeline*

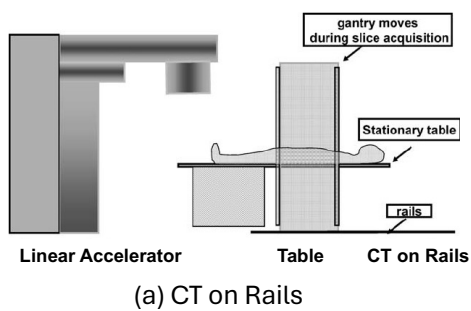
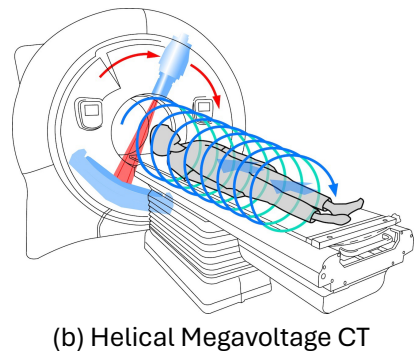

(a) CT on Rails　　　　　　　　　　　　(b) Helical Megavoltage CT

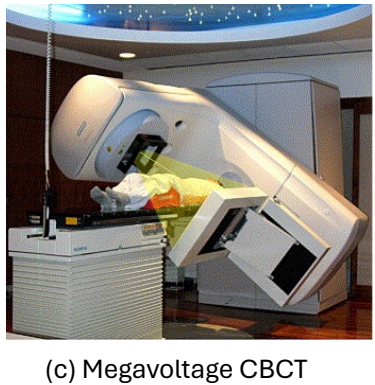
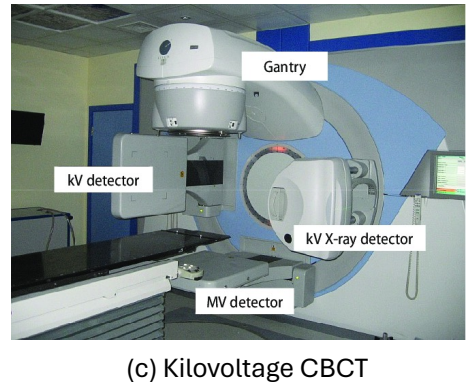

(c) Megavoltage CBCT　　　　　　　　　(c) Kilovoltage CBCT

*Figure 1-8 Illustration on four types of in-room CT guided IGRT.*

### 1.4.2　Computed Tomography Reconstruction

CT reconstruction has been conventionally performed using filtered back-projection (FBP). Sufficient projection sampling conditions, e.g., Tuy's data sufficient condition[60], are required for mathematical rigorous CT reconstruction. Violating Tuy's condition leads to geometry and intensity distortion in the reconstructed images (referred to as limited-angle artifacts in the following). Besides the sampling trajectory, a minimal sampling density is required to avoid streak artifacts that can severely corrupt the image with sparse samplings, e.g., <100 views.

On the other hand, data-sufficient conditions may not always be met due to practical limitations, including imaging dose considerations, limited gantry freedom, and the need for continuous image guidance for radiotherapy[61,62]. For instance, the total ionizing radiation exposure in mammograms is kept low to protect the sensitive tissue[63]. However, 2D mammograms without depth differentiation can be inadequate with dense breast tissues. Digital breast tomosynthesis (DBT), a limited-angle tomographic breast imaging technique, was introduced to overcome the problem of tissue superposition in 2D mammography while maintaining a low dose level. In DBT, limited projection views are acquired while the X-ray source traverses along a predefined trajectory, typically an arc spanning an angular range of 60° or less. The acquired limited angle samplings are then reconstructed as the volumetric representation with improved depth differentiation but still inferior quality to CT[64,65]. In different applications, the acquisition angles are not restricted. Still, the density of projection is reduced to lower the imaging dose[66,67] or to capture the dynamic information in retrospectively sorted 4DCT[68,69], resulting in sparse views in each sorting bin.

Over the past few decades, substantial effort has been made to advance the development of sparse-view CT reconstruction from two general avenues. One line of research lies in developing regularized iterative methods based on the compressed sensing (CS) [70] theory. For instance, Sidky et al. proposed the adaptive steepest descent projection onto convex sets (ASD-POCS) method by minimizing the expected CT volume's total variation (TV) from sparsely sampled views[71]. Following that, the adaptive-weighted TV (awTV) model was introduced by Liu et al. for improved edge preservation with local information constraints[72], while an improved TV-based algorithm named TV-stokes-projection onto Convex sets (TVS-POCS) was proposed immediately after to eliminate the patchy artifacts and preserving more subtle structures[73]. Apart from the TV-based methods, the prior image-constrained CS (PICCS)[74], patch-based nonlocal means (NLM)[75], tight wavelet frames[76], and feature dictionary learning[77] algorithms were introduced to further improve the reconstruction performance in representing patch-wise structure features. More recently, deep learning (DL) techniques were explored for improved CT reconstruction quality in the image or sinogram domain. The image domain methods learn the mapping from the noisy sparse-view reconstructed CT to the corresponding high-quality CT using diverse network structures such as feed-forward network[78,79], U-Net[80], and ResNet[81]. The sinogram domain methods work on improving/mapping the FBP algorithm[82–85] or interpolating the missing information in the sparse-sampled sinograms[86–89] with DL techniques. These and other deep learning-based sparse view CT reconstruction studies are comprehensively reviewed in Podgorsak et al. and Sun et al.[90,91].

*Table 1-2 Summary of CT reconstruction Algorithms*

|  | **Analytical Methods** | **CS Methods** | **DL Methods** |
|---|---|---|---|
| **Examples** | FBP/ART/SART | ASD-POC[71]/awTV[72]/ TVS-POCS[73]/PICCs[74]/NLM[75] | Image domain methods (Noisy to HD image mapping)/Sinogram domain methods (neural net (NN) enhanced FBP/ NN interpolation on sparse-sampled sinograms) |
| **Comments** | Suffers from limited angle artifacts that worsen with sparser projection | High-quality image reconstruction from ultra-sparse projection referencing is unachievable. | More efficient than CS. But high-quality image reconstruction from ultra-sparse projection referencing is still unachievable. |

### 1.4.3   Computed Tomography Multi-material Decomposition

The contrast of CT is determined by a combination of X-ray material interactions, including photoelectric and Compton, while the CT number primarily reflects electron density with a component dependent on the material composition. CT numbers are degenerate because multiple materials can produce similar HU values, leading to ambiguities in tissue differentiation. This degeneracy arises due to the underlying X-ray interactions and their dependence on electron density and material composition. Conventional CT is obtained using a single source detector without explicit energy separation in the single energy CT (SECT) mode[92]. However, the information preserved in SECT is inadequate to resolve the individual subcomponents contributing to the CT number, where inaccuracies in the conversion of CT HU to electron density/stopping power ratio is a main source of uncertainty for proton therapy. In pursuing better material differentiation and contrast, dual-energy CT (DECT) was invented[93] to acquire attenuation information with different energies or energy spectra. Several different mechanisms have been used to acquire DECT, including X-ray tubes operating at different voltages[94], filters to modify the X-ray spectra[92], detectors with different sensitivities to X-ray energies[95,96], and photon-counting detectors[97]. DECT has shown promise in various clinical applications, including iodine mapping[98], gout classification[99], fat and essential trace metal quantification[100,101], bone separation [102], and virtual unenhanced and monochromatic image generation[103].

While DECT offers additional information for multi-material decomposition (MMD), the problem itself is still non-linear and ill-posed, particularly for separating > 2 materials[104,105]. Additional priors of the materials are often assumed to achieve acceptable decomposition results. Existing MMD ( ≥ 2 ) typically follows two approaches: theory-integrated methods that conduct decomposition with the assistance of a physical model[106], or analytical methods that perform direct attenuation decomposition in dual-energy (DE) image or projection domain[106,107]. Among them, the theory-integrated algorithms and analytical methods directly process projections and are thus robust to beam hardening artifacts, which introduce non-linearity into the attenuation measurements, making the decomposition problem (especially >2 MMD) more challenging.[106,107]. Theory-integrated techniques requiring iterative operations of forward and backward projections are extremely computationally expensive[106]. Projection-domain MMD has been conditioned for 2 basis materials[107] that are inadequate for more complex patient MMD problems. Though analytical methods based on image domain have been demonstrated for >2 material decomposition, the decomposed results highly depend on strong priors, which limit its generalizability, and are susceptible to image quality[108–111].

In the past decade, as new annotated data and powerful computational resources emerged, deep neural networks (DNN) started to realize their potential for learning complex relationships and incorporating existing knowledge into the inference model[112]. Multiple studies have demonstrated the advantage of DNN-inspired DECT reconstruction models for fast MMD while suppressing noise and artifacts[113]. For instance, Zhang et al. proposed a butterfly network to acquire image-domain decomposition of two basis materials[114]. A visual geometry group (VGG) network with an enlarged receptive field was trained by Chen et al. to perform MMD on five simulated basis materials[115]. Badea et al. deployed U-Net for the tasks of prediction in the photoelectric effect, Compton scattering, and material density maps[116]. Nevertheless, these algorithms performed MMD in the image domain suffered from information loss caused by converting DE transmission to image and were suboptimal with the data-driven nature of DL methods. They were limited in learning efficiency and MMD performance for challenging problems, including the separation of sparse structures such as calcification from noise and material boundary definition between tissues with similar atomic compositions.

Lately, Zhu et al. performed DNN-based MMD of more than two basis materials directly from the dual energy projections[117]. Specifically, they introduced a framework termed Triple-CBCT, which consisted of a cascade of three sub-networks: 1) SD-Net - a seven-layer convolutional neural network (CNN) with sparse skip-connection and mixed sizes of the convolutional filter designed to generate the sinograms of numerous virtual monochromatic images (VMI) from the domain of DE transmission; 2) Domain transform Net - a fix-parameter predictor with the integration of standard Feldkamp-Davis-Kress image reconstruction algorithm[118] to map sinogram signals to image domain; 3) ID-Net - an eight-layer CNN with dense inter-layer skip-connections and 3x3 convolutional filter for denoising image, suppressing artifact- and grouping multiple VMIs generated from 2) to the final base material images. Triple-CBCT shows the feasibility of performing >2 MMD from DE projections. Yet, the downsides of Triple-CBCT are its excessive and uninterpretable network architectures and heavy training parameters.

## 1.5     Magnetic Resonance Imaging

### *1.5.1    Magnetic Resonance Imaging Application in Radiation Therapy*

Unlike CT, magnetic resonance imaging (MRI) can produce different image contrasts depending on the specific sequences used for different imaging purposes. For instance, T1-weighted images are sensitive to the longitudinal relaxation time of tissues. They are particularly useful for visualizing the anatomy, providing high contrast between soft tissues. T1 weighted images are often used with Gd contrast for perfusion measurements. T2-weighted images, on the other hand, are sensitive to transverse relaxation time and are suited for detecting fluid and pathological changes, such as tumors, cysts, abscesses, and areas of inflammation.

Conventionally, CT was the main imaging tool used for treatment planning and dose delivery. However, considerable efforts have been devoted to integrating MRI into the stage of RT planning as well as treatment monitoring, termed MRI-guided radiation therapy (MRgRT), owing to its superior soft-tissue contrast, better organ motion visualization, and capability of providing tumor and tissue physiological change monitoring. In particular, offline MRgRT, using MRI at the stage of treatment planning, has already been applied in clinical practice.  4D MRI, which is a respiratory-resolved volumetric imaging technique, is especially powerful in quantifying tumor motion[21,22,119]. For instance, in a clinical planning workflow for free-breathing liver radiotherapy treatments, the liver tumor delineated in individual 4D MR images forms an internal target volume (ITV). Under or over-estimating ITV can cause tumor underdose or normal tissue injury during radiation. MRI-integrated LINACs have recently entered clinical practice, facilitating in-treatment monitoring and guided beam adaptation. Unique opportunities are offered by online MRgRT for real-time inter/intra-fraction adaptive dose delivery that continuously tracks the anatomy and modifies the radiation delivery accordingly and thus can offset continuous respiratory motion in the lung, kidneys, liver, and pancreas regions, as well as the sporadic motion in the head and neck, prostate and rectum without fiducial implantation. However, being a relatively slow modality, MRI acquisition often has to compromise among acquisition time, resolution, and image quality. Currently, real-time MRI imaging achieved on MR linacs is limited to two-dimensional (2D), unable to capture through-plane motion and the sophisticated 3D organ and tumor deformation information. Alternatively, 4D MRI can be acquired. K-space data are placed in separate motion bins according to their respiratory phases, and then reconstructed individually. Since 4D MRI collects data over multiple breathing cycles (up to 100 cycles), it reflects a composite view of a patient's respiratory motion but not an instantaneous motion state. 2D cine MRI and 4D MRI have already utilized acceleration techniques, including parallel imaging, view sharing, and compressed sensing. Yet, ~50X acceleration is still needed to achieve real-time 3D cine image acquisition, which can provide real-time volumetric motion tracking, superior soft-tissue contrast, and enhanced online treatment adaptation, leading to more precise, personalized, and effective radiation therapy for mobile tumors but will require novel k-space data acquisition and reconstruction approaches.

## *1.5.2 Magnetic Resonance Imaging Reconstruction*

The inverse fast Fourier transform (FFT) algorithm has served the MR community very well as the conventional state-of-the-art (SOTA) image reconstruction methods from k-space domain with fully Cartesian grid signal sampling (Figure 1-9 (a)). For more efficient k-space acquisition, non-Cartesian sampling trajectories (Figure 1-9 (b-c)), such as radial trajectories and spiral trajectories, were explored to oversample the central region of k-space (low-frequency information/direct current component; general shapes and structures) to mitigate motion artifact. For densely sampled non-Cartesian k-space signals, signal re-gridding with appropriate density compensation factors followed by inverse FFT is still robust and effective[120].

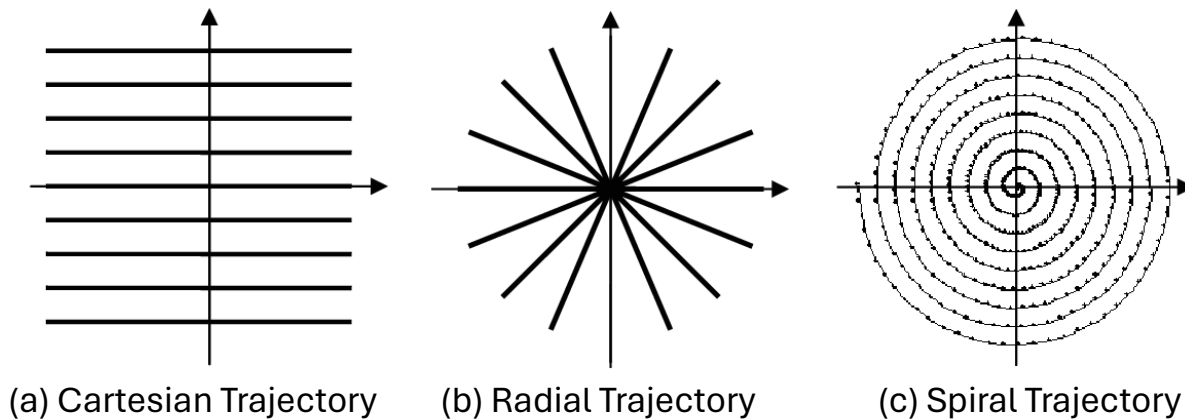

(a) Cartesian Trajectory　　(b) Radial Trajectory　　(c) Spiral Trajectory

*Figure 1-9 Representation of k-space signal acquisition trajectories*

However, to expedite the MRI imaging (mainly 4D MRI) speed in radiotherapy, undersampling of k-space signals was extensively investigated, where inverse FFT fails to reconstruct images with satisfactory quality. Earlier on, the main direction for MR acceleration was rooted in the development of ultrafast imaging[121], parallel imaging[122,123] and massive coil arrays[124], which have increased scanning speed by several orders of magnitude. Followed by that, methods like view sharing[125,126] and reduced field-of-view (FOV)[127,128] emerged which enable acceleration through under-sampling k-space measurements along the spatial or spatial-temporal dimensions to reduce signal redundancy. Since only a fraction of raw information is acquired, those methods heavily rely on effective and efficient reconstruction methods to recover qualitative images from under-sampled k-space data in a timely manner[129,130].

In the past decade, CS approaches have heavily dominated in the field of 4D MRI reconstruction under the premise that CS theoretically guarantees a satisfactory recovery of specified signals from fewer measurements than the quantity calculated from the Nyquist limit[131]. Most existing CS-based frameworks includes prior information, such as sparsity in transformed domains[132], total variation penalties[133], low-rank property[134] or a combination of several constraints[135] into their optimization schemes to stabilize the solution. For cohesion in the time-axis, some studies also considered combining motion estimation/compensation (ME/MC) in 4D MRI reconstruction[136,137]. For example, Zhao et al. proposed to integrate an intensity-based optical flow prior into the traditional CS mechanism so that 4D MRI sequence recovery and MC through the optical flow estimated motion field can be jointly constructed[136]. Nevertheless, CS embedded algorithms mostly are offline approaches due to long processing time and rely on case-wise heuristic hyper-parameter tuning, which largely deviates from the clinical need of a real-time, automatic and generalizable solution[138].

Recent DL advances have offered a data-driven approach for 4D MRI reconstruction. In contrast to the model-based CS reconstruction, DL learns reconstruction mapping from the rich information in the training data representation, matching or exceeding the CS quality and is significantly faster[91,139,140]. Previous works have explored 4D MRI reconstruction using customized convolutional neural networks (CNNs), recurrent neural networks (RNNs), and Transformers. For instance, Schlemper et al. proposed integrating K nearest neighbor (KNN) enabled temporal data-sharing mechanism into a cascade 3D CNN architecture for 4D MRI reconstruction trained with L2 loss objective[139]. Their network demonstrates the capability of 2D frame recovery but is suboptimal in capturing complex dynamic relationships. Adapting from the UNet architecture, Dracula[141] and Moivenet[142] proposed methods to accelerate 4D MR reconstruction. However, there is still room for improvement in the reduction of reconstruction time (Dracula at 28 s and Moivenet at 0.69 s). Moreover, Huang et al. introduced a motion-guided framework using RNN-inspired Conv-GRU for initial 2D frame reconstruction and U-FlowNet for motion estimation in the optical flow field. The overall architecture is trained with regularized L1 loss[143]. Their pipeline reconstructed a cardiac dataset 5 and 8 times accelerated (5x and 8x), but detail loss was evident at a high acceleration ratio. Additionally, their RNN-based architecture confines its tensor processing to be sequentially frame-by-frame and takes ~5s for inference of a volume, which limits its application for MRI-guided real-time interventions[144].

*Table 1-3 Summary of MRI reconstruction Algorithms*

|  | **Analytical Methods** | **CS Methods** | **DL Methods** |
|---|---|---|---|
| **Examples** | Inverse FFT | Various prior information-based constraints are included in the optimization objectives. | Draculad[141]/MovieNet[142]/U-FlowNet[143] |
| **Comments** | Cannot directly apply on under-sampled (non-Cartesian) k-space data. | • Not generalizable<br>• Ultra-sparse undersampling unachievable. | • More efficient than CS.<br>• But ultra-sparse undersampling is not robust. |

## 1.6 Overview

The following contents will be organized as following. Chapter 2 describes a novel method proposed for ultra-sparse view CT reconstruction using generative radiance field (GRAF). Section 2.1 introduces the motivation of the proposed algorithm, section 2.2 elaborates on the method and theory of the proposed algorithm, section 2.3 describes the results from comprehensive experiments and section 2.4 as well as 2.5 discusses and summarizes the proposed methods. It is a version of the manuscript titled "TomoGRAF: A Robust and Generalizable Reconstruction Network for Single-View Computed Tomography" under review in the Machine Learning with Applications journal[145].

Chapter 3 illustrates a novel method proposed for projection domain multi-material decomposition for dual-energy CT using DL approach. Section 3.1 overviews the background of this problem, section 3.2 elaborates on the methods, data cohort and benchmarks of the proposed methods, section 3.3 presents the results of the evaluation and section 3.4 as well as 3.5 discusses and conclude the proposed algorithm. It is a version of the manuscript tilted "A Two-Step Framework for Multi-Material Decomposition of Dual Energy Computed Tomography from Projection Domain" published in the EMBC 2024 conference proceedings[146].

Chapter 4 presents a series of DL based accelerated 4D MRI reconstruction algorithms for enhanced imaging quality in guiding radiotherapy treatment planning or delivery. In specific, section 4.1 introduces a transformer-based reconstruction algorithm, and the method is evaluated on dynamic lung MRI sequence. It is a version of the manuscript titled "Learning Dynamic MRI Reconstruction with Convolutional Network Assisted Reconstruction Swin Transformer" published in the MICCAI 2023 conference proceedings[147]. Section 4.2 introduces a GAN-based reconstruction algorithm for 4D liver MRI reconstruction. It is a version of the manuscript titled "Paired conditional generative adversarial network for highly accelerated liver 4D MRI" published in the Physics in Medicine and Biology journal[148]. Section 3.3 introduces a Diffusion-based reconstruction algorithm extremely accelerated 4D liver MRI reconstruction. It is a version of the manuscript titled "Rapid Reconstruction of Extremely Accelerated Liver 4D MRI via Chained Iterative Refinement" accepted by the SPIE Medical Imaging 2025 Proceedings[149].

# 2. Ultra-sparse View CT Reconstruction

## 2.1 Introduction

Image reconstruction from sparse-view and limited-angle samplings is an ill-posed inverse problem. Yet, little progress was made in reconstructing high-quality tomographic imaging using less than ten projections, a practical problem in real-time radiotherapy or interventional procedures. For the former, an onboard X-ray imager orthogonal to the mega-voltage (MV) therapeutic X-ray provides the most common modality of IGRT. However, a trade-off must be made between slow 3D cone beam CT (CBCT) and fast 2D X-rays[150,151]. A similar trade-off exists in interventional radiology[152]. When real-time interventional decisions need to be made in the time frame affording one to two 2D X-rays, yet 3D visualization of the anatomy is desired, a unique class of ultra-sparse view CT reconstruction problems combining extremely limited projection angles *and* sparsity is created.

With the advancement of deep learning and more powerful computation hardware, several recent studies proposed harnessing inversion priors through training data-driven networks for single/dual-view(s) image reconstruction. Specifically, Shen et al. designed a three-stage CNN trained on patient-specific 4DCT to infer CT of a different respiratory phase using a single or two orthogonal view(s)[153]. Ying et al. built a generative adversarial network framework (X2CT-GAN) with a 2D to 3D CNN generator to predict tomographic volume from two orthogonal projections[154]. Though promising, their generalization and robustness to external datasets have not been demonstrated and may be fundamentally limited by two factors: First, they generate volumetric predictions purely from 2D manifold learning. As a result, these networks are incapable of comprehending the 3D world and the projection view formation process[155]. Second, a prerequisite for deep networks with complex enough parameters to implicitly represent 2D to 3D manifold mapping is large and diverse training data, a condition difficult to meet for medical imaging[156].

An effective perspective to mitigate those problems is to leverage intermediate voxel-based representation in combination with differentiable volume rendering for a 3D-aware model, which requires smaller data to generalize. The Neural Radiance Fields (NeRF)[157] model successfully implemented this principle for volumetric scene rendering. NeRF proposed synthesizing novel views of complex scenes by optimizing an underlying continuous volumetric scene function using a sparse set of input views. NeRF achieved this by representing a scene with a fully connected deep network with the input of 5D coordinates representing the spatial location, view direction, and the output of the volume density and view-dependent emitted radiance at the spatial location. The novel view was synthesized by querying 5D coordinates along the camera rays and using volume rendering techniques to project the output color and densities onto an image[157]. NeRF was designed to generate unseen views from the same object and typically required fixed camera positions as supervision. As an improvement, GRAF, a 3D-aware generative model 2D-supervised by unposed image patches, introduced a conditional radiance field generator trained within the Generative Adversarial Network (GAN) framework[158] that is capable of rendering views of novel objects from given sparse projection views[155].

The success of NeRF and GRAF motivated their applications to solve the 3D tomography problems. MedNeRF[159] was proposed by Corona-Figueroa et al. for novel view rendering from a few or single X-ray projections. MedNeRF inherits the general GRAF framework, remains 2D-supervised, and assumes visible-light photon transportation configuration in the generator with an addition of self-supervised loss to the discriminator. However, there are distinct differences between CT volume reconstruction and "natural object" 3D representation rendering in terms of the available choices of training supervision, imaging setup, and properties of the rays. Specifically, 3D training supervision (object mesh with information on surface color) is often hard to acquire for "natural objects." In contrast, existing CT scans are an ideal volumetric training ground truth (GT) for fitting a 3D tomographic representation learning model. Moreover, optical raytracing works by computing a path from an imaginary camera (eye) through each pixel in a virtual screen and calculating the color of the object visible through the virtual screen via simulating ray reflection, shading, or refraction on the object surface (Figure 2-1 (b)). The solving target of optical ray tracing is the object surface color $(r, g, b)$ and density $\sigma$ in a 3D location $(x, y, z)$. Meanwhile, X-rays are transported from the focal spot through an object to the detector plane, accounting for scattering and attenuation (Figure 2-1 (c)). The goal of CT reconstruction is voxel-wise material densities $\delta$ at a 3D location $(x, y, z)$. Because of these major differences between natural scenes and 3D medical images, the direct application of NeRF has not resulted in usable CT with ultra-sparse views.

To overcome the challenges in ultra-sparse view CT reconstruction while maintaining the superior NeRF 3D structure representation efficiency, we introduced an X-ray-aware tomographic volume generator, termed TomoGRAF, to simulate CT imaging setup and use CT and its projections for 3D- and 2D-supervised training. TomoGRAF is further enhanced with a GAN framework and computationally scaled with sub-volume and image patch GTs training. To the best of our knowledge, this is the first pipeline that informs the NeRF simulator with X-ray physics to achieve generalizable high-performing CT volume reconstruction with ultra-sparse projection representation.

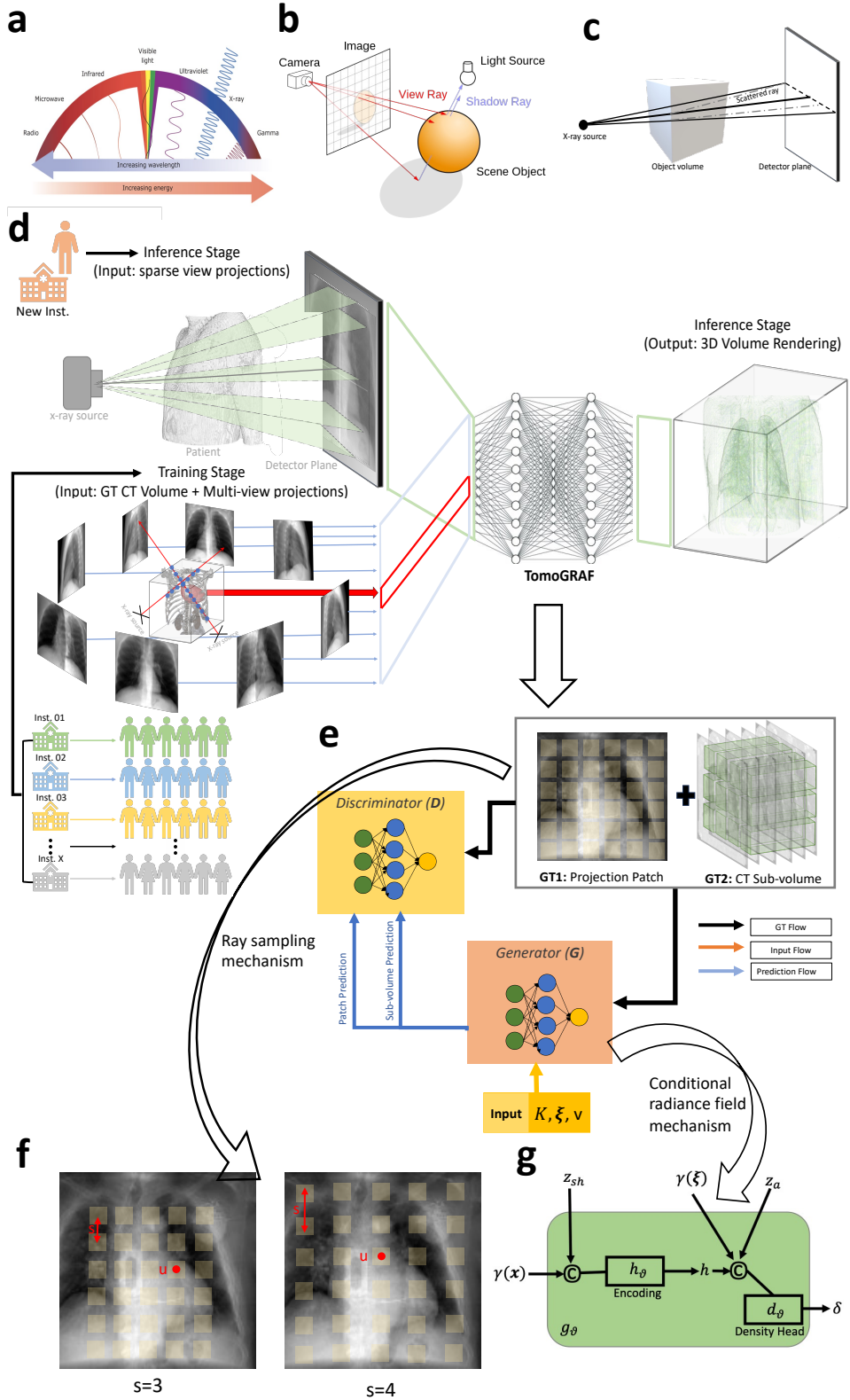

*Figure 2-1 TomoGRAF Framework image panels.*

## 2.2 Methods

### 2.2.1 Problem Formulation

As illustrated in Figure 2-1 (d), we formulated the problem of 3D image reconstruction from 2D projection(s) into a GAN-based DL framework, including modules of generator G and discriminator equation, given a series of 2D projections $X$ denoted as $\{X_1, X_2, \cdots, X_N\}$ where $X_i \in \mathbb{R}^2 = \mathbb{R}^{H \times W}$ for $i \in [1, N]$, $N$ is the number of available 2D projections, $H$ and $W$ is the projection height and width. Our modeling target was to form a $G$ that can predict the 3D volume $\hat{Y} \in \mathbb{R}^3 = \mathbb{R}^{D \times H \times W}$ ($D$ represents volume depth) where $G$ is supervised by $X$ and 3D volume GT $Y \in \mathbb{R}^3 = \mathbb{R}^{D \times H \times W}$, and is penalized by $D$ to encourage optimal convergence.

### 2.2.2 TomoGRAF Framework

TomoGRAF does not aim to optimize densely posed projections for rendering a single patient volume. Instead, it targets fitting a network for synthesizing new patient volume by learning on various unposed projections. Note that the generator and discriminator work on image patches and sub-volumes during training for better efficiency, whereas a complete patient volume is rendered at inference time. The detailed components in TomoGRAF are shown in Figure 2-1 (e, f, g). In what follows, we elaborate on the model architecture.

### 2.2.3 Generator

Adapted from GRAF[155], Our generator consists of three main components: ray sampling, conditional generative radiance field, and projection rendering. Ray sampling modules render the X-ray paths in 3D that are associated with truth patch/sub-volume, and the conditional generative radiance field module predicts the material density from a 3D location along the rendered X-ray paths. Lastly, the projection rendering module obtains the 2D composition from the predicted volumetric material densities. Overall, the generator $G$ takes X-ray source setup matrix $K$, view direction (pose) $\xi = (\theta, \phi)$, 2D sampling pattern $v$ and shape, and appearance codes $z_{sh} \in \mathbb{R}^{M_s}$ and $z_a \in \mathbb{R}^{M_a}$ as input, and predicts a size $D \times M \times M$ CT sub-volume $V' \in \mathbb{R}^3 = \mathbb{R}^{D \times M \times M}$ as well as the associated size $M \times M$ CT projection patch $P' \in \mathbb{R}^2 = \mathbb{R}^{M \times M}$ ($M$ is a hyper-parameter defined by user; $M = 32$ is used in our experiments). $K$ consists of $d = (d_1, d_2)$, $d_1$ is distance of source to patient, $d_2$ is distance of source to detector. The detector resolution $S \in \mathbb{R}^2$ and $M$ is bounded by $(H, W)$.

**Ray Sampling**: The rendered rays $r \in \mathbb{R}^3$ are constrained by $\xi$, $K$ and $P'$. The X-ray pose $\xi$ is sampled from a predefined pose distribution $p_\xi$ collected from projection angles in training data with the X-ray source facing towards the origin of the coordinate system all the time. As shown in Figure 2-1 (f), $v = (u, s)$ determines the center $u = (x', y') \in \mathbb{R}^2$ and scales $s \in \mathbb{R}^+$ of $P'$ that we target to predict, where $V'$ is formed with all the corresponding 3D coordinates that form $P'$. At the training stage, $u$ and $s$ are uniformly drawn from $u \sim U(\Omega)$ and $s \sim U[1, S]$ where $\Omega$ defined the 2D projection domain and $S = \min(H, W)/M$. Noteworthily, the coordinates in $P'$ and $V'$ are real numbers for the purpose of continuous radiance field evaluation. Following the stratified sampling approach in Mildenhall et al.[157], $Q$ number of points are sampled along each $r$. The number of rays $R = M \times M$ and R= $H \times W$ per training and inference, respectively.

**Conditional Generative Radiance Field**: Adapted from GRAF[155], the radiance field is represented by a deep fully connected coordinate network $g_\vartheta$ with the input of the positional encoding of a 7D vector $(x, y, z, \theta, \phi, z_{sh}, z_a)$ consisting of 3D location $x = (x, y, z)$ and projection pose $\xi$. The output is the material density $\delta$ in the corresponding $x$, where $\vartheta$ represents the network parameters and $z_{sh} \sim p_{sh}$ and $z_a \sim p_a$ with $p_{sh}$ and $p_a$ drawn from standard Gaussian distribution.

$$g_\vartheta: \mathbb{R}^{\mathcal{L}_x} \times \mathbb{R}^{\mathcal{L}_\xi} \times \mathbb{R}^{M_{sh}} \times \mathbb{R}^{M_a} \to \mathbb{R}^3 \times \mathbb{R}^+ \tag{2-1}$$

$$(\gamma(x), \gamma(\xi), z_{sh}, z_a) \to \delta \tag{2-2}$$

Where $\mathcal{L}_x$ and $\mathcal{L}_\xi$ represent the latent codes of $x$ and $\xi$, $M_{sh}$ and $M_a$ define the shape and appearance codes with $z_{sh} \in \mathbb{R}^{M_{sh}}$ and $z_a \in \mathbb{R}^{M_a}$, and $\gamma(\cdot)$ represents positional encoding.

The architecture of $g_\vartheta$ is visualized in Figure 2-1 (g) with Equation (2-3~2-6). First, shape encoding $h_\vartheta$ is conducted with the input of $\gamma(x)$ and $z_{sh}$. Second, $h_\vartheta$ is concatenated with $\gamma(\xi)$ and $z_a$ and then sent to the density head $d_\vartheta$ to predict $\delta$.

$$h_\vartheta: \mathbb{R}^{\mathcal{L}_x} \times \mathbb{R}^{M_{sh}} \to \mathbb{R}^H \tag{2-3}$$

$$(\gamma(x), z_{sh}) \to h \tag{2-4}$$

$$d_\vartheta: R^H \times \mathbb{R}^{\mathcal{L}_\xi} \times \mathbb{R}^{M_a} \to \mathbb{R}^1 \tag{2-5}$$

$$(h, \gamma(\xi), z_a) \to \delta \tag{2-6}$$

where the encoding was implemented with a fully connected network with ReLU activation.

**Projection Rendering**: Lastly, given the material density $\{\delta_i^r\} = V' \in \mathbb{R}^3$ where $1 \leq i \leq Q$ of all points along the rays $\{r_j\}$ where $1 \leq j \leq M \times M$ in training, we used a CT projection algorithm[160] to synthetic 2D radiograph patch $P'$ given preset $K$ and $\xi$.

### 2.2.4 Discriminator

Following the discriminator architecture defined in MedNeRF[159], two self-supervised auto-encoded CNN discriminators,[41] $D_{1,\emptyset}$ and $D_{2,\emptyset}$ compare predicted sub-volume $V'$ to real sub-volume $V$ extracted from real volume $Y$ and predict projection patch $P'$ to real projection patch $P$ extracted from real projection $I$, respectively. $D_{1,\emptyset}$ is defined to convolve in 3D while $D_{2,\emptyset}$ is defined to convolve in 2D to align with the dimension of its discrimination targets. For extracting $V$ and $P$, we first extracted $P$ from a real projection $I$ given $\boldsymbol{v}$ and $s$ randomly drawn from their corresponding distributions, and then located the coordinates of $V$ from $Y$ based on $\boldsymbol{\xi}$ and $\boldsymbol{K}$. $D_{1,\emptyset}$ and $D_{2,\emptyset}$ are backpropagated separately with their respective weights, while we defined them to share weights while discriminating different sub-volume/patch locations.

### 2.2.5　Loss function

**Training stage**: Our loss objective consists of discrimination towards patch as well as sub-volume predictions. First, the global structures in intermediate decoded patches of $D_{1,\emptyset}$ and $D_{2,\emptyset}$ were separately assessed by Learned Perceptual Image Patch Similarity (LPIPS)[161] (denoted in Equation (20-21)).

$$L_{r,V} = E_{f_v \sim D_1(v), v \sim V}[\frac{1}{whd} \left\|\emptyset_i(\mathcal{G}(\boldsymbol{f_v})) - \emptyset_i(\mathcal{T}(v))\right\|_2] \qquad (2\text{-}7)$$

$$L_{r,P} = E_{f_p \sim D_2(p), p \sim P}[\frac{1}{whd} \left\|\emptyset_i(\mathcal{G}(\boldsymbol{f_p})) - \emptyset_i(\mathcal{T}(p))\right\|_2] \qquad (2\text{-}8)$$

Where $\emptyset_i(\cdot)$ denotes the output from the $i$th layer of the pretrained VGG16[162] network, $\boldsymbol{f_v}$ and $\boldsymbol{f_p}$ represent the feature maps from $D_{1,\emptyset}$ and $D_{2,\emptyset}$, $w, h$ and $d$ stands for the width, height and depth of the corresponding feature space, $\mathcal{G}$ is the pre-processing on $\boldsymbol{f}$, and $\mathcal{T}$ is the processing on truth sub-volumes/patches.

Second, hinge loss was selected to classify $P'$ from $P$ and $V'$ from $V$ with formulas listed in Equation (2-9~2-10).

$$L_{h,V} = E_{v' \sim V'}[f(D_{1,\emptyset}(v'))] + E_{v \sim V}[f(-D_{1,\emptyset}(v))] \qquad (2\text{-}9)$$

$$L_{h,P} = E_{p'\sim P'}[f(D_{2,\emptyset}(p'))] + E_{p\sim P}[f(-D_{2,\emptyset}(p))] \tag{2-10}$$

Where $f(t) = \max(0, 1+t)$.

Lastly, data augmentation, including random flipping and rotation, was implemented to V' and P' prior to sending into $D_{1,\emptyset}$ and $D_{2,\emptyset}$, following the theory proposed by Data Augmentation Optimized for GAN (DAG) framework[163]. $D_{1,\emptyset}/D_{2,\emptyset}$ share weights while discriminating multiple augmented sub-volumes/patches. Therefore, we have the overall loss objective formulated as Equation (2-11~2-12).

$$L(\vartheta, \{\emptyset_{1,k}\}, \{\emptyset_{2,k}\}) = L(\vartheta, \emptyset_{1,0}, \emptyset_{2,0}) + \frac{\lambda_1}{n-1} \sum_{k=1}^{n} L(\vartheta, \emptyset_{1,k}, \emptyset_{2,k}) \tag{2-11}$$

$$L(\vartheta, \emptyset_{1,k}, \emptyset_{2,k}) = L_{r,V,k} + L_{h,V,k} + \lambda_2(L_{r,P,k} + L_{h,P,k}) \tag{2-12}$$

Where $n = 4$, $\lambda_1 = 0.2$ and $k = 0$ corresponding to the identity transformation follows the definition in Trans et al[163]. $\lambda_2 = 0.5$ to give the model more attention on conformal $V'$ rendering.

**Inference Stage**: A fully trained $G_\vartheta$ was further fine-tuned with $z_{sh}$ and $z_a$ using the referenced sparse view projection(s) prior to rendering the final volumetric prediction $\hat{Y}$. Since we conducted moderate optimization with limited iterations, $G_\vartheta$ was tuned with fully size $I$ instead of patches. Depending on the available views, a referenced projection was randomly drawn for each iteration until $G_\vartheta$ reached the convergence criteria. In our experiments, peak signal-to-noise ratio (PSNR)=25 was set as the stopping threshold. The inference loss objective is defined in Equation (25) with a combination of LPIPS, PSNR, and the negative log-likelihood loss (NLL).

$$L_{G_\vartheta} = \lambda_1 L_{r,I} + \lambda_2 L_{PSNR,I} + \lambda_3 L_{NLL,I} \tag{2-13}$$

Where $\lambda_1 = \lambda_3 = 0.3$ and $\lambda_2 = 0.1$ were set in our experiments.

### 2.2.6 Implementation Details

During training, the RMSprop optimizer[164] with a batch size of 4 (4 × 1), learning rate of 0.0005 for the generator, learning rate of 0.0001 for the discriminator, and 40000 iterations were performed. Per inference fine-tuning, RMSprop[164] optimizer with a batch size of 1, learning rate of 0.0005, stopping threshold of PNSR=25 (mostly under 1000 iterations) was implemented towards the generator. All the experiments were carried out on a NVDIA RTX 4×A6000 cluster.

### 2.2.7 Evaluation Metrics

We evaluate the predicted CT volume $\hat{Y}$ and projection $\hat{I}$ corresponding to the reference view of our TomoGRAF generator using PSNR, structure similarity index measurement (SSIM) and rooted mean squared error (RMSE) as Equation (2-14~2-16).

$$PSNR = 20 \cdot \log_{10} \frac{MAX_I}{RMSE} \qquad (2\text{-}14)$$

$$SSIM = \frac{(2\mu_{G_\vartheta}\mu_y + c_1)(2\sigma_{G_\vartheta y} + c_2)}{(\mu_{G_\vartheta}^2 + \mu_y^2 + c_1)(\sigma_{G_\vartheta}^2 + \sigma_y^2 + c_2)} \qquad (2\text{-}15)$$

$$RMSE = \sqrt{\frac{\sum_{i=1}^{N} ||y(i) - \hat{y}(i)||^2}{N}} \qquad (2\text{-}16)$$

Where $MAX_I$ is the max possible pixel value in a tensor, RMSE stands for rooted mean squared error, $\mu_{G_\vartheta}$ and $\mu_y$ is the pixel mean of $G_\vartheta$ and $y$ and $\sigma_{G_\vartheta y}$ is the covariance between $G_\vartheta$ and $y$, $\sigma_{G_\vartheta}^2$ and $\sigma_y^2$ is the variance of $G_\vartheta$ and $y$. Lastly, $c_1 = (k_1 L)^2$ and $c_2 = (k_2 L)^2$, where $k_1 = 0.01$ and $k_2 = 0.03$ in the current work and $L$ is the dynamic range of the pixel values ($2^{\#\ bits\ per\ pixel} - 1$).

### 2.2.8 Baseline Algorithms

A CNN-based method X2CT-GAN[154], and a NeRF-based method MedNeRF[159] were included as our benchmarks with both the performance in projection inference and CT volume rendering compared. X2CT-GAN and MedNeRF are evaluated using the open-sourced codes and network weights released by authors, with the input of our in-house test set arranged following their data organization guidelines.

### 2.2.9 Data Cohorts

1011 CT scans were selected from LIDC-IDRI[165] thoracic CT database for organizing the training set. Digital reconstructed radiographs (DRRs) were generated as projections for training supervision. Several scanner manufacturers and models were included (GE Medical Systems LightSpeed scanner models, Philips Brilliance scanner models, Siemens Definition, Siemens Emotion, Siemens Sensation, and Toshiba Aquilion). The tube peak potential energies for scan acquisition include 120 kV, 130 kV, 135 kV, and 140 kV. The tube current is in the range of 40-627 mA. Slice thickness includes 0.6 mm, 0.75 mm, 0.9mm, 1.0 mm, 1.25 mm, 1.5 mm, 2.0 mm, 2.5 mm, 3.0 mm, 4.0 mm and 5.0 mm. The in-plane pixel size ranges from 0.461 to 0.977 mm[165]. 72 DRRs that cover a full 360° (generated each of 5° rotations) vertical rotations were generated for each scan. All the DRRs and CT volumes were black-border cropped out. DRRs were resized with a resolution of 128 times 128, and CT volumes were interpolated with a resolution of $128 \times 128 \times 128$ for model learning preparation. All the data were normalized to [0, 1].

The test data was organized under IRB approval (IRB # 20-32527) and included 100 de-identified CT scans from lung cancer patients who underwent robotic radiation therapy. All patients were scanned by Siemens Sensation with tube peak potential energy of 120 kV, tube current of 120 mA, slice thickness of 1.5 mm, and in-plane pixel size of 0.977 mm. The anterior-posterior (AP) and lateral views were generated for inference reference, with 1-view-based inference solely referencing the AP view projection. All the DRRs and CT volumes had the black border cropped out. Additionally, DRRs were resized with a resolution of 128 × 128, and CT volumes were interpolated with a volume size of 128 × 128 × 128. All the data were normalized to [0, 1] prior to being fed into models for inference.

### *2.2.10 Model performance as a function of the number of views and 3D supervision*

The model baseline performance was established using a single AP view. 1, 2, 5, and 10 view reconstructions were also performed to determine the model performance. The view angles are specified as follows: for 1-view-based reconstruction, the AP view is used for referencing. For 2-view-based reconstruction, the AP and lateral views are used for inferencing. For 5-view-based reconstruction, a full 360° is covered with rotation every 72°, starting from the AP view. For 10-view-based reconstruction, a full 360° is covered with rotation of every 36°, starting from the AP view.

## 2.3   Results

The results of TomoGRAF, MedNeRF, and X2CT-GAN with 1 or 2 views for 2D projection and volume rendering are visually demonstrated in Figure 2-2 and Figure 2-3, with accompanying statistics reported in Table 2-1 and Figure 2-4. TomoGRAF consistently outperforms MedNeRF and X2CT-GAN in both tasks with the most evident advantage in 1-view volume reconstruction.

For 2D projection rendering, TomoGRAF achieves marginally better results than MedNeRF. Both models maintain the overall critical body shapes of GT, and TomoGRAF visualizes more detailed morphology, such as heart, spine, and vascular structures. In comparison, the projection results of X2CT-GAN show visible distortion and significantly worse quantitative performance.

For 3D volume reconstruction, as shown in Figure 2-3, with 1-view, TomoGRAF depicts rich and correct anatomical details with visible tumors and a pacemaker consistent with GT. The results are further refined with the second orthogonal X-ray view, improving fine details' recovery. In comparison, MedNeRF and X2CT-GAN fail to render patient-relevant 3D volumes with 1 or 2 views. MedNeRF loses most anatomical details; X2CT-GAN deforms 3D anatomies that do not reflect patient-specific characteristics, such as lung tumors and the pacemaker.

As shown in Table 2-1, TomoGRAF is vastly superior in quantitative imaging metrics, achieving SSIM and PSNR of $0.79 \pm 0.03$ and $33.45 \pm 0.13$, respectively, vs. MedNeRF (SSIM at $0.37 \pm 0.05$ and PSNR at $7.68 \pm 0.10$) and X2CT-GAN (SSIM at $0.31 \pm 0.012$ and PSNR at $14.39 \pm 0.19$). There is a similar reduction in RMSE. Additionally, we can observe from Fig. 4 that the SSIM distribution of TomoGRAF is highly left skewed and leptokurtic in both 1 and 2-view-based volume rendering, with the majority clustering tightly towards the higher end, while that of MedNeRF and X2CT-GAN tends to be normal and moderately right-skewed (values lean towards the lower end).

Table 2-2 shows the 3D reconstruction performance with or without 3D supervision using 1, 2, 5, and 10 views as input. Using more views improved both volume and projection inference performance. 3D GT training markedly boosted the model performance in 3D volume rendering only. Figure 2-5 shows line profile comparisons for varying view inputs. 1 view TomoGRAF recovered major structures but missed fine details, which were better preserved with more X-ray views.

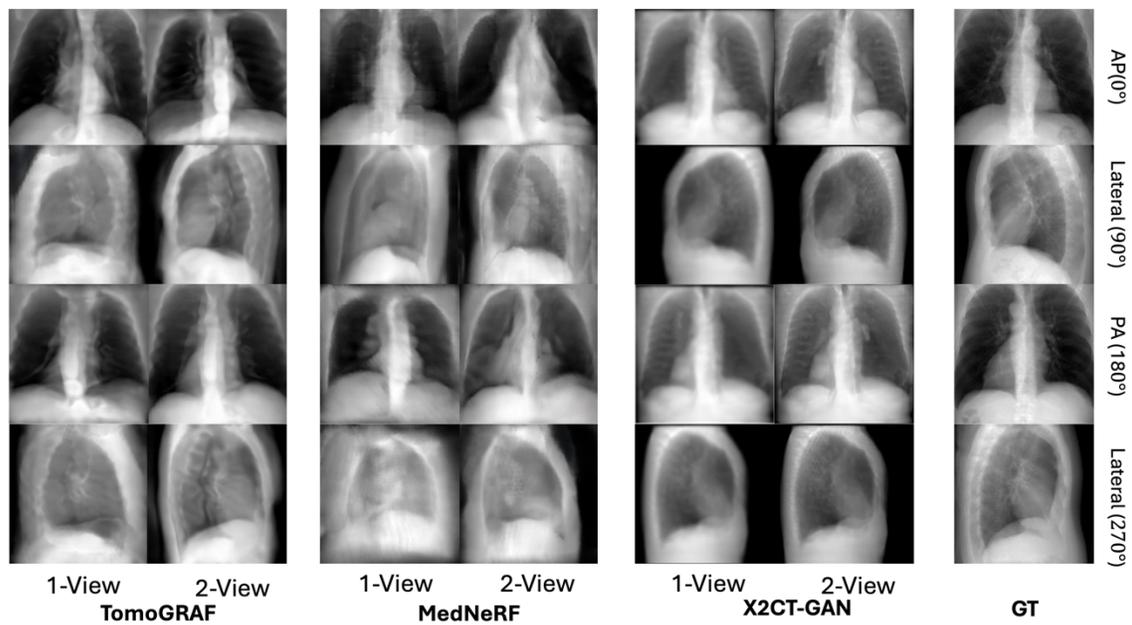

*Figure 2-2 Lung projection rendering in four 360°-clockwise-rotated (visualized every 90° of rotation) views from a patient in a test set.*

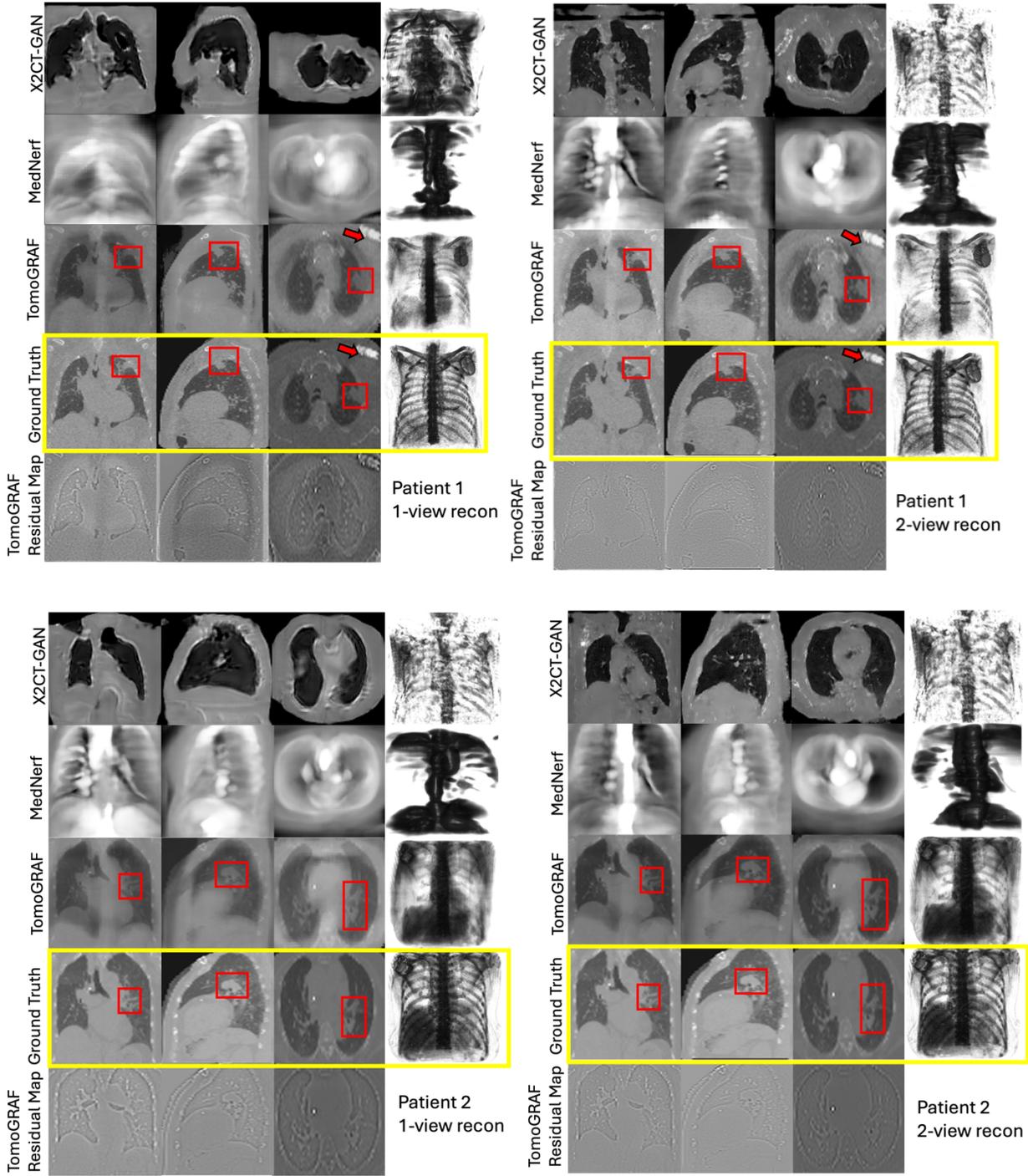

*Figure 2-3 CT Reconstruction results for two representative patients in the test set.*

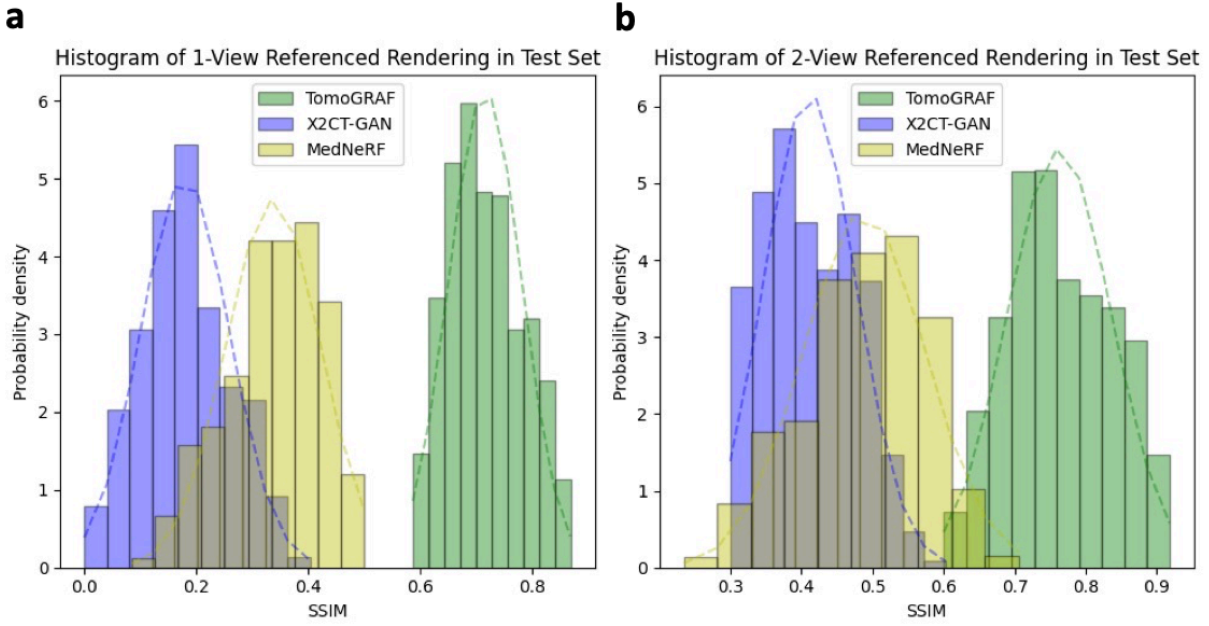

*Figure 2-4 Patient-wise evaluated SSIM of 1/2-view-based volume rendering results of TomoGRAF, X2CT-GAN, and MedNeRF in the test set.*

*Table 2-1 Statistical results evaluated on the test set.*

|  | Modality | 1-View | | | | 2-Views | | | |
|---|---|---|---|---|---|---|---|---|---|
|  |  | SSIM↑ | PSNR(dB)↑ | RMSE(HU)↓ | Inference Time (s)↓ | SSIM↑ | PSNR (dB)↑ | RMSE↓ | Inference Time (s)↓ |
| CT Volume | X2CT-GAN | 0.31 ± 0.12 | 14.39 ± 0.19 | 386.69 ± 27.43 | 0.27 | 0.48 ± 0.06 | 17.35 ± 0.21 | 347.76 ± 25.46 | 1.31 |
|  | MedNeRF | 0.37 ± 0.08 | 7.68 ± 0.10 | 321.87 ± 22.87 | 527.78 ± 15.48 | 0.50 ± 0.09 | 18.21 ± 0.09 | 299.49 ± 21.58 | 865.46 ± 30.81 |
|  | TomoGRAF | 0.79 ± 0.03 | 33.45 ± 0.13 | 175.48 ± 10.47 | 344.25 ± 10.32 | 0.85 ± 0.04 | 35.89 ± 0.13 | 146.73 ± 9.63 | 719.46 ± 26.78 |
| Projection | X2CT-GAN | 0.34 ± 0.11 | 11.88 ± 0.19 | 51.96 ± 7.98 | - | 0.51 ± 0.09 | 18.23 ± 0.26 | 47.64 ± 7.32 | - |
|  | MedNeRF | 0.67 ± 0.07 | 25.02 ± 0.15 | 36.48 ± 4.36 |  | 0.69 ± 0.08 | 27.31 ± 0.14 | 33.42 ± 4.01 |  |
|  | TomoGRAF | 0.69 ± 0.03 | 25.43 ± 0.14 | 34.37 ± 4.58 |  | 0.71 ± 0.04 | 27.99 ± 0.13 | 31.22 ± 4.12 |  |

*Table 2-2 Statistical results of TomoGRAF ablation study evaluated on test set.*

|  | Reference View | 3D Supervised Training | SSIM↑ | PSNR (dB)↑ | RMSE(HU)↓ | Inference Time (s)↓ |
|---|---|---|---|---|---|---|
| **CT Volume** | 1 | Y | 0.79 ± 0.03 | 33.45 ± 0.13 | 175.48 ± 10.47 | 344.25 ± 10.32 |
|  |  | N | 0.66 ± 0.05 | 26.76 ± 0.16 | 197.47 ± 11.24 |  |
|  | 2 | Y | 0.85 ± 0.04 | 35.89 ± 0.13 | 146.73 ± 9.63 | 719.46 ± 26.78 |
|  |  | N | 0.69 ± 0.06 | 29.87 ± 0.19 | 168.35 ± 10.21 |  |
|  | 5 | Y | 0.88 ± 0.03 | 37.23 ± 0.13 | 138.45 ± 9.12 | 987.35 ± 37.89 |
|  |  | N | 0.72 ± 0.04 | 30.15 ± 0.18 | 147.56 ± 9.79 |  |
|  | 10 | Y | 0.93 ± 0.01 | 39.98 ± 0.11 | 127.68 ± 8.78 | 1238.81 ± 46.72 |
|  |  | N | 0.75 ± 0.01 | 31.86 ± 0.18 | 138.98 ± 9.54 |  |

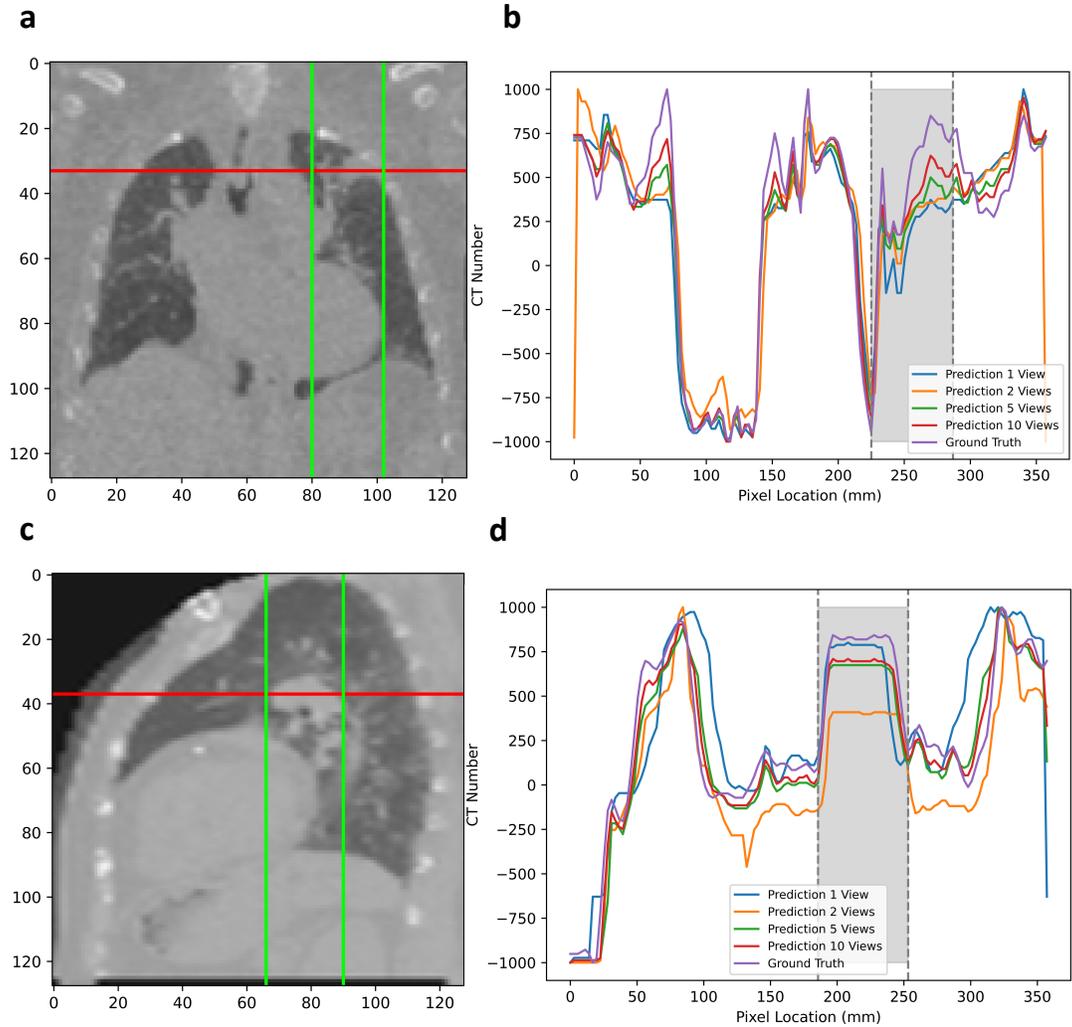

*Figure 2-5 Line profile comparison of the two patients shown in Fig. 3.*

## 2.4 Discussion

The current chapter presents a GAN-embedded NeRF generator (TomoGRAF) for volumetric CT rendering from ultra-sparse X-ray views. TomoGRAF extends radiance fields into medical imaging reconstruction with a CT imaging-informed ray casting/tracing simulator. Also, TomoGRAF leverages the availability of 3D volumetric information at the training stage to enable an effective generator trained with full volumetric supervision. The robustness of TomoGRAF is demonstrated on an external dataset independent of the training set. TomoGRAF vastly improves 1-view 3D reconstruction performance yet scales well with variable views for a wide range of image-guided clinical settings.

Reconstruction of 3D CT volume from ultra-sparse angular sampling is an ill-posed inverse problem that is extremely underconditioned to solve. On the other hand, such 3D reconstruction is practically desirable and widely applicable when a full gantry rotation is prevented by mechanical limitations or the dynamic process of interest is significantly faster than CT acquisition speed[64,65,68]. Therefore, there has been a consistent effort to reconstruct 3D images with extremely sparse views that circumvent these mechanical and temporal restrictions. Although CS-powered iterative methods and some earlier DL methods were able to reconstruct 3D images with as few as 20 views[74], the resultant image quality noticeably degraded. Still, they were unable to meet the challenges of many aforementioned practical scenarios where only one or two views were available for a given anatomical instance. Reconstruction with even more sparse views cannot be achieved without stronger priors and statistical learning. DL methods marched further in realizing ultra-sparse sampling (1 view) reconstruction using state-of-the-art (SOTA) networks, two of which are compared in this study.

TomoGRAF is distinctly superior to two SOTA methods for ultra-sparse view CT reconstruction in the following aspects. 1) In comparison to CNN-based networks with an extremely large number of parameters, such as X2CT-GAN[154], the radiance field-based generators train a lighter model with significantly fewer parameters (X2CT-GAN: ~4.28G FLOPS vs. TomoGRAF ~0.9G FLOPS, FLOPS: floating point operation per second) to represent the interaction process between photons and objects, which formalizes a better-defined goal for the network to reach and can effectively reduce the amount of training data required for achieving global robustness. Plus, CNN frameworks generally lack flexibility in view referencing. The number and angle of views at the inference stage must align with the input at the training stage. Meanwhile, TomoGRAF, besides 1-view referencing, can leverage additional X-ray views at arbitrary angles. It is worth noting that X2CT-GAN performance in the current study is markedly worse than the original report[154]. To determine the correctness of our implementation, we tested the X2CT-GAN code on the LIDC-IDRI data with the same data split and arrived at a similar performance. We believe the sharp decline in performance from internal LIDC-IDRI data testing to external in-house organized data testing is due to the differences in the training and testing data. CT images in LIDC-IDRI are cropped to keep only the thoracic organs, while our in-house test data are intact CT with the complete patient's chest wall, arms, and neck. Such variation or domain shift is common and expected in practice: the patients can vary in size and be set up with different immobilization devices or arm positions. In stark contrast to X2CT-GAN, TomoGRAF is robust to such variation.

2) Compared to MedNeRF, we adapt NeRF with a physically realistic volume rendering mechanism based on the X-ray transportation properties, where photons pass through the body, and the volumetric photon attenuation along the ray path within the body is the focus of reconstruction. In other words, TomoGRAF learns 3D X-ray image formation physics, whereas MedNeRF assumes visible light transportation physics and is intended for 2D manifold learning of object surfaces. The limitation is evident in both low quantitative imaging metrics and orthogonal cuts of MedNeRF reconstructed patients: there is better retention of outer patient contour than internal anatomical details, which are largely lost in MedNeRF images. Additionally, TomoGRAF employs paired 3D CT supervision at the training stage to maximize the prior knowledge exposed to the network, which contributed to the model robustness in volume rendering at the inference stage. As a result, TomoGRAF successfully leverages the efficient object representation capacity of NeRF while overcoming the intrinsic limitations due to its lack of X-ray transportation physics and 3D volume comprehension. To our knowledge, TomoGRAF is the first truly generalizable single-view 3D X-ray reconstruction pipeline robust to substantial domain shifts.

At the practical level, TomoGRAF provides a unique solution for applications where only one or a few X-ray views are available, but 3D volumetric information is desired. The applications include image-guided radiotherapy, interventional radiology, and angiography. For the former, 2D kV X-rays can be interlaced with MV therapeutic X-ray beams to provide a real-time view of the patient during treatment[166]. However, the 2D projection images do not describe the full 3D anatomy, which is critical for adapting radiotherapy to the real-time patient target and surrounding tissue geometry. Similarly, 4D CT digitally subtracted angiography (DSA) better describes dynamics of the contrast for enhanced diagnosis than single-phase CT DSA[167], but fast helical and flat panel-based 4D-DSA requires repeated scans of the subject, increasing the imaging dose and leading to compromised temporal resolution for intricate vascular structures[168]. TomoGRAF can be potentially used to infer real-time time-resolved 3D DSA with significantly reduced imaging dose. Our results show that TomoGRAF is flexible in incorporating more views for further improved inference performance. Dual views with fixed X-ray systems are widely used in radiotherapy for stereotactic localization[151,169,170], but the modality is limited to triangulating bony anatomies or implanted fiducials. TomoGRAF can utilize the same 2D stereotactic views to provide rich 3D anatomies for soft tissue-based registration and localization. Besides mechanical and imaging dose constraints, inexpensive portable 2D X-rays are more readily available for point-of-care and low-resource settings where a CT is impractical. The ability to reconstruct 3D volumes using a single 2D view would markedly increase the imaging information available for clinical decisions. Our study also shows the feasibility of using more views in TomoGRAF for further improved performance and broader applications, including 4D CBCT and tomosynthesis with sparse or limited angle views.

At the theoretical level, TomoGRAF validates the extremely high data efficiency of neural field representation of 3D voxelized medical images. TomoGRAF, for the first time, materializes high data efficiency, achieving good quality (SSIM=0.79-0.93) 3D reconstruction of CT images with 1-10 views, which is a major stride in comparison to existing research using NeRF or GRAF. The work thus has significant implications in 3D image acquisition, storage, and processing, which are currently voxel-based. Voxelized 3D representation does not provide intrinsic structural information regarding the relationships among voxels and thus can be expensive to acquire and reconstruct. Previous compressed sensing research explored some of the explicit structural correlations, such as piece-wise smoothness, for reduced data requirements. TomoGRAF indicates a new form of data representation that exploits implicit structural information with higher efficiency than conventional methods or neural networks without encoded physics.

Nevertheless, the current study leaves several areas for future improvement. First, TomoGRAF requires further fine-tuning at the inference stage, which increases the reconstruction time (1-view at 344.25±10.32 s and 2-view at 719.46±26.78 s). The time further increases with inference using more views. Significant acceleration is desired for online procedures such as motion adaptive radiotherapy[171]. A larger dataset with a more diverse distribution will eliminate or reduce the burden of the fine-tuning step and bring TomoGRAF inference to a sub-second level, which would be essential for interventional procedures. Second, TomoGRAF is developed and tested on CT-synthesized DRR, which differs from kV X-rays obtained using an actual detector in image characteristics due to simplification of the physical projection model, detector dynamic ranges, noise, pre and postprocessing[152]. The current model may need to be adapted based on actual X-ray projections. Third, TomoGRAF reconstruction results with 1-view are geometrically correct but lose fine details and CT number accuracy, which is partially mitigated with increasing views up to 10. Therefore, in its current form, TomoGRAF is suited for object detection and localization tasks, but its appropriateness for quantitative tasks such as radiation dose calculation needs to be further studied. Moreover, the recovery of detail should also improve with reconstruction resolution, which is currently limited in rendering a maximum of 128 times 128 times 128 resolution due to GPU memory constraints. This limitation, however, is expected to be overcome soon with rapidly increasing GPU memory capacity.

## 2.5　　　Summary

TomoGRAF, a novel GAN-based NeRF generator, is presented in the current work. TomoGRAF is trained on a public dataset and evaluated on 100 in-house lung CTs. TomoGRAF reconstructed good quality 3D images with correct internal anatomies using 1-2 X-ray views, which SOTA DL methods fail to accomplish. TomoGRAF performance further improves with more views. The superior TomoGRAF performance is attributed to novel X-ray physics encoding in the radiance field training and paired 3D CT supervision.

# 3. Dual Energy Computed Tomography Multi-Material Decomposition from Projection Domain

## 3.1 Introduction

Sidky et al. reported the top-ranked algorithms in the 2022 AAPM DL-spectral-challenge[172], where most pipelines decomposed materials from the sinogram domain and achieved remarkably low RMSE. However, they are limited to offline applications due to the dependence on conventional and slow iterative reconstruction as a sub- or main component. The iterative forward and backward projections are extremely computationally expensive in training and testing stages and incapable of generating real-time predictions. For instance, Hauptmann et al. reported that an iterative DL model (5 forward and backward iterations) are around five times slower than non-iterative learning methods[173]. Similar comparison were reported by other authors[174,175].

Thus, we believe that the encouraging statistical results from Zhu et al.[117] and the 2022 AAPM challenge[172] warrant further development for improvement in model applicability. Specifically, a workflow using a lightweight and interpretable network structure is desirable without losing quality control. We realize such a vision in the novel rFast-MMDNet, which is elaborated as follows.

The proposed rFast-MMDNet includes 1) SinoNet – a UNet[176] based architecture for the decomposition of DE projection to basis material sinograms; 2) DenoiseNet – a customized denoising CNN consisting of sixteen ResNet[177] blocks and an integrated one-time FBP[178] for fast conversion of signals from sinogram to image. Notably, the training of SinoNet and DenoiseNet is conducted separately for fast and stable convergence of feature extraction in each stage.

## 3.2 Methods

### 3.2.1 Data Cohort

The 2022 American Association of Physicists in Medicine (AAPM) DL-spectral-CT challenge released a set of simulated DECT scans with GT tissue phantoms for a breast model[179] containing three materials: adipose, fibroglandular, and calcification (assumed to be composed of hydroxyapatite[180]). The DL-spectral-CT dataset modeled ideal fast kVp-switching (50 and 80 kVps) acquisition of circular fan-beam X-rays projected on a flat panel detector with 1024 pixels. The 50-/80-kVp transmission ($I_w$) is generated under the assumption of a spectral CT model formulated as Equation (3-1) with assumed spectral sensitivities ($S$) and linear attenuation ($\mu_m$) distributed as the upper row of Figure 3-1. The dataset has 1000 pairs of training, 10 pairs of validation, and 100 pairs of testing images. Each of the inputs of low(50)/high(80)-kVp acquisitions has 256 projection views. The GTs are all 2- 2D and have a H × W of $512 \times 512$. Additionally, both the input and GT signals are normalized to between 0-1. The bottom row of Figure 3-1 shows the data processing pipeline.

$$I_w = \int S_w(E) \exp[-\mu_a(E)\mathcal{P}x_a - \mu_f(E)\mathcal{P}x_f - \mu_c(E)\mathcal{P}x_c]\, dE \qquad (3\text{-}1)$$

Where $w$ index is either 50 or 80 kVp, $m$ index is ether adipose ($a$), fibroglandular ($f$), or calcification ($c$), $x_m$ is tissue phantom, and $\mathcal{P}$ is short for forward projection.

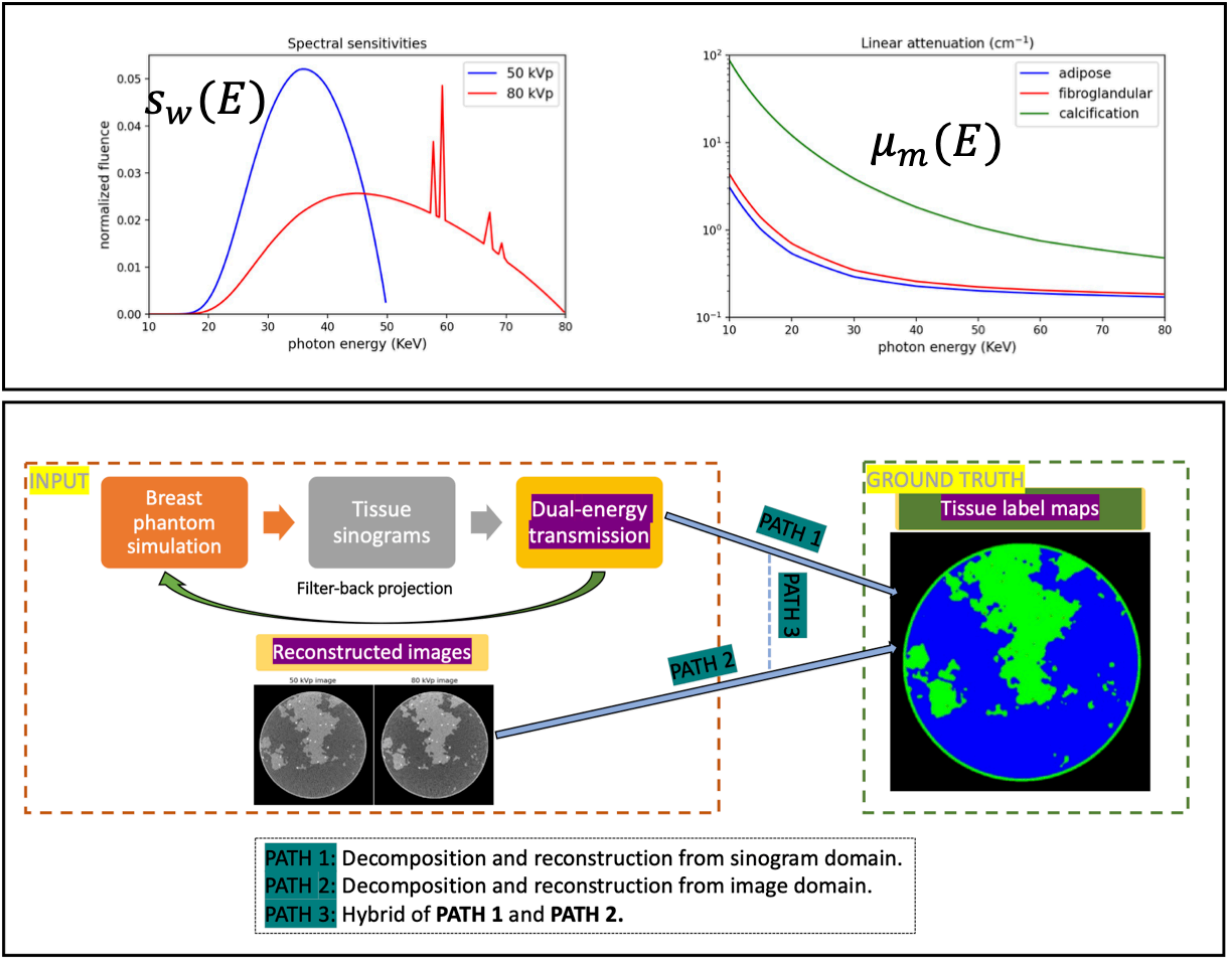

*Figure 3-1 The spectral sensitivity and linear attenuation distribution utilized for simulating the low-/high-energy transmission and simulation process of the DL-Spectral-CT dataset.*

Regarding the input of SinoNet, we first created two zero-filling $1 \times 512 \times 1024$ matrices named zero-matrix-high and zero-matrix-low, respectively. To process the high-kVp transmission, we arranged the sequence from each view of a raw projection into the odd rows of zero-matrix-high with no modification in the even rows. Vice versa for low-KVp, for which we placed the projections in the even rows of zero-matrix-low and left odd rows unchanged. The matrix structure reproduced the alternating pattern of fast kVp switching low- and high-energy X-ray data collection. Next, we stacked the dilated 2D high- and low-kVp signals channel-wise to form the input volume with the dimension of Channel (C) $\times$ H $\times$ W in $2 \times 512 \times 1024$. To process GT, the forward projection was performed first to generate separate sinograms for individual materials in the phantom, and then the forward projection sinograms were channel-wise concatenated for the three basis materials with a dimension of $C \times H \times W = 3 \times 512 \times 1024$. Lastly, resizing (scales in $[0.7, 0.9, 1.0, 1.1, 1.2]$) was applied to augment the training data five-fold while perserving the projection geometry within the input volume.

DenoiseNet combined both input (FBP images) and GT tissue phantoms to form a new matrix ($C \times H \times W = 3 \times 512 \times 512$). Data augmentation, including random rotation (angles in $[-30°, -15°, 0°, 15°, 30°]$), random resizing (scales in $[0.7, 0.9, 1.0, 1.1, 1.2]$), random gaussian blur (kernels in $[1, 3, 5]$), random cropping, random brightness, and mirroring was implemented.

### 3.2.2  rFast-MMDNet

Figure 3-2 shows the complete architecture of rFast-MMDNet. As mentioned in Section 3.1, the overall pipeline is divided into two stages. In the first stage, SinoNet learns decomposed raw DE projection features in the sinograms. In the second stage, rough tissue patterns are generated via FBP-based domain adaptation, further refined by a DenoiseNet to suppress noise and remove artifacts. Detailed illustrations and justification regarding each module are elucidated as follows.

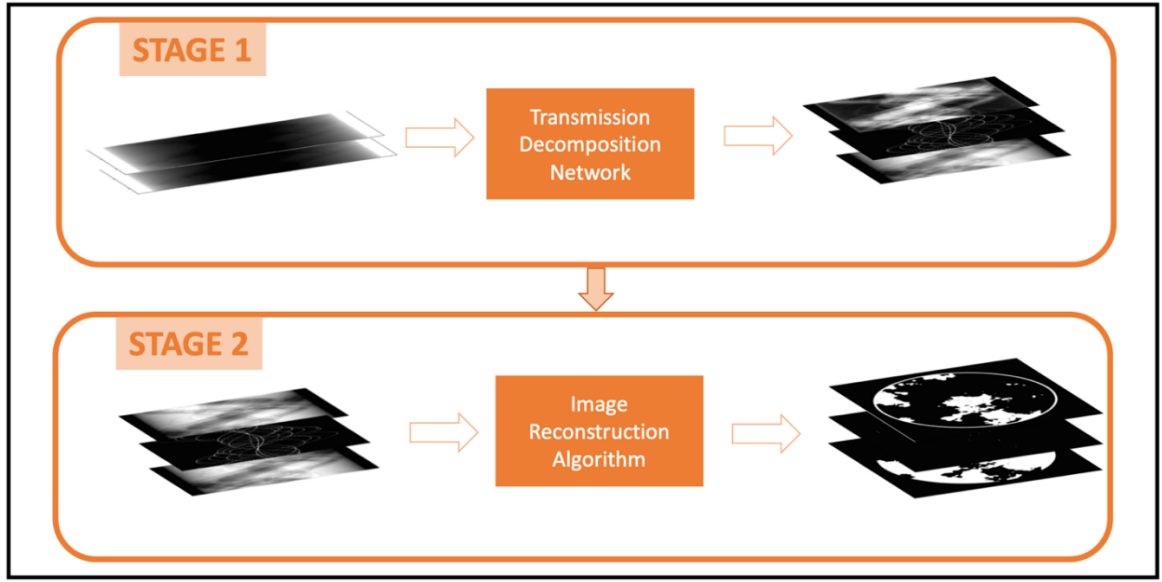

*Figure 3-2 The complete pipeline of rFast-MMDNet.*

3.2.2.1 SinoNet

Figure 3-3 shows the structure of SinoNet for DE projection decomposition. In particular, SinoNet was inspired by Xu et al.[181] designed using a 3D UNet-based encoder-decoder combined with skipped information feeding from each encoding to its corresponding decoding layer via channel-wise concatenation. The $3 \times 3$ convolution and $2 \times 2$ max pooling filters were applied to all layers in SinoNet. Altogether, SinoNet has one bottleneck block for structure balancing and 7 separate encoding and decoding layers. The number of spatial filter channels doubles per layer from the base of 32 to a maximum of 320, and the feature map dimension halves per individual step of encoding convolution. Vice versa, the decoding blocks reduce the filter channel number and increase the figure map dimension by 2. The rationale of such a network structure can be understood as follows. The encoding stage casts the DE transmissions to a higher dimension depicting various VMI projections at distinct energy levels. The decoding stage then combines virtual monochromatic projections into the sinograms of adipose, fibroglandular, and calcification, respectively. The architecture reduces the network complexity, facilitates efficient training

convergence, and maintains model robustness. As discussed, the input of SinoNet was dilated and channel-wise concatenated DE projections, while the corresponding GTs were channel-wise stacked sinograms forward-projected from tissue phantoms.

The residual discrepancy between the SinoNet predicted sinograms and the theoretical material sinograms was minimized using DenoiseNet in the second stage.

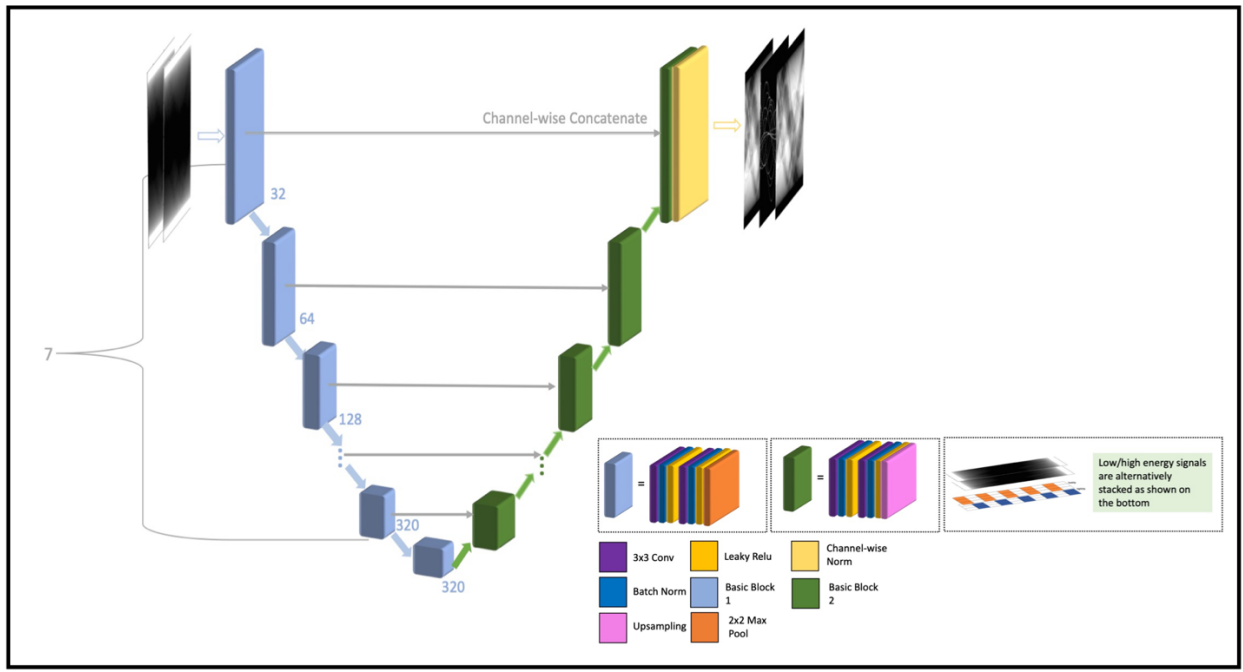

*Figure 3-3 Architecture of SinoNet for decomposing material sinograms from DE transmission singals.*

### 3.2.2.2 FBP + DenoiseNet

As demonstrated in the first row of Figure 3-4, the image reconstruction scheme entails two sub-components: 1) FBP image generator to characterize the coarse morphology of tissue phantoms in the image domain; 2) DenoiseNet for detail refinement. DenoiseNet (second row in Figure 3-4) consists of 16 standard ResNet blocks (structure shown in the third row of Figure 3-4), each having a convolution kernel of size $3 \times 3$, a stride of 1, and spatial filter channels in an increase-to-decrease form. Pooling layers are not applied in DenoiseNet to maintain the resolution of feature maps.

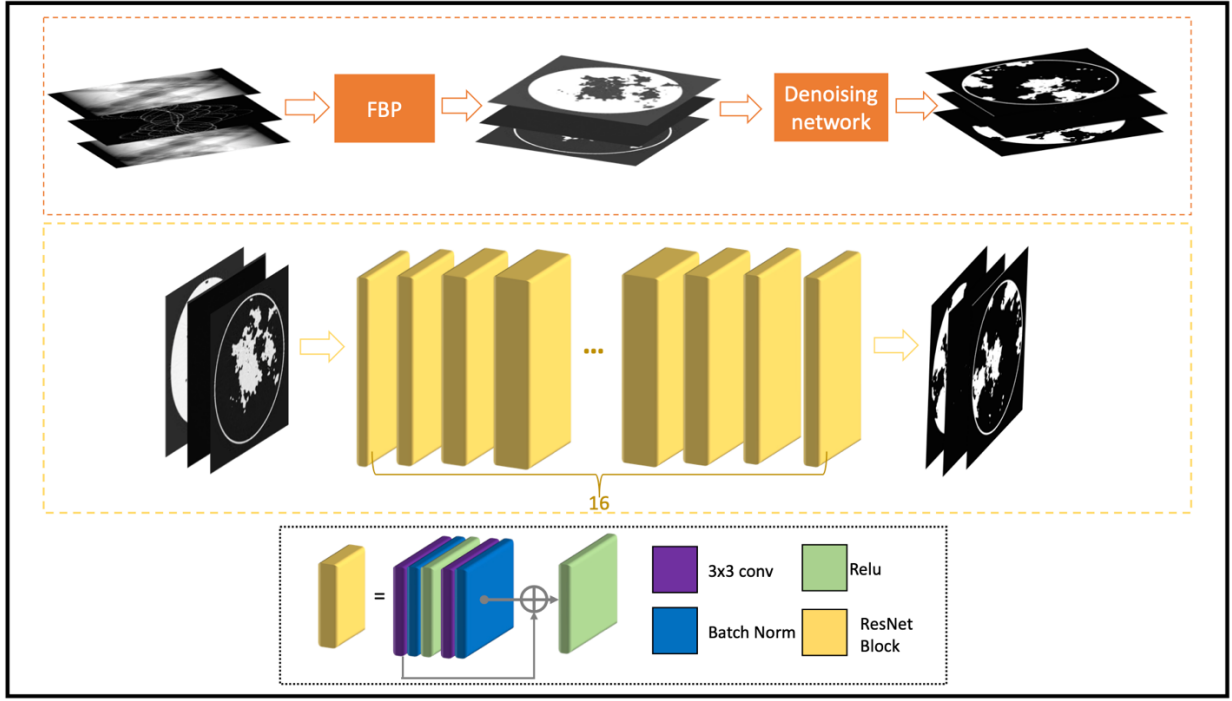

*Figure 3-4  Framework of DenoiseNet.*

### 3.2.3    Loss Function

SinoNet and DenoiseNet share the same cost function combining negative PSNR, Multi-scale-SSIM, and $L_1$ deviation measurement.

$$L = -\alpha \cdot L_{PSNR} + (1-\alpha) \cdot [\beta \cdot (1 - L_{MS-SSIM}) + (1-\beta) \cdot L_1] \quad (3\text{-}2)$$

Where $\alpha$ and $\beta$ in Equation (3-2) are hyperparameters set to 0.5 and 0.25, respectively.

### 3.2.4    Model Training

Both SinoNet and DenoiseNet were implemented in PyTorch. The training was performed on a GPU cluster with $4 \times$ RTX A6000.

For SinoNet training, we set the maximum number of epochs as 1k and observed the model convergence at ~450 epochs. Adam optimizer with an initial learning rate (LR) of 0.001 and batch size of 4 × 2 was applied during learning.

Regarding training of DenoiseNet, under the assumption that limited noises and artifacts exist in the FBPed images, we set the max-epoch number as 200 and found that the validation loss reached plateau at ~70 epochs. Adam optimizer with an initial learning rate (LR) of 0.01 and batch size of 4 × 4 was implemented during model tuning.

It is worth noting that all the data augmentations described in Sec. 3.2.1 were performed on the fly along with DNN training.

### 3.2.5 Benchmarks

We compare our in-house rFast-MMDNet with representative MMD methods in four distinct categories. 1) Classical image-domain MMD: We selected an analytical algorithm MMD (AA-MMD) based on the assumption of volume conservation ≤ 3 materials in an individual voxel. For each voxel, AA-MMD loops over a material triplet library, identify the best-fit triplet, and perform MMD via direct matrix inversion[182]; 2) DL image-domain MMD: We chose a UNet-based image-domain MMD (ID-UNet), which feeds the low- and high-energy images into a 3D UNet and predicts out decomposed materials[116]; 4) image-domain MMD followed by in-house DenoiseNet: to fairly compare the overall efficacy of image and our in-house two-stage pipeline, we also build up two algorithms which run our proposed DenoiseNet after AA-MMD and ID-UNet; 4) DL projection-domain MMD: We selected the previously mentioned Triple-CBCT[117].

## 3.3  Experiments

The results from SinoNet and our rFast-MMDNet are reported. For performance comparison, the outcomes from the benchmark algorithms (AA-MMD, ID-UNet, Triple-CBCT) along with the intermediate results from our in-house algorithm (SinoNet + FBP) and that from the rFast-MMDNet pipeline are evaluated. The effectiveness of our DenoiseNet post-processing can be seen from the SinoNet + FBP results. Deviation metrics, including RMSE and mean absolute error (MAE), and similarity measurements, including negative PSNR and SSIM, are incorporated as evaluation metrics. Both quantitative and qualitative outcomes are reported.

### *3.3.1  Results of SinoNet*

As presented in Figure 3-5, GT and SinoNet results are visually indistinguishable. Quantitative results also show nearly negligible deviation from GT with validation RMSE, MAE, negative PSNR, and SSIM of $0.002 \pm \sim 0$, $0.001 \pm \sim 0$, $-55.738 \pm 0.314$, and $0.001 \pm \sim 0$, and test of $0.002 \pm \sim 0$, $0.001 \pm \sim 0$, $-54.872 \pm 0.347$, and $0.001 \pm \sim 0$ of averaged mean values among three tissues.

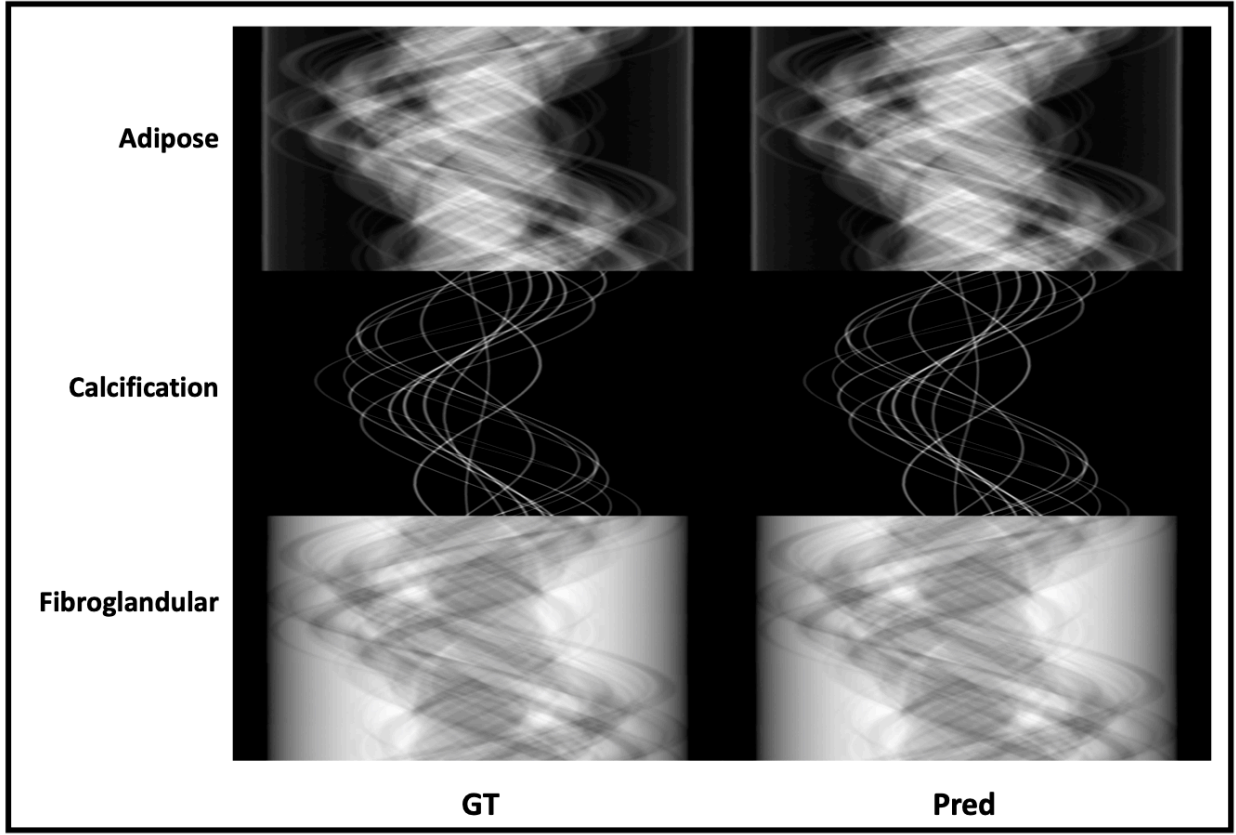

*Figure 3-5 Results illustration in sinogram domain.*

### 3.3.2  Results of rFast-MMDNet

In Figure 3-6, we can observe that the predictions from AA-MMD show visible losses of information, including poor image resolution, incorrect boundaries of subtle structures (e.g., adipose and fibroglandular), and noisy predictions of calcification. ID-UNet mitigates the noise in calcification predictions and is slightly better at preserving anatomical details but still shows discrepancies of inaccurate micro-pattern delineation with GT (see the visualization of ID-UNet in Figure 3-6). Next, SinoNet + FBP can roughly characterize the anatomical components but is quantitatively inaccurate due to beam hardening artifacts. Finally, we found that Triple-CBCT and our rFast-MMDNet are qualitatively best performers, which produce visually similar decomposition results.

Quantitatively, we can see from Table 3-1 and Tabl3 3-2 that rFast-MMDNet outperforms Triple-CBCT in all cases, showing $(0.001, 0.003)$ lower error metrics and $\sim 0.006$ and $\sim -3$ better in similarity metrics. Both rFast-MMDNet and Triple-CBCT are markedly better than AA-MMD and SinoNet-FBP by two orders of magnitude in deviation from GT and $(10, 30)$ in negative PSNR. Additionally, even though running DenoiseNet on top of AA-MMD and ID-UNet, image-domain MMD algorithms are still far worse than sinogram domain MMD. Besides mean values, rFast-MMDNet has the lowest standard deviation (std) $\sim 0$ standard deviation in RMSE, MAE, and SSIM, and 0.453/0.542 in negative PSNR for both validation and test sets in Table 3-1. The observation is consistent with individual materials analysis In Table 3-2. It is worth noting that the calcification prediction errors for all MMD methods are lower than adipose and fibroglandular tissue predictions, likely due to its high contrast and sparse representation in the phantoms.

Table 3-1 shows that all image domain methods make sub-second predictions. In comparison, Triple-CBCT is significantly slower, taking up to 174 seconds to finish a decomposition prediction. Our proposed rFast-MMDNet achieved comparable or faster speed with image-domain methods while maintaining the advantages of being a sinogram-domain method. Among all DL models, Triple-CBCT has the heaviest model parameters and computation demand and image domain algorithms have roughly half of the model parameters and MACs than sinogram domain methods.

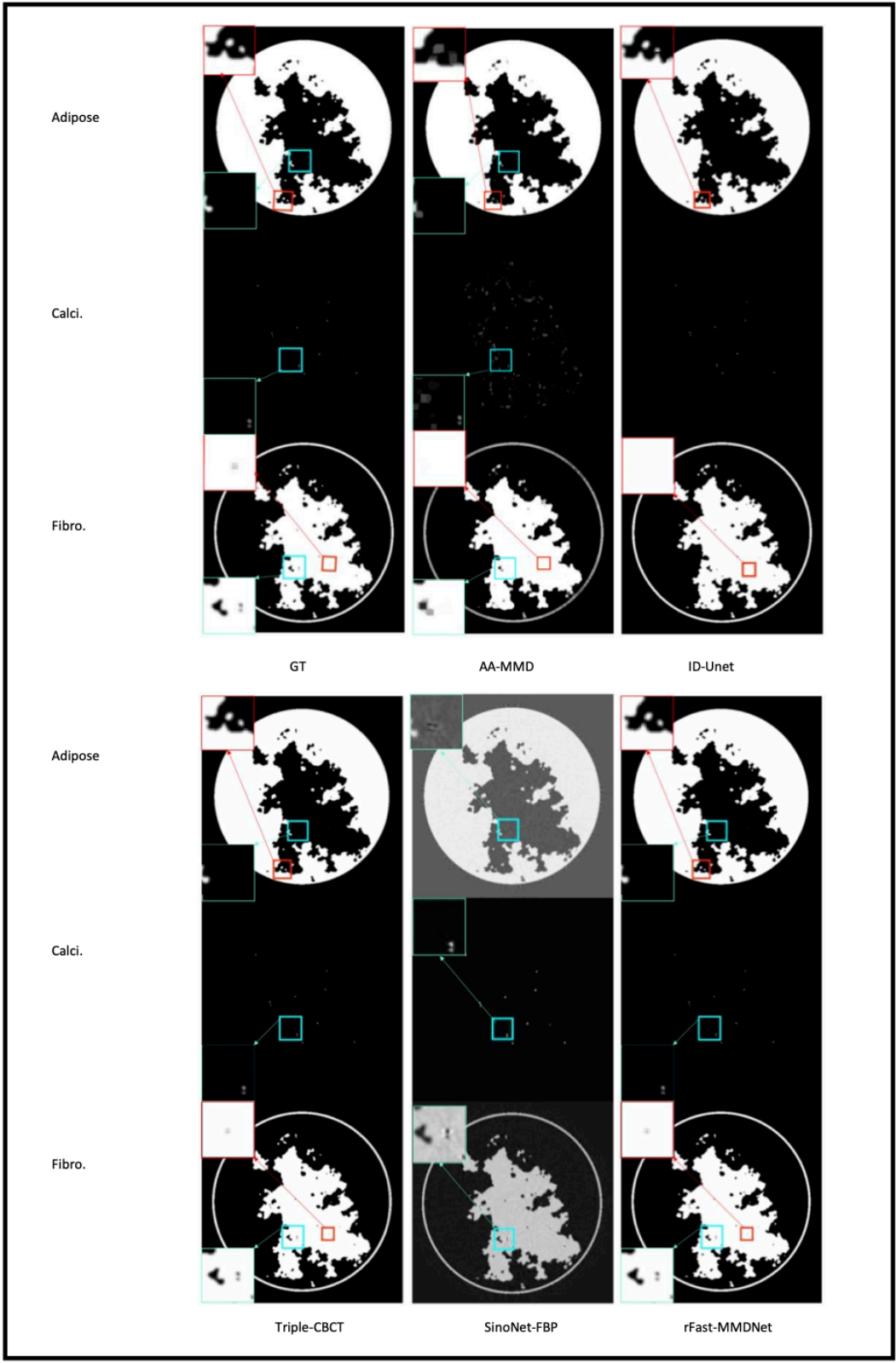

*Figure 3-6 Results illustration in image domain.*

*Table 3-1 The measured mean values and stds of difference between GTs and predictions among all tissue phantoms*

| Domain | Dataset | Method | RMSE | MAE | -PSNR | SSIM | EXEC (ms) | Parameters (M) | MACs (G) |
|---|---|---|---|---|---|---|---|---|---|
| Image | Validation | AA-MMD[182] | 0.195 ± 0.107 | 0.186 ± 0.014 | −13.453 ± 1.874 | 0.472 ± 0.003 | | | |
| | | AA-MMD[182]+DenoiseNet | 0.023±0.007 | 0.031±0.012 | -24.335±0.945 | 0.023±0.042 | | | |
| | | ID-UNet[116] | 0.013 ± 0.003 | 0.015 ± 0.017 | −27.994 ± 0.731 | 0.021 ± 0.027 | - | | |
| | | ID-UNet[116] + DenoiseNet | 0.010±0.003 | 0.011±0.009 | -36.732±0.631 | 0.009±0.004 | | | |
| | Test | AA-MMD[182] | 0.183 ± 0.112 | 0.175 ± 0.019 | −15.774 ± 1.457 | 0.421 ± 0.004 | 204 | - | - |
| | | AA-MMD[182] + DenoiseNet | 0.021±0.006 | 0.028±0.011 | −26.543±0.902 | 0.022±0.039 | 291 | 44.790 | 67.520 |
| | | ID-UNet[116] | 0.010 ± 0.004 | 0.012 ± 0.020 | −29.827 ± 0.871 | 0.018 ± 0.020 | **67** | 35.210 | 60.330 |
| | | ID-UNet[116] + DenoiseNet | 0.009±0.003 | 0.010±0.010 | -37.374±0.624 | 0.008±0.003 | 150 | 80.000 | 80.000 |
| Projection | Validation | Triple-CBCT[117] | 0.011 ± 0.005 | 0.0106 ± 0.012 | −35.143 ± 0.532 | 0.010 ± ~0 | | | |
| | | SinoNet-FBP | 0.195 ± 0.072 | 0.134 ± 0.011 | −13.717 ± 1.286 | 0.410 ± 0.003 | - | | |
| | | rFast-MMDNet | **0.005 ± ~0** | **0.002 ± ~0** | **−43.792 ± 0.453** | **0.004 ± ~0** | | | |
| | Test | Triple-CBCT[117] | 0.007 ± 0.006 | 0.008 ± 0.015 | −38.571 ± 0.631 | 0.008 ± ~0 | 174000 | 118.590 | 147.030 |
| | | SinoNet-FBP | 0.173 ± 0.095 | 0.120 ± 0.095 | −15.616 ± 1.378 | 0.381 ± 0.003 | 101 | **32.870** | **57.360** |
| | | rFast-MMDNet | **0.004 ± ~0** | **0.005 ± ~0** | **−45.027 ± 0.542** | **0.002 ± ~0** | 184 | 77.660 | 124.880 |

*Table 3-2 The measured mean values and stds of difference between GTs and predictions of individual tissues.*

| Material | Dataset | Method | RMSE | MAE | -PSNR | SSIM |
|---|---|---|---|---|---|---|
| ***Adipose*** | Validation | AA-MMD[182] | 0.207 ± ~0 | 0.189 ± ~0 | −12.712 ± 0.779 | 0.489 ± 0.004 |
| | | AA-MMD[182] + DenoiseNet | 0.032 ±~0 | 0.033 ±~0 | -21.702±0.873 | 0.005±0.023 |
| | | ID-UNet[116] | 0.015 ± ~0 | 0.019 ± ~0 | −30.829 ± 0.612 | 0.003 ± 0.018 |
| | | ID-UNet[116] +DenoiseNet | 0.012±~0 | 0.017 ± ~0 | −32.801 ± 0.617 | 0.005 ± 0.019 |
| | | Triple-CBCT[117] | 0.012 ± ~0 | 0.004 ± ~0 | −33.345 ± 0.491 | 0.016 ± 0.006 |
| | | SinoNet-FBP | 0.137 ± 0.003 | 0.128 ± 0.005 | −16.153 ± 0.532 | 0.017 ± 0.007 |
| | | rFast-MMDNet | **0.008 ± ~0** | **0.003 ± ~0** | **−40.134 ± 0.490** | **0.011 ± 0.002** |
| | Test | AA-MMD[182] | 0.197 ± ~0 | 0.178 ± ~0 | −13.224 ± 0.892 | 0.471 ± 0.005 |
| | | AA-MMD[182] + DenoiseNet | 0.029 ±~0 | 0.032 ±~0 | -22.515±0.794 | 0.004±0.019 |
| | | ID-UNet[116] | 0.013 ± ~0 | 0.015 ± ~0 | −32.187 ± 0.781 | 0.002 ± 0.021 |
| | | ID-UNet[116] + DenoiseNet | 0.011±~0 | 0.013 ± ~0 | −33.072 ± 0.604 | 0.005 ± 0.016 |
| | | Triple-CBCT[117] | 0.010 ± ~0 | 0.004 ± ~0 | −35.271 ± 0.562 | 0.013 ± 0.009 |
| | | SinoNet-FBP | 0.136 ± 0.003 | 0.123 ± 0.008 | −17.402 ± 0.605 | 0.013 ± 0.009 |

| Material | Dataset | Method | RMSE | MAE | -PSNR | SSIM |
|---|---|---|---|---|---|---|
| | | rFast-MMDNet | **0.007 ± ~0** | **0.002 ± ~0** | **−43.193 ± 0.657** | **0.008 ± 0.003** |
| *Calcification* | Validation | AA-MMD[182] | 0.084 ± ~0 | 0.061 ± ~0 | −18.893 ± 0.412 | 0.572 ± 0.004 |
| | | AA-MMD[182] + DenoiseNet | 0.035±~0 | 0.010±~0 | -27.468±0.573 | 0.012±0.001 |
| | | ID-UNet[116] | 0.011 ± ~0 | 0.007 ± ~0 | −30.548 ± 0.459 | 0.004 ± ~0 |
| | | ID-UNet[116] +DenoiseNet | 0.011 ± ~0 | 0.006 ± ~0 | −31.764 ± 0.394 | 0.003 ± ~0 |
| | | Triple-CBCT[117] | 0.007 ± ~0 | 0.003 ± ~0 | −38.893 ± 0.391 | 0.005 ± ~0 |
| | | SinoNet-FBP | 0.043 ± ~0 | 0.041 ± 0.001 | −26.879 ± 0.535 | 0.510 ± 0.001 |
| | | rFast-MMDNet | **0.003 ± ~0** | **0.001 ± ~0** | **−53.658 ± 0.732** | **0.001 ± ~0** |
| | Test | AA-MMD[182] | 0.068 ± ~0 | 0.058 ± ~0 | −21.782 ± 0.534 | 0.543 ± 0.004 |
| | | AA-MMD[182] + DenoiseNet | 0.031±~0 | 0.008±~0 | -28.723±0.501 | 0.010±0.001 |
| | | ID-UNet[116] | 0.009 ± ~0 | 0.006 ± ~0 | −34.476 ± 0.653 | 0.002 ± ~0 |
| | | ID-UNet[116] + DenoiseNet | 0.007 ± ~0 | 0.005 ± ~0 | −35.871 ± 0.632 | 0.002 ± ~0 |
| | | Triple-CBCT[117] | 0.006 ± ~0 | 0.001 ± ~0 | −41.778 ± 0.432 | 0.004 ± ~0 |
| | | SinoNet-FBP | 0.037 ± ~0 | 0.036 ± 0.001 | −28.724 ± 0.631 | 0.462 ± 0.003 |
| | | rFast-MMDNet | **0.002 ± ~0** | **~0 ± ~0** | **−55.708 ± 0.941** | **0.002 ± ~0** |
| *Fibroglandular* | Validation | AA-MMD[182] | 0.223 ± ~0 | 0.195 ± ~0 | −10.753 ± 0.782 | 0.543 ± 0.002 |
| | | AA-MMD[182] + DenoiseNet | 0.017±~0 | 0.019±~0 | -27.452±0.701 | 0.012±0.006 |
| | | ID-UNet[116] | 0.014 ± ~0 | 0.017 ± ~0 | −31.001 ± 0.681 | 0.004 ± 0.004 |
| | | ID-UNet[116] +DenoiseNet | 0.012 ± ~0 | 0.014 ± ~0 | −32.178 ± 0.632 | 0.003 ± 0.001 |
| | | Triple-CBCT[117] | 0.011 ± ~0 | 0.004 ± ~0 | −35.343 ± 0.721 | 0.003 ± ~0 |
| | | SinoNet-FBP | 0.279 ± ~0 | 0.237 ± 0.016 | −9.631 ± 0.578 | 0.312 ± 0.003 |
| | | rFast-MMDNet | **0.008 ± ~0** | **0.002 ± ~0** | **−40.654 ± 0.591** | **0.002 ± ~0** |
| | Test | AA-MMD[182] | 0.201 ± ~0 | 0.186 ± ~0 | −12.457 ± 0.889 | 0.482 ± 0.004 |
| | | AA-MMD[182] + DenoiseNet | 0.016±~0 | 0.017±~0 | -28.525±0.619 | 0.010±0.004 |
| | | ID-UNet[116] | 0.012 ± ~0 | 0.015 ± ~0 | −32.375 ± 0.794 | 0.002 ± 0.005 |
| | | ID-UNet[116] + DenoiseNet | 0.010 ± ~0 | 0.013 ± ~0 | −33.785 ± 0.513 | 0.002 ± 0.001 |
| | | Triple-CBCT[117] | 0.009 ± ~0 | 0.002 ± ~0 | −36.784 ± 0.741 | 0.002 ± ~0 |
| | | SinoNet-FBP | 0.264 ± ~0 | 0.229 ± 0.016 | −11.619 ± 0.682 | 0.293 ± 0.006 |
| | | rFast-MMDNet | **0.007 ± ~0** | **0.001 ± ~0** | **−43.349 ± 0.692** | **0.001 ± ~0** |

### 3.3.3 Ablation Studies

Ablation studies regarding the components of rFast-MMDNet and Triple-CBCT are conducted as seen in Table 3-3. Since the overall setup between Triple-CBCT and rFast-MMDNet is different, we slightly modified the network input/output channels per model adaption and keep the rest architecture unaltered. In specific, to adapt SD-Net and ID-Net into our rFast-MMDNet pipeline, we modified the output channel of SA-Net and the input channel SDNet from k (k=16) to 3. For adaption of SinoNet and DenoiseNet into Triple-CBCT framework, we changed the output channel of SinoNet and the input channel of DenoiseNet from 3 to 16. Since we propose to use loss combined with PSNR, SSIM, and $L_1$ deviation while Triple-CBCT used mean squared error (MSE) for model training, we also cross trained all the networks in Table 3-3 with the two loss functions. We can observe from Table 3-3 that models trained with combination loss can achieve better results than those trained with MSE loss. Additionally, both SinoNet and DenoiseNet can perform marginally better than SD-Net as well as ID-Net in the current dataset.

Table 3-3 Ablation Study of rFast-MMDNet with Triple-CBCT on test set.

| Method | Loss Function | RMSE | MAE | -PSNR | SSIM |
|---|---|---|---|---|---|
| SD-Net-FBP | Combination | 0.228 ± 0.125 | 0.153 ± 0.109 | −12.422 ± 1.452 | 0.402 ± 0.011 |
| | MSE | 0.231 ± 0.129 | 0.174 ± 0.134 | −10.205 ± 1.878 | 0.487 ± 0.014 |
| SinoNet-FBP | Combination | **0.173 ± 0.095** | **0.120 ± 0.095** | **−15.616 ± 1.378** | **0.381 ± 0.003** |
| | MSE | 0.187 ± 0.103 | 0.143 ± 0.018 | −11.734 ± 1.675 | 0.404 ± 0.007 |
| SD-Net-FBP + DenoiseNet | Combination | 0.006 ± 0.003 | 0.006 ± 0.006 | −41.078 ± 0.498 | 0.006 ± ~0 |
| | MSE | 0.006 ± 0.005 | 0.007 ± 0.015 | −39.843 ± 0.595 | 0.007 ± ~0 |
| SinoNet-FBP + DenoiseNet (rFast-MMDNet) | Combination | **0.004 ± ~0** | **0.005 ± ~0** | **−45.027 ± 0.542** | **0.002 ± ~0** |
| | MSE | 0.006 ± 0.003 | 0.006 ± 0.007 | −42.108 ± 0.664 | 0.005 ± ~0 |
| SinoNet-FBP + ID-Net | Combination | 0.005 ± ~0 | 0.006 ± ~0 | −43.073 ± 0.587 | 0.004 ± ~0 |
| | MSE | 0.006 ± 0.004 | 0.007 ± 0.008 | −41.878 ± 0.547 | 0.006 ± ~0 |
| SD-Net-FBP + ID-Net (Triple-CBCT) | Combination | 0.006 ± 0.004 | 0.007 ± 0.008 | −40.032 ± 0.515 | 0.007 ± ~0 |
| | MSE | 0.007 ± 0.006 | 0.008 ± 0.015 | −38.571 ± 0.631 | 0.008 ± ~0 |

## 3.4 Discussion

Accurate electron density (ED) estimation is crucial for precise dose calculations in breast radiation therapy. Conventional single-energy CT maps Hounsfield Units to electron density using empirical calibration curves, which can introduce inaccuracies due to beam hardening and patient-specific variations. Being an under-determined problem, the additional attenuation information presented in the second energy channel of DECT enables a more direct extraction of electron density value[183]. On the other hand, the additional information in DECT may still be inadequate to completely solve the MMD problem with more than two materials commonly involved in breast imaging and diagnosis, which hampers precise dose distribution calculations, increases errors in treatment planning, especially for techniques such as IMRT and VMAT (Photon IMRT/VMAT: Dose errors of 2-5%, potentially 5-10% in high-Z regions due to material decomposition inaccuracies; Proton Therapy: Stopping power ratio errors causing 1-2 mm per 1% error of stopping power ratio, with possible 3-5 mm range shifts in breast cancer cases.)[184]. Thus far, existing >2 breast DECT MMD methods mostly decomposed on the reconstructed image using algebraic/physical-theory integrated hand-crafted models, which have been shown inaccurate, sensitive to domain shift, and slow to optimize for each incoming image[185–187]. A recent learning method on sinogram Triple-CBCT showed the feasibility of >2 material MMD but used a network structure that is difficult to train, generalize, and impose quality control. Therefore, we present a robust two-stage rFast-MMDNet trained directly on dual-energy CT projections for multiple material decompositions. The network comprises two 3D CNNs (SinoNet and DenoiseNet) trained independently for DE projection decomposition and image post-processing. SinoNet and DenoiseNet share the same loss objective combining negative PSNR, L1, and MS-SSIM with two tunable hyper-parameters to enforce model optimization. Forward and filtered back projections were also integrated into rFast-MMDNet for domain adaptation. Quantitative and qualitative evaluations were carried out with our pipeline on the 2022 AAPM DL-Spectral-CT dataset. Previously reported DECT MMD algorithms, including AA-MMD, AA-MMD + DenoiseNet, ID-UNet, ID-UNet + DenoiseNet, and Triple-CBCT, were selected for comparison. In both visual and numeric experimental results, rFast-MMDNet was competitive in all evaluation metrics and outperformed all comparison methods. Besides average error and similarity metrics, rFast-MMDNet resulted in the lowest and close to zero standard deviation, indicating superior model robustness.

We compared rFast-MMDNet with four representative methods, including hand-crafted image-domain direct inversion (AA-MMD), DL-based image-domain decomposition (ID-UNet), image domain MMD (AA-MMD/ID-UNet) + DenoiseNet, and DL based projection-domain MMD (Triple-CBCT) [117]. Unsurprisingly, without a learning component, AA-MMD performed worst, showing high noise level and poor boundary fidelity. Both AA-MMD and ID-UNet, which operated in the image domain, were hampered by the information loss in domain transfer. Triple-CBCT and rFast-MMDNet hold a theoretical advantage due to their direct access to raw information in the sinograms. The theoretical advantage is confirmed by the superior performance of the two sinogram-based methods vs. image domain MMD + DenoiseNet methods on the breast data.

We attribute the superior performance of rFast-MMDNet in comparison to SinoNet + FBP to the synergy of SinoNet + DenoiseNet. Because the composition of a decomposed image ($I'$) in theory can be seen as information ($I$) + noise ($N$), both factors influence the predicted image quality. The difference between MMD in image and sinogram domains lies in $I$, where properly designed sinogram MMD theoretically can achieve better $I$ than that of image MMD. Accordingly, though in the case of rFast-MMDNet, FBP as an imperfect backprojection option that introduces streaking and cupping artifacts, those artifacts can be practically treated as noise and factored into $N$. Hence, it is mostly the artifacts introduced by FBP in $N$ that lead to to significantly worse performance of SinoNet + FBP than AA-MMD and ID-UNet. Thus, we ran DenoiseNet right after the step of SinoNet + FBP to suppress $N$ from $I'$. Whereas, in image-domain MMD, $N$ is already suppressed in the reconstruction process with accompanying information loss from the sinogram space to the imaging space. In comparison, sinogram based MMD methods keep richer information and more noise. Subsequently as shown in Tab. 1 and Tab. 2, DenoiseNet improves SinoNet + FBP results more than those of AA-MMD and ID-UNet.

Of the two DL methods on sinograms, rFast-MMDNet is superior to Triple-CBCT due to the following attributes. 1) Clearer pipeline design to match network with task: Triple-CBCT splits the task of projection decomposition between SD-Net and ID-Net for transmission to VMI sinograms and VMI grouping, respectively, while multitasking on image refinement. The unclear task distribution among network modules limited model performance and interpretability. In contrast, rFast-MMDNet divides projection decomposition and image post-processing into separate SinoNet and DenoiseNet, each with its own objective. The clear separation reduces the training burden and encourages fast convergence. 2) More robust Network Design in two fold: First, the SD-Net in Triple-CBCT uses mixed spatial filter sizes of $1 \times 3$ and $3 \times 3$, and randomly plugs several skip-connection branches into the network flow, which causes inconsistent information to be passed throughout convolutions. Also, SD-Net consists of only seven layers, which limits the number of trainable parameters and model complexity. SinoNet, as the first stage of rFast-MMDNet, has a balanced U-shape with consistent skip-information flow between encoding and decoding. The balanced network combines 15 layers (7 encoding + 1 bottleneck + 7 decoding), with all layers having 3 × 3 convolution and 2 × 2 max-pooling filters, to achieve higher network learning and generalization capacity. Second, instead of directly stacking eight convolutional layers with inter-layer skip-connection for image domain processing in ID-Net of Triple-CBCT, DenoiseNet improves image denoising efficiency by stacking 16 ResNet blocks, where each ResNet Block has consistent intra-layer skip connection to ensure denser residual information feeding and effectively avoids gradient vanishing in backpropagation. 3) Reduced training load: Compared to the domain transfer net of Triple-CBCT that requires additional training, rFast-MMDNet relaxes the requirement of accurate domain adaption and leaves the final polish of the image quality to DenoiseNet. As a result, rFast-MMDNet requires the training of fewer networks while improving the prediction accuracy.

The top-ranked methods from the AAPM challenge achieved near-zero errors, further improvement of which is neither meaningful nor practical. While achieving comparable performance to these reported methods, we believe rFast-MMDNet holds a theoretical advantage in speed without incorporating the time-consuming iterative reconstruction as part of the network training and inference in most top performers[172].

Despite the encouraging results of automatic MMD frameworks demonstrated by rFast-MMDNet, they are several limitations. First, the supervision for SinoNet is forward projection sinograms, which are generated assuming ideal projection conditions. Specifically, a continuous object is discretized into a finite number of parameters per ray tracing modeling[188] for a fast and memory-efficient implementation of forward projection. Alternatively, more accurate and realistic sinograms can be generated using methods such as the pixel-basis/Kaiser-Besel basis functions or distance-driven forward projections[188], at a significantly higher computational cost. Second, FBP is used as the back projection algorithm in rFast-MMDNet for fast domain transformation, which could introduce unwanted artifacts in decomposed images. In the present solution, DenoiseNet is implemented to mitigate, not eliminate, the issue since limited data is available for model training. In future study, while more private/public data is available, we will consider combining transfer and federated learning to train a more robust and versatile denoise network through sufficient expose to diverse kinds of noise for better suppression of the artifacts introduced by FBP[189]. Thirdly, we used a synthesis dataset to evaluate rFast-MMDNet, where the GT is available and "noise-free". Yet often times GTs are missing or noisy in practice. Therefore, rFast-MMDNet needs to be further tested on real-world data. Fourthly, the proposed rFast-MMDNet requires raw material sinograms training supervision, which is hard to obtain without accurate geometry of the DECT scanner. Fifthly, we were unable to directly compare rFast-MMDNet with AAPM Grand Challenge Top Performers due to insufficient details to reproduce their work[172]. The comparison is needed to quantify if the speed difference is relevant in the clinical setting. Lastly, although rFast-MMDNet has an architecture conducive to generalization, the point needs to be further validated on additional datasets, particularly patient datasets acquired from different scanners and protocols. Currently, both our training and testing are limited by the available data. Data scarcity is a common problem for machine learning and deep learning research. Efforts have been made to increase the data availability via auto-/semi-auto annotation and mitigation of confidentiality concerns via techniques such as federated learning[190].

## 3.5    Summary

A robust and efficient two-stage DECT MMD pipeline to achieve an accurate decomposition of three materials directly using projection sinograms is presented in this study. The method was tested on digital breast phantom images with ground truth classification of the fibroglandular, adipose and calcification materials. Both qualitative and quantitative results show the advantage of using our network for fast, robust, and accurate MMD on DECT.

# 4. Accelerated 4D Magnetic Resonance Imaging Reconstruction

## 4.1 Dynamic MRI Reconstruction with Convolutional Network Assisted Reconstruction Swin Transformer

### *4.1.1 Introduction*

Lately, Transformer were introduced in the context of machine translation at the aim of avoiding recursion so as to allow parallel computation (decreased training and inference time) and meanwhile to reduce drops in performance due to long sequential dependencies [30]. The vision community is witnessing a modelling shift from CNNs to Transformers which have attained top accuracy in the major video processing benchmarks[36]. Apart from the preeminence in positional-embedding, Video Swin Transformer (VST), as one of the frameworks beneficial to video understanding, uses a novel hierarchical shifted window mechanism, which perform efficient self-attention in the spatial-temporal domain using non-overlapping local windows and cross-window connection[191]. Broadly speaking, 4D MRI essentially is a subcategory of video data. Considering the superiority of VST in computational efficacy and self-attention based spatial-temporal domain understanding, we propose to extend the architecture of VST to the task of DMRI reconstruction.

To this end, we developed a novel NN called Reconstruction Swin Transformer (RST) specialized in real-time and qualitative 4D MR recovery. To accelerate the training speed and reduce the complexity of RST, we employed a CNN based 2D reconstruction network, named SADXNet[181], for rapid recovery of MR frames prior feeding into RST. To the best of our knowledge, this is the first work that introduces self-attention and positional embedding mechanisms to the task of 4D MRI reconstruction.

### *4.1.2    Methods*

#### 4.1.2.1    SADXNet Assisted Reconstruction Swin Transformer

As mentioned, our 4D MRI reconstruction framework consists of two stages. 1) Initial individual 2D MR frame reconstruction using SADXNet[181]. 2) 4D tensor learning in the spatial-temporal domain with RST. Noted that, RST theoretically can yield comparable performance without SADXNet initialization conditioned upon unlimited network parameters, training/prediction time and GPU capacity. Nevertheless, numerous studies showed that training on high-resolution 4D tensors requires significantly more network parameters to fully generalize the feature representation, which takes overly long training time to converge, deviates from the aim of sub-second prediction, and is beyond most of the commercial GPU memories[192,193]. Thus, we proposed to first use SADXNet for 2D spatial information recovery and then followed by a comparable lighter-weighted RST for collective spatial-temporal data reconstruction. We will elaborate on the structures of SADXNet and RST in the following subsections.

**SADXNet.** SADXNet was first introduced as an image-denoising network for rib-suppression in chest X-ray images[181]. Since tasks of image denoising and reconstruction fundamentally both work on pixel intensity recovery, we applied SADXNet for reconstruction of 2D MR frames. SADXNet is designed to be fully convoluted and densely connected as shown in Figure 3-1. In specific, first, there are 7 densely connected layers in SADXNet. The dense connectivity of SADXNet is designed to be channel-wise concatenation where feature maps produced from a specific layer and all its preceding layers will be concatenated in the channel dimension and fed into all its subsequent layers. Second, for each convolution layer, three consecutive operations are conducted, including batch normalization (BN), rectified linear unit (ReLU), and convolution with filter size of 3×3. Third, no up- or down-sampling layer is implemented in the overall network design to preserve the spatial dimension of feature maps through the overall network flow. Fourth, in each layer, the channel size of its corresponding kernel is organized in an increase-to-decrease setting for balancing between the model complexity and time efficacy. Lastly, the cost function for SADXNet is organized as a combination of negative PSNR, muti-scale SSIM (MS-SSIM), and $L_1$ deviation measurement as shown in Equation (4-1)[181].

$$L = -\alpha \cdot L_{PSNR} + (1-\alpha) \cdot [\beta \cdot L_{MS-SSIM} + (1-\beta) \cdot L_1] \tag{4-1}$$

Where $\alpha$ and $\beta$ are tunable hyperparameters.

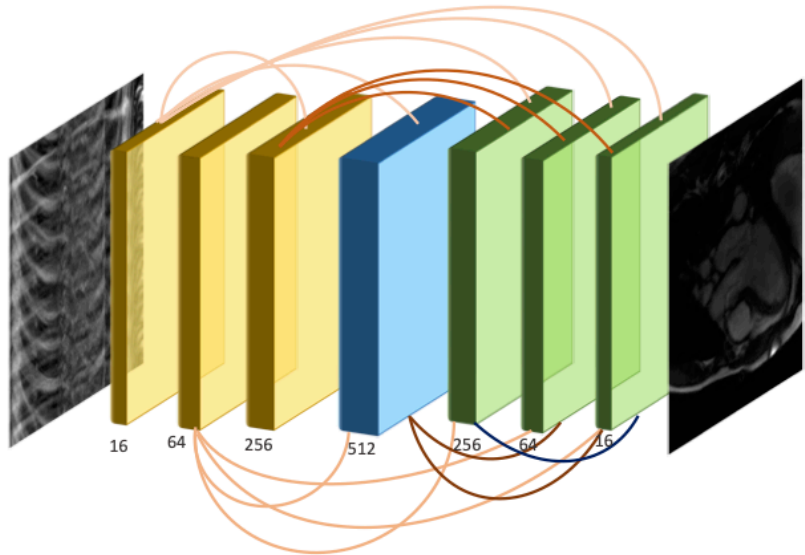

*Figure 4-1 Network architecture of SADXNet*

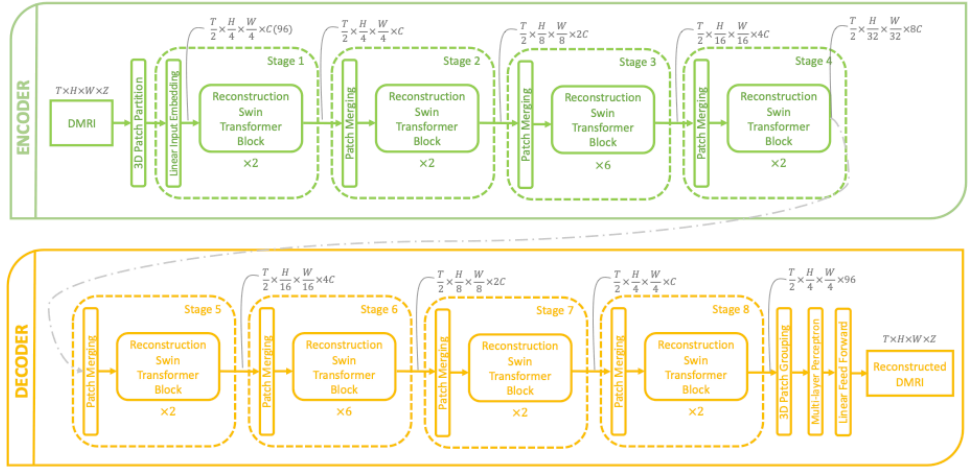

*Figure 4-2 The complete architecture of RST-T*

**Reconstruction Swin Transformer.** RST strictly follows the hierarchical design of the backbone (encoder) of the original VST but introduces a novel reconstruction head (decoder) to fulfill the goal of sequence reconstruction. Similar to the variants of VST, RST also has four architectures with distinct model complexities. 1) RST-T: $C=96$, block numbers $= \{2,2,6,2,2,6,2,2\}$; 2) RST-S: $C=96$, block numbers $= \{2,2,18,2,2,18,2,2\}$; 3) RST-B: $C=128$, block numbers $= \{2,2,18,2,2,18,2,2\}$; 4) RST-L: $C=192$, block numbers $= \{2,2,18,2,2,18,2,2\}$, where $C$ represents the channel number of the first RST block[191].

We illustrate the architecture of RST-T in Figure 3-2. In specific, firstly, the encoder of RST rigorously inherits the hierarchical structure of the original VST backbone, which consists of four stages and performs $2\times$ down-sampling solely in the spatial dimension of the patch merging layer in each block [191]. The RST blocks of encoder and decoder are symmetrically arranged, except the difference that all the patch merging layers in decoder performs spatial $2\times$ up-sampling. Secondly, regarding tokenizing the 4D DMRI, under the definition that the input 4D tensor of dimension of $T\times H\times W\times Z$, we treat each patch of size $2\times 4\times 4\times Z$ as a 3D token, where the Z-axis of DMRI is considered as the color channel in natural images. Thus, after processing through the 3D patch partition layer in the beginning of encoder, we obtain $\frac{T}{2} \times \frac{H}{2} \times \frac{W}{2} \times Z$ 3D tokens. Meanwhile, a linear embedding layer is applied right after 3D patch partition layer to project the channel dimension from $Z$ to pre-specified $C$ for the destinated RST variant, which is 96 in RST-T.

The critical component of the overall architecture is the RST block, where its design and difference from the standard transformer block can be seen from Figure 3-3. Specifically, the structure of RST block is identical to that of VST, which is built through replacing the muti-head self-attention (MSA) layer in standard transformer block with the 3D shifted window based muti-head self-attention (SW-MSA) layer and have the rest sub-components unaltered [191]. Explicitly, a RST block starts with layer normalization, followed by 3D SW-MSA, and finalized with another layer normalization and then 2-layer muti-layer perceptron with GELU[194] nonlinearity in between.

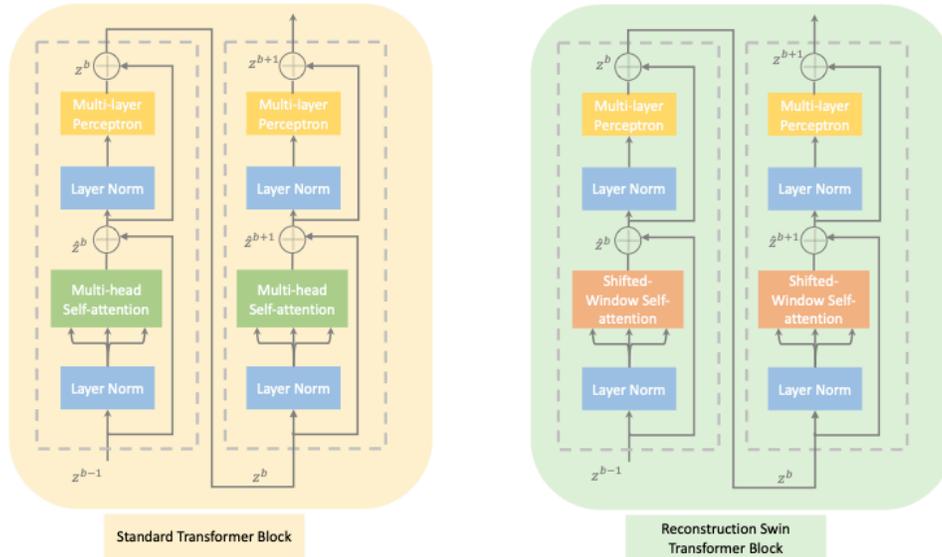

Figure 4-3 The components of Reconstruction Swin Transformer block and comparison with standard Transformer block.

Our design of the SW-MSA follows that of VST, which introduces a locality inductive bias to the self-attention module to accommodate additional information intake resulted from the $t$-axis. SW-MSA mechanism is realized with two consecutive W-MSA modules. In specific, in the first W-MSA, given an initial input composed of $T' \times H' \times W'$ 3D tokens and a window of size $P \times M \times M$, we will arrange the windows in a manner that can non-overapply partition the input tokens. Next, to establish cross-connections among various windows, 3D shifted window mechanism is introduced in the immediately following W-MSA layer. Given the $\frac{T'}{P} \times \frac{H'}{M} \times \frac{W'}{M}$ windows obtained from the first W-MSA having size of $P \times M \times M$, the partition configuration of the second W-MSA will shift the windows along the temporal and spatial axis by $\frac{P}{2} \times \frac{M}{2} \times \frac{M}{2}$ tokens from the partition coordinates in the first W-MSA. In Figure 3-4, given tokens and window of size $8 \times 8 \times 8$ and $4 \times 4 \times 4$ respectively, W-MSA-1 will partition the tokens into $\frac{T'}{P} \times \frac{H'}{M} \times \frac{W'}{M} = 2 \times 2 \times 2 = 8$ windows. Next, the windows are shifted $\frac{P}{2} \times \frac{M}{2} \times \frac{M}{2} = 2 \times 2 \times 2$ tokens, which results in $3 \times 3 \times 3 = 27$ windows in W-MSA-2.

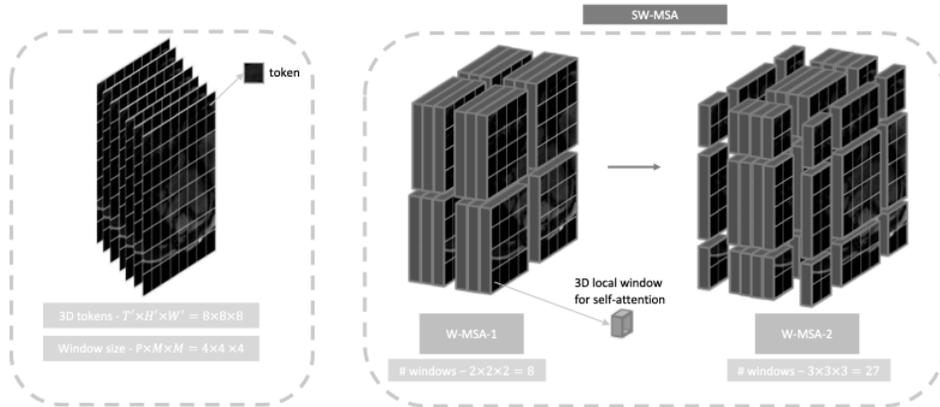

*Figure 4-4 An instance to illustrate the SW-MSA mechanism*

The cost function of RST is designed to be identical to that of SADXNet, which is a combination of PSNR, MS-SSIM, and $L_1$ deviation.

**Experimental Setup.** Both SADXNet and RST were implemented in Pytorch, and the training was performed on a GPU cluster with $4 \times RTX\ A6000$. For training of SADXNet, data augmentation including random rotation, random resizing, random cropping, and random Gaussian blurring was implemented. Adam optimizer with initial learning rate of 0.001, epoch of 1000 and batch size of 2×4 was applied during learning. For training of RST-T, data augmentation including random cropping, random resizing, and random flipping was implemented. Adam optimizer with initial learning rate of 0.001 and batch size of 1×4 was applied during training.

**Baseline Algorithms.** We compared with four state-of-the-art approaches including k-t SLR[135], k-t FOCUSS+ME/MC[195], DC-CNN (3D)[139], and MODRN (e2e)[143]. The first two are conventional CS-based methods, where solely k-t FOCUSS+ME/MC includes ME/MC procedures, whereas the last two are DL methods. DC-CNN (3D) explores spatial-temporal information using 3D convolution and MODRN (e2e) includes ME as well as MC into their RNNs. Three qualitative metrics are used for evaluation including RMSE, PSNR and SSIM.

#### 4.1.2.2  Data Cohort

**Coronal and Sagittal Lung Data (CSLD).** The CSLD dataset consists of DMRI sequences imaged from 18 subjects - 10 healthy volunteers and 8 lung tumor patients with approval of local Institutional Review Board (IRB). All studies were performed without breath coaching on a 1.5-Tesla whole-body clinical MRI scanner (Avanto, Siemens Medical Solution, Erlangen, Germany), using a four-element phased-array body coil and a spinal coil. The imaging parameters were set as repetition time/echo time 3.1/0.93 ms, FOV 300 × 206 mm, flip anger 52°, slice thickness 7 mm, and phase sharing 120. Scans from coronal and sagittal panes are available with matrix size of 192 × 192 as well as 144 × 192 respectively and each of frames 120. Each subjects underwent 2-3 times of scans within two weeks with both coronal and sagittal view scans available. More details can be found in[20,136,196–198]. For preparing the training data, in order to simulate k-space data, we transferred the data into k-space under the assumption of Cartesian-sampling. Next, we used VISTA[199] to acquire under-sampled sequences at AR of 5 × and 8 ×. Then, we transferred the under-sampled measurements into image domain to prepare the training and validation dataset. Lastly, the dataset split was conducted subject-wise with ratio of training/test = 14/4 to evenly balance out healthy volunteers and patients in each set.

*4.1.3    Results*

For the CSLD dataset, performance comparison was made among SADXNet initialized RST-T, solely RST-S, and the selected benchmark algorithms - k-t-SLR, k-t-FOCUSS-ME/MC, DC-CNN, and MODRN (e2e). We respectively trained SADXNet+RST-T and RST-S for around 2000 and 8000 epochs till observing training convergence.

Quantitative results and visualization of the validation set are reported in Table 4-1 and Figure 4-5 (a-b). Overall, statistical results and visual outcomes consistently show that DL models, including DC-CNN, MODRN (e2e), RST-S, and RST-T, significantly outperform CS methods (k-t-SLR and k-t-FOCUSS+ME/MC). In addition to the promising performance illustrated by the quantitative metrics in Table 4-1, we can also observe that the DL predictions in Figure 4-5 have fewer residuals and streaking artifacts, sharper edge delineation, detailed morphology (arrows pointed), and better concentration of optical flow estimated motion on the cardiac region of interest (ROI) than the ones made by CS methods. Those imply that DL methods have better dynamic motion understanding than the selected CS baselines. Meanwhile, all the DL approaches take considerably less time to reconstruct incoming dynamic imaging than CS algorithms, as seen in Table 4-1.

Within the two CS-based methods, the predictions from k-t-FOCUSS+ME/ME are marginally better than those from k-t-SLR. Among all listed neural networks, the performance ranking can be summarized as SADXNet initialized RST-T > RST-S > MODRN (e2e) > DC-CNN with SADXNet+RST-T, RST-S, and DC-CNN able to make a sub-second prediction. Additionally, though trained significantly longer for 8000 epochs, RST-S still has difficulty capturing finer morphologies and subtle dynamic movements compared with SADXNet + RST-T, which further substantiates the importance of CNN 2D frame initialization prior to RST training in practice. Finally, we observe that predicted frames from SADXNet+RST are consistent and steady along the T axis in Figure 4-5 (b), where the robustness of our proposed framework is demonstrated.

*Table 4-1 The statistical results on CSLD test sets with 5x and 8x ARs.*

| AR | Algorithm | RMSE↓ | PSNR↑ | 1-SSIM↓ | Prediction Time (s) | Device |
|---|---|---|---|---|---|---|
| 5x | k-t SLR | 0.0249±0.00835 | 28.281±5.183 | 0.238 ±0.101 | 183.82 | CPU |
|  | K-t FOCUSS + MC/ME | 0.0173±0.00996 | 32.551 ± 5.732 | 0.195±0.0893 | 226.891 | CPU |
|  | DC-CNN | 0.0135±0.00973 | 39.892±5.986 | 0.149±0.0423 | 0.501 | GPU |
|  | MODRN (e2e) | 0.00999±0.00602 | 42.013±6.892 | 0.104±0.0332 | 4.987 | GPU |
|  | RST-T | 0.00954±0.00598 | 42.783±6.489 | 0.0956±0.0293 | 0.299 | GPU |
|  | SADXNET + RST-T | **0.00743±0.00524** | **45.785±5.854** | **0.0687±0.0231** | **0.574** | GPU |
| 8x | k-t SLR | 0.0271±0.00921 | 27.329±4.998 | 0.251 ±0.113 | 183.82 | CPU |
|  | K-t FOCUSS + MC/ME | 0.0189±0.00982 | 31.187 ± 5.525 | 0.215±0.00994 | 226.891 | CPU |
|  | DC-CNN | 0.0151±0.00932 | 38.234±5.687 | 0.138±0.0435 | 0.501 | GPU |
|  | MODRN (e2e) | 0.0121±0.00634 | 40.977±6.995 | 0.195±0.0385 | 4.987 | GPU |
|  | RST-T | 0.0107±0.00586 | 41.378±6.985 | 0.109±0.0331 | 0.299 | GPU |
|  | SADXNET + RST-T | **0.00873±0.0569** | **42.547±5.214** | **0.0753±0.0242** | **0.574** | GPU |

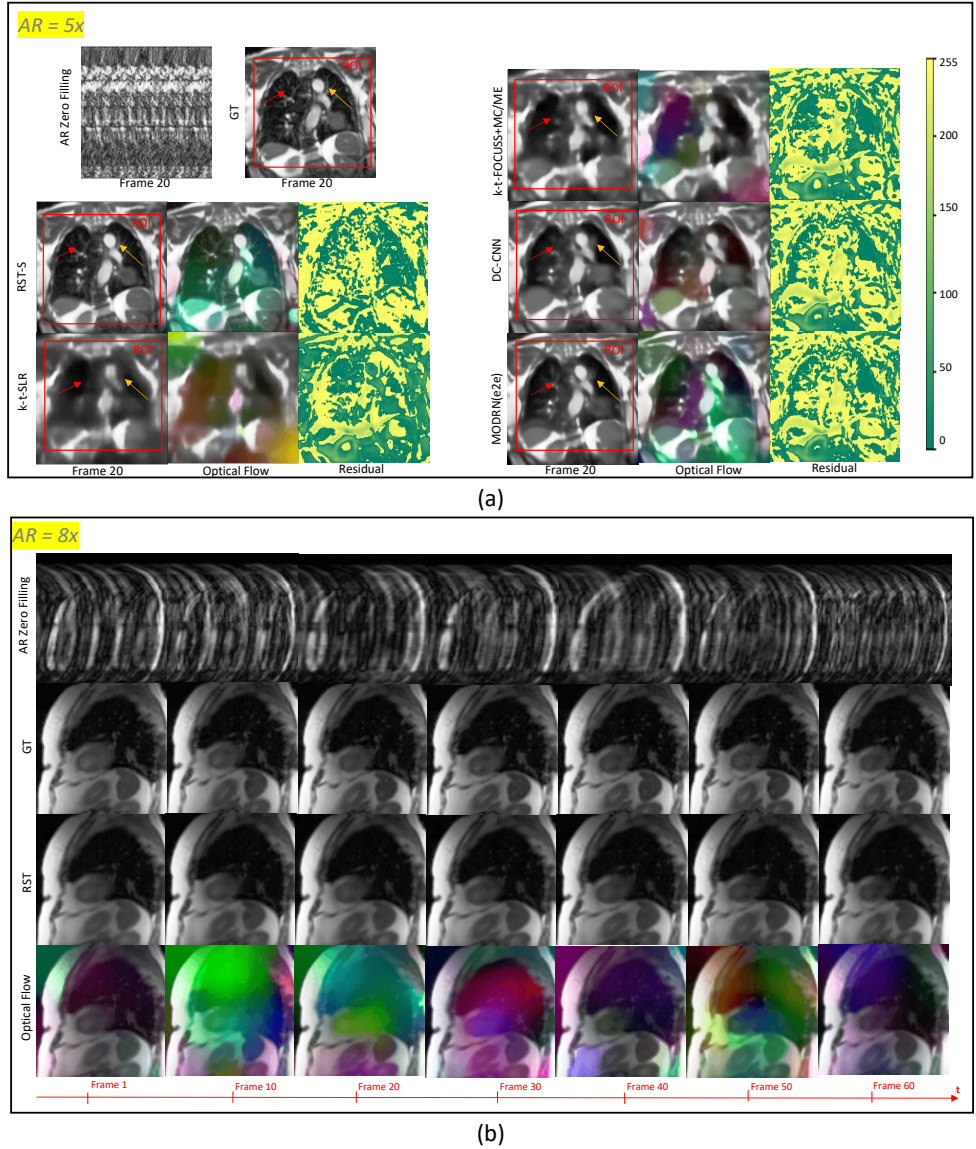

*Figure 4-5 Visualization of reconstruction results on a coronal-view sequence in the test set and Visualization of frames 1-60 prediction from a sagittal-view test sequence*

## 4.1.4 Discussion

The study in the current section presents a novel RST architecture adapted from VST for DMRI reconstruction. RST inherits the backbone of VST and introduces a novel reconstruction head to fulfill the demand for pixel-wise intensity prediction. Compared with existing DL-based reconstruction approaches[139,143], RST markedly improved efficiency and efficacy. The efficiency stems from the positional embedding scheme of the Transformer, which enables fully distributable processing in the spatial and temporal domains, resulting in enormously reduced training and prediction time. The efficacy is boosted via SW-MSA, which conducts self-attention learning in both spatial and temporal dimensions. Additionally, the SADXNet for the initial restoration of 2D MR frames substantially reduced the training burden, model parameters, and GPU memory footprint. RST-based algorithms outperformed the comparison SOTA CS and DL reconstruction methods by a large margin in the experiments reconstructing the unsampled OCMR dataset. Compared to training RST-S on raw under-sampled images, SADXNet-initiated RST-T delivered an additional performance boost in statistical and visual outcomes while significantly reducing training and model complexity. The results thus support the importance of 2D frame initialization prior to official training from the 4D domain.

Our method is not without room for improvement. SADXNet was introduced to reduce the computation demand of RST, but the current framework still requires substantial GPU memory restricted to higher-end non-consumer grade hardware. As SADXNet demonstrates the promise of RST parameter reduction with spatial domain initialization, we will explore temporal domain initialization before conducting joint 4D dimensional learning in future studies.

### *4.1.5 Summary*

A dynamic MR reconstruction framework, SADXNet-assisted RST, is proposed with accuracy enabled through the "swin window" mechanism and efficiency guaranteed from Transformer positional embedding and CNN spatial initialization. Validation from the CSLD dataset substantiates the superior performance of SADXNet+RST to SOTA CS and DL methods.

## 4.2 Highly Accelerated Liver 4D MRI Reconstruction with Paired Conditional Generative Adversarial Network

### *4.2.1 Introduction*

Liver 4D MRI data is usually acquired in a continuous free-breathing scan followed by data sorting based on respiratory motion surrogates. One common option is a 3D golden angle stack-of-stars sequence with self-navigation to acquire 4D MRI datasets[119]. In stack-of-stars acquisition, radial sampling is employed in the $kx - ky$ plane, which enables reduced motion sensitivity[200] and incoherent k-space under-sampling if acceleration is desired[201]. Cartesian sampling is used in the kz dimension, which allows for a flexible selection of volumetric coverage/slice resolution[202]. However, this technique has a few limitations. Scan time is usually long (8-10 min[119]), and slice resolution is often sacrificed to maintain sufficient volumetric coverage and in-plane resolution, increasing the inaccuracy of small malignancy contouring[202]. Streak artifacts caused by under-sampling, sampling trajectory deviation, or nonuniform k space coverage can be challenging to mitigate with conventional constrained reconstruction. Parallel imaging[122,203] and compressed sensing[131,204,205] have been employed to accelerate both static and dynamic MRI.

It is well known that the success of NN training, apart from exploring various architectures, hinges on the loss function design[206,207]. However, designing an effective loss function that encourages NNs to precisely converge towards the target often requires balancing conflicting constraints such as sharpness vs. streak artifacts reduction. GANs[158] took an alternative approach – rather than explicitly specifying all components of the loss function, the discriminator network implicitly guides the generator loss reduction by distinguishing real from synthesized images. This adversarial dynamic can lead to outputs that are closer to the ground truth target. Inconsistent and blurry predictions are discriminated against in GANs. Additionally, since GANs only require the generator model at inference time, the adversarial training process does not add computational burden during reconstruction. This makes real-time 4D MRI reconstruction more tractable[158].

Several previous studies have demonstrated superior MRI reconstruction using GANs. For Cartesian sampling, Yang et al.[208] demonstrated that their UNet-based conditional GAN could provide better reconstruction with preserved perceptual imaging details than non-adversarial CNN methods on 3D T1-weighted brain and cardiac MRI dataset. Mardani et al.[209] built a least squares conditional GAN, demonstrating competitive performance in pediatric contrast-enhanced 3D MRI reconstruction. For non-Cartesian sampling, Liu et al.[210] presented a robust performance by cycle-GAN trained with varying under-sampling patterns on 3D golden-angle radial sampled liver imaging. Gao et al.[211] also demonstrated the feasibility of using a conditional GAN framework for 3D stack-of-radial Liver MRI reconstruction. However, the previous work has not explored the capability of GANs in 4D MRI temporal profiling and reconstruction.

The current section explores the feasibility of using GANs for 4D MRI reconstruction. We have developed a novel architecture termed Reconstruct paired Conditional GAN (Re-Con-GAN), specifically for 4D MRI reconstruction. The proposed framework is designed to learn 2D+time image series from under-sampled data. Experiments on an in-house 4D liver MRI dataset demonstrate the superior performance of Re-Con-GAN compared to conventional compressed sensing and supervised deep learning reconstruction models. To further validate the robustness of Re-Con-GAN's reconstructed images, we evaluate the impact on downstream tasks of liver tumor detection and segmentation using a Mask R-CNN[212] pipeline.

## 4.2.2 *Methods*

### 4.2.2.1 Data Cohort

The study was approved by the local Institutional Review Board at UCSF (#14-15452). 48 patients were scanned on a 3T MRI scanner (MAGNETOM Vida, Siemens Healthcare, Erlangen, Germany) after injection of hepatobiliary contrast (gadoxetic acid; Eovist, Bayer) for each patient. A prototype free-breathing T1-weighted volumetric golden angle stack-of-stars sequence was used for 4D MRI acquisition. The scanning parameters were - TE=1.5 ms, TR=3 ms, matrix size = 288x288, FOV = 374 mm x 374 mm, in-plane resolution=1.3 mm × 1.3 mm, slice thickness=3 mm, radial views (RV) per partition=3000, number of slices or partitions = 64-75, acquisition time = 8-10 min. The pulse sequence ran continuously over multiple respiratory cycles. Images reconstructed from the entire space data of 3000 radial spokes (RV-3000) were treated as the fully sampled ground truth reference (Based on Nyquist sampling theorem, fully sampled radial images require sampling points $\times \frac{\pi}{2}$ spokes, resulting in 452 spokes for a matrix size of 288 × 288. After motion binning, each of the 8 bins has on average 375 spokes with RV3000, which is close to 452 spokes and could well preserve imaging quality). Retrospective under-sampling was performed by keeping the first 1000, 500, and 300 spokes from the 3000 spokes, respectively, corresponding to acceleration rates of 3x, 6x, and 10x. For initial image reconstruction, data sorting based on a self-gating signal was performed to divide the continuously acquired k-space data into 8 respiratory phases. nonuniform fast Fourier transform (nuFFT) algorithm was applied to reconstruct each phase individually.

Only regular breathers (48 patients) were included in the current project. Breathing regularity was quantified using the self-gating signal waveform (No Biofeedback)[210,213,214]. The peak-to-trough range and mid-level amplitude (A), i.e., (peak-A + trough-A)/2, were calculated for each respiratory cycle. The average mid-level amplitude across all respiratory cycles normalized with the average peak-to-trough range was used as the regularity measurement. Patients with a score greater than 20% were classified as irregular breathers and excluded.

To augment the sample size, we organized the 48 3D+t data as 12332 2D+t images with images from an individual patient sorted in one subset. The data was split into training (37 patients with 10721 2D+t images) and testing (11 patients with 1611 2D+t images), where patients with various profiles (body mass index and breathing regularity score) are balanced in each split. The images were resized to 256 × 256 and normalized using Z-score normalization. Data augmentation was employed, including random rotation, flipping, and cropping.

### 4.2.2.2 Paired Conditional GANs

#### 4.2.2.2.1 Network Architecture

The architecture of GANs can be designed in unconditional or conditional settings. Unconditional GANs learn a mapping from a random noise vector $z$ to output image $y$, $G: z \rightarrow y$, whereas conditional GANs (cGANs) learn a mapping from an observed image $x$ as well as a random noise vector $z$ to output image $y$, $G: \{x, z\} \rightarrow y$. cGANs can be further classified into paired and unpaired architectures. Paired cGANs learn a one-to-one mapping of input to output, while unpaired cGANs only conduct domain-level supervision with input and output randomly selected from its domain data corpus. The current work employed a cGAN structure to perform the image-domain 4D MRI reconstruction as a paired image-to-image translation task.

Paired cGANs consist of two major components – generator $G$ and discriminator $D$. On the one hand, $G$ is trained to generate a "fake" reconstructed image series that cannot be differentiated from their corresponding "real" fully sampled ground truth (GT) image series by $D$. On the other hand, $D$ is trained to classify between "fake", $G$ synthesized image series $\hat{G}(x)$ and $x$, and "real", fully sampled image series $y$ and $x$, tuples. In cGANs, both $G$ and $D$ can observe input under-sampled image series. Details of the discriminator training workflow is diagrammed in Figure 3-6.

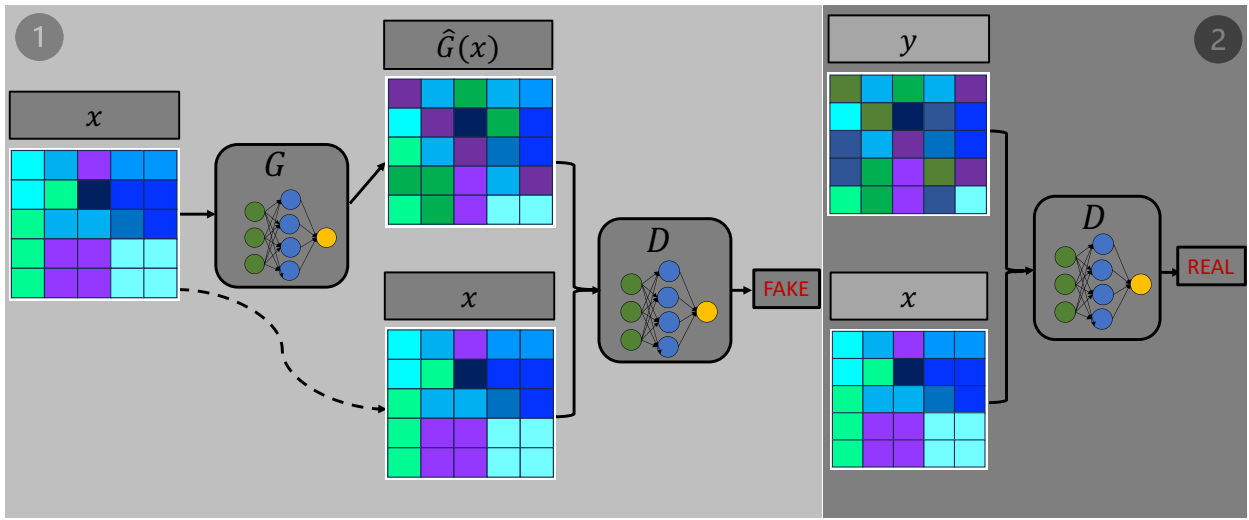

*Figure 4-6 Training cGANs Discriminator D to map input images x to target domain y.*

Our design of the generator and discriminator is improved from Isola et al.[158]. Re-Con-GAN is a versatile architecture with plug-and-playable generator, discriminator, and loss objective. A couple of examples for each sub-component are experimented with and demonstrated in this paper. Details of the model architectures can be visualized in Figure 4-7 and are elaborated as follows.

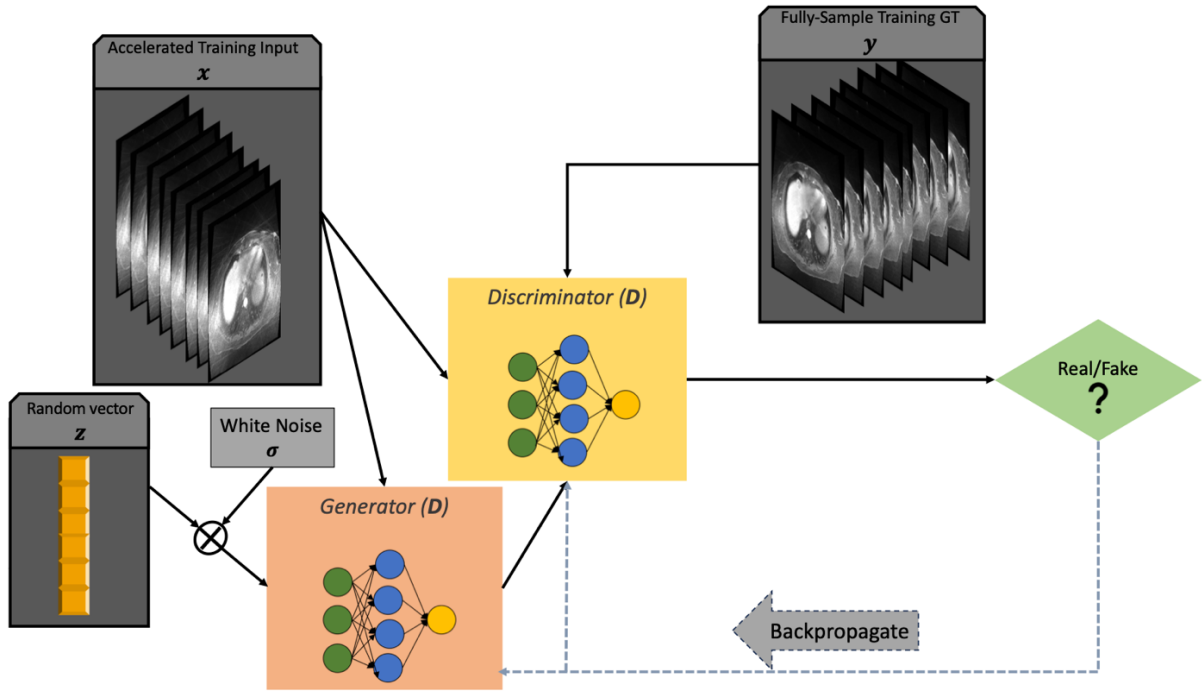

*Figure 4-7 The design of our proposed Re-Con-GAN.*

### 4.2.2.2.1.1   Generator

The defined task for the generator of 4D MRI reconstruction is to map a low-resolution input grid with noise and artifacts to a high-resolution noise- and artifact-reduced output grid. The input and output image series differ in morphological details and a few surface structures. But they intrinsically both render the same underlying general structures. Several previous DL approaches solve the 4D MRI reconstruction problem with encoder-to-decoder NNs[139,144]. Such architectures conduct a series of down-sampling operations until they reach the bottleneck layer, and then reverse the operations to up-sampling to gradually recover the dimension of the feature map from that of input. The edge of the encoder-to-decoder architecture is that numerous low-level information shared between input and output is revealed per progressively down-sampling while the difference between output and input is parameterized with the up-sampling layers.

**3D ResNet9**: The 3D ResNet9 is adopted from Johnson et al.[215]. Let $c7s1 - k$ denote a residual block consisting of a $7 \times 7$ convolutional layer with $k$ number of filters and stride of 1, Instance normalization and ReLU operations. $dk$ denotes a residual block consisting of a $3 \times 3$ convolutional layer with $k$ number of filters and stride of 2, instance normalization and ReLU operations. $Rk$ denotes a residual block consisting of two of $3 \times 3$ convolutional layers with the same $k$ number of filters and stride of 2, instance normalization, and ReLU operations. $uk$ denotes a fractional-stride residual block consisting of two of $3 \times 3$ convolutional layers with the same $k$ number of filters and stride of $\frac{1}{2}$, instance normalization, and ReLU operations. Reflection padding is used per convolution to reduce artifacts. The 3D ResNet9 is structured as Equation (4-2).

$$c7s1 - 64, d128, d256, R256, R256, R256, R256, R256, R256, R256, R256, R256, R256, u64, c7s1 - 8 \quad (4\text{-}2)$$

Where $c7s1 - 8$ is used to map the prediction to the expected number of output channels (8 in the current work).

**3D UNet**: The 3D UNet structure follows the design of Isola et al.[158]. Let $Ck$ denote a UNet block consisting sequentially of a $4 \times 4$ convolutional layer with $k$ number of filters and stride of 2, batch normalization and ReLU operations. Let $CDk$ denote a UNet block consisting sequentially of a $4 \times 4$ convolutional layer with $k$ number of filter and stride of 2, batch normalization, dropout at 50%, and ReLU operations. Convolutions in the encoder stage down-sample by a factor of 2 at each block, whereas those in the decoder stage up-sample by a factor of 2. The 3D UNet is structured as Equation (4-3).

$$\begin{aligned} encoder &: C64, C128, C256, C512, C512, C512, C512, C512 \\ decoder &: CD512, CD1024, CD1024, CD1024, CD1024, CD512, CD256, CD64, CD8 \end{aligned} \quad (4\text{-}3)$$

Where $\mathcal{C}\mathcal{D}8$ is used to map the prediction to the expected number of output channels (8 in the current work).

**3D RST**: The 3D RST structure follows the design of Xu et al.[144]. RST-Tiny (RST-T) is employed in the current work. Let $\mathcal{R}k$ denote an RST block with k number of filters consisting of a window multi-head self-attention layer (W-MSA) followed by a shifted window MSA (SW-MSA). A W-MSA unit consists sequentially of layer normalization, window self-attention, layer normalization, and multi-layer perception operations. The SW-MSA unit duplicates W-MSA, except that window self-attention is substituted with shifted window self-attention. $X \times \mathcal{R}k$ represents X number of identical $\mathcal{R}k$ blocks. The Encoder down-samples the feature by a factor of 2 at each block, whereas the decoder up-samples by a factor of 2. Skip connections between the encoder and decoder are not included to avoid GPU memory overflow at the decoding stage. The RST-T is structured as Equation (4-4).

$$encoder: 2 \times \mathcal{R}96, 2 \times \mathcal{R}192, 6 \times \mathcal{R}384, 2 \times \mathcal{R}768$$
$$decoder: 2 \times \mathcal{R}768, 6 \times \mathcal{R}384, 2 \times \mathcal{R}192, 2 \times \mathcal{R}96, \mathcal{R}8$$
(4-4)

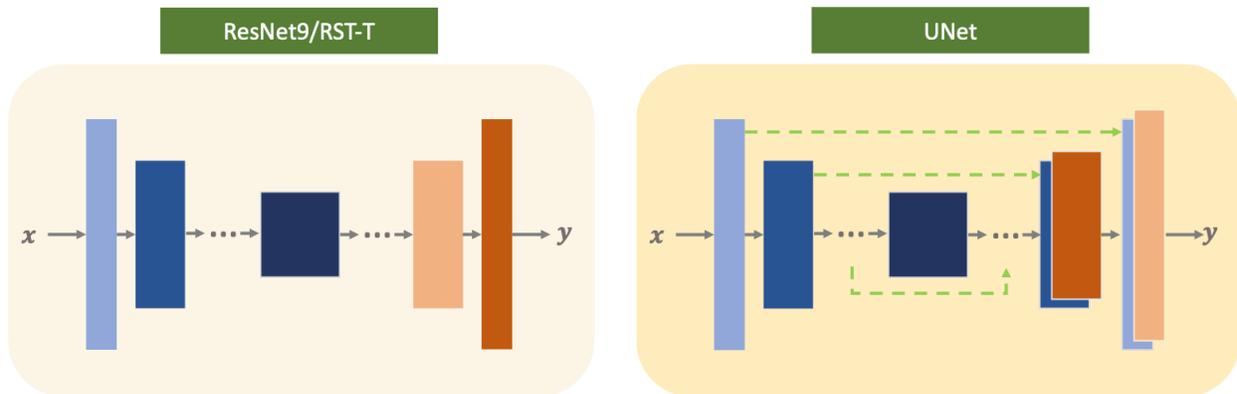

*Figure 4-8 three encoder-to-decoder network designs, including ResNet9[215], UNet[176] and RST[144].*

4.2.2.2.1.2   Discriminator

Following Isola et al.[158], a $N \times N$ PatchGAN is used as the discriminator in Re-Con-GAN. PatchGAN penalizes image series structure at the scale of patches. Specifically, PatchGAN works on identifying if each $N \times N \times t$ image series patch is a "real" or "fake" image series. The discriminator is run temporal-patch-based across the entire spatial dimension, averaging all the corresponding patch predictions from an image series to generate the final discrimination output. Assuming independence among pixels divided by more than a patch coverage, the PatchGAN discriminator essentially models the input image series as a Markov random field to understand the style/texture difference between "real" and "fake" image series[216,217]. The patch size $N$ is a tunable hyperparameter. As Isola et al.[37] discussed, smaller patch sizes have fewer parameters, thus running faster but potentially increasing tiling artifacts. Relatively larger patch sizes sacrifice the running speed but reduce the artifacts. $N$ is set as 70 across the entire current experiments. The $70 \times 70$ PatchGAN is structured as Equation (4-5) with $C'k$ denoting a convolutional block consists sequentially of a $4 \times 4$ convolutional layer with $k$ number of filter and stride of 2, batch normalization and leaky ReLU operations.

$$C'64, C'128, C'256, C'512, C'1 \qquad (4\text{-}5)$$

After the last layer, a single-channeled smaller feature map is generated. An exception in Equation (33) is that Batch normalization is skipped in $C'64$.

*4.2.2.2.2   Loss Objective*

Earlier on, GANs formulated the discriminator as a classifier with a sigmoid cross entropy loss function, as shown in Equation (4-6)[158].

$$L_{cGAN(G,D)} = \mathbb{E}_{x,y}[\log D(x,y)] + \mathbb{E}_{x,z}[\log(1 - D(x, G(x,z)))] \qquad (4\text{-}6)$$

However, later studies show that the sigmoid cross entropy loss function is susceptible to vanishing gradient and has poor training stability during the learning process[218]. Therefore, least square GANs (LSGANs)[218] are proposed to remedy the issues. LSGANs modify the objective function as Equation (4-7~4-8), penalizing sample feature maps based on their pixel distance to the corresponding decision boundary. In this way, more gradients are generated to update the generator.

$$L_{LScGAN(D)} = \frac{1}{2}\sum \mathbb{E}_{x,y}[(D(x,y)-b)^2] + \frac{1}{2}\sum \mathbb{E}_{x,z}[(D(G(x,z))-a)^2] \quad (4\text{-}7)$$

$$L_{LScGAN(G)} = \frac{1}{2}\sum \mathbb{E}_{x,z}[(D(G(x,z))-c)^2] \quad (4\text{-}8)$$

Where $a$ and $b$ are the labels for fake and real data, and $c$ is the value that $G$ wants $D$ to believe for fake data.

Although LSGANs address the gradient vanishing as well as training instability issues, they only consider pixel-wise differences between feature map $D(x)$/ $D(G(x,z))$ and its corresponding label. On the contrary, SSIM considers the changes in overall structural information between the feature map and its target, providing a more holistic comparison and improved perceptual quality of reconstructed images. Taking another step forward, MS-SSIM generalizes single-scale SSIM to incorporate the variations of image resolution and viewing conditions[219]. Therefore, we propose to extend the objective of LSGANs with the addition of MS-SSIM for training of Re-Con-GAN. The loss function is designed as Equation (4-9~4-10).

$$L_{LScGAN(D)} = \frac{1}{2}\left[\sum \mathbb{E}_{x,y}[(D(x,y)-b)^2] + (1-MSSSIM(D(x,y),b))\right] + \\ \left[\frac{1}{2}\sum \mathbb{E}_{x,z}[(D(G(x,z)-a)^2] + (1-MSSSIM(D(G(x,z),a))\right] \quad (4\text{-}9)$$

$$L_{LScGAN(G)} = \frac{1}{2}\sum \mathbb{E}_{x,z}[(D(G(x,z)-c)^2](1-MSSSIM(D(G(x,z),c)) \quad (4\text{-}10)$$

Additionally, previous studies have demonstrated that mixing the GAN objective with a more traditional loss, such as $L_2$ or $L_1$ distances is more beneficial to the convergence of the generator. Both $L_2$ and $L_1$ distances have been explored by pioneers with $L_1$ distance proved to encourage less blurring over $L_2$[158]. Thus, our final objective for $L_{LScGAN(G)}$ is designed as Equation (4-11~4-12).

$$L_{LScGAN(G)} = \frac{1}{2}\sum \mathbb{E}_{x,z}[(D(G(x,z)-c)^2](1-MSSSIM(D(G(x,z),c))) + \lambda L_{L_1}(G) \quad (4\text{-}11)$$

$$L_{L_1}(G) = \mathbb{E}_{x,y,z}[||G(x,z)-y||_1] \quad (4\text{-}12)$$

Where $\lambda$ is a hyperparameter and is set as 1 across all training.

*4.2.2.2.3 Model Training*

The pipeline was implemented with Pytorch, and the training was performed on a GPU workstation with $2 \times RTX$ 4090. All the models were trained for 200 epochs, with the gradient linearly decayed after epoch 100. Adam optimizer with a learning rate of 0.0002 and batch size of $2 \times 4$ was applied.

4.2.2.3   Baseline Algorithms

Conventional CS and non-GAN DL approaches were included as benchmarks. DL baselines consist of 3D UNet (U256)[176], 9 blocks ResNet (ResNet9)[215], and RST-T[144].

Ablation studies that solely tune generators (ResNet9, U256 and RST-T) without generative adversarial training are also conducted to underpin the improvement made by Re-Con-GAN.

### 4.2.2.4 Evaluation

Image evaluation consists of two parts. First, we performed quantitative quality assessment against fully sample nuFFT using the following metrics: root mean squared error (RMSE), PSNR, SSIM, and inference time.

Second, a radiotherapy-specific task was performed to test the accuracy of liver tumor detection and segmentation using an automated liver tumor segmentation network trained on a separate static 3D MR data cohort with a similar imaging protocol. Specifically, 70 patients (excluding the 48 4D MR data cohort) containing 103 manual GTV contours were scanned on the same 3T MRI scanner (MAGNETOM Vida, Siemens Healthcare, Erlangen, Germany) after injection of hepatobiliary contrast (gadoxetic acid; Eovist, Bayer) for each patient. A prototype free-breathing T1-weighted volumetric Cartesian sequence was used for 3D MRI acquisition. The scanning parameters were – TE=1.35 ms, TR=4.05 ms, matrix size = 260x320, in-plane resolution=1.09 mm × 1.09 mm and slice thickness=3 mm.

Mask-RCNN[212] has been used for various types of tumor detection and segmentation[40,48,220]. We employed this framework for the current task of liver GTV detection + segmentation with reconstructed accelerated images. ResNet50[35] without weight-frozen per training stage was used as backbone. ImageNet[221] pretrained weights followed by 3D static MR dataset fine-tuning was implemented for yielding the model convergence. Both mask as well as detection heads were turned on during network training[212]. The pipeline was implemented with Pytorch, and the training was performed on a GPU workstation with $4 \times RTX\ A6000$. All the models were trained for 80k iterations, with a learning rate 0.02 which is decreated by 10 at the 50k and 70k iterations. Stochastic gradient descent optimizer with a batch size of 32 (4 × 8), weight decay of 0.0001 and momentum of 0.9 was used. The final training loss decreased to ~0.03. The 70 3D MR patients were used as the training set for tuning the detection + segmentation network. Images with positive GTV annotations were 3 times augmented in the training set to balance the ratio between positive and negative images. The training data is geometrically augmented using random resizing (image largest width to 640-800), horizontal flipping (p=0.5), and random rotation (angle 0-180°) and morphologically augmented using random gaussian noise (p=0.5, kernel=5, sigma=1) and random brightness (p=0.5).

The trained network segmented liver tumor in the 3x, 6x, and 10x images from the validation set of 4D MR (11 patients; 14 GTVs) processed by Re-Con-GAN, U256, ResNet9, RST-T, CS along with FS nuFFT and 3x, 6x and 10x US nuFFT validation images. Since the detection+segmentation network was designed to detect region of interest in 2D, we ignored the inter-z-dimension and inter-temporal-dimension relationship and organized all the images from 3D training and 4D test sets as 2D frames. All the images were z-score normalized, black border cropped out, and resized to $512 \times 512$. Both images with and without positive GTV annotations were included during the training and test stages.

Image-wise object detection (intersection over union threshold=0.5) and segmentation precision, recall and Dice score as well as 2D segmentation 95% Hausdorff distance ($d_{H95}$) were used to evaluate the model performance as shown in Equation (4-13~4-16).

$$Precision = \frac{TP}{TP + FP} \tag{4-13}$$

$$Recall = \frac{TP}{TP + FN} \tag{4-14}$$

$$Dice = \frac{TP}{TP + \frac{1}{2}(FP + FN)} \tag{4-15}$$

$$d_{H95}(X,Y) = MAX_{95}[d_{XY}, d_{YX}]$$
$$= MAX_{95}\big[MAX_{95,x\in X}MIN_{95,y\in Y}d(x,y), MAX_{95,y\in Y}MIN_{95,x\in X}d(x,y)\big] \quad (4\text{-}16)$$

Where TP, FP and FN stand for true positive, false positive and false negative, $MAX_{95}$ and $MIN_{95}$ represents the 95$^{th}$ percentile of the distances between boundary points in $X$ and $Y$.

### *4.2.3 Experiment*

Reconstruction statistical results and visualization of the validation set are reported in Table 4-2, Figure 4-9, and Figure 4-10.

Visually, Figure 4-9 shows that as the acceleration ratio increases from 3x to 10x and the under-sampled nuFFT input degraded, Re-Con-GAN architectures gradually lost prediction sharpness while showing increasing streaking and tiling artifacts. Nonetheless, Re-Con-GAN with ResNet9 and U256 generators recovered sharper and more detailed morphologies than RST-T. Regarding the quantitative metrics of Re-Con-GAN, the architecture with the ResNet9 generator performed slightly better than that with the U256 generator, while the prediction of Re-Con-GAN with RST-T generator vastly degraded in comparison to the other two. Two different loss objectives ($L_1 + L_2$ and $L_1 + L_2$ +MS-SSIM) were compared during Re-Con-GAN training, with the addition of MS-SSIM encouraging slightly better model convergence. The per-patient inference speed of Re-Con-GAN with ResNet9 and U256 is 150 ms and 160 ms, respectively, meeting the requirements of real-time 4D MR reconstruction (<500 ms)[222].

Re-Con-GAN with the ResNet9 generator slightly outperformed CS, which is comparable to Re-Con-GAN with the U256 generator and substantially better than Re-Con-GAN with RST-T generator. As shown in Figure 4-9 and Figure 4-10, CS reconstruction results for 3x and 6x acceleration show minimal artifacts and good detail retention. CS shows more obvious streaking artifacts than Re-Con-Gan when increasing the acceleration to 10x. CS reconstruction time of 120 s is ~700X longer than Re-Con-Gan.

GTV detection and segmentation statistical results and visualization of an example from the validation set are reported in Table 4-3 and Figure 4-11. The liver tumor can be reliably segmented using images acquired with up to 5x acceleration, but the performance dropped sharply with 10x. All the images reconstructed from different models (proposed and benchmarks) can, to varying degrees, improve the detection and segmentation results than US nuFFT images. From Table 4-3, we can observe that Re-Con-GAN ResNet9 achieved slightly inferior outcomes than FS nuFFT but was consistently superior to other benchmarks, including CS, Re-Con-GAN with U256 and RST-T generators and 3D non-adversarial trained networks. All models achieved better precision than recall, indicating a systematic under-segmentation/detection using the network.

From Figure 4-11, we can see that the GTV was still detectable at a 100% confidence score on a 3x US nuFFT image frame, where the confidence score dropped to 79% on the 6x frame, and the model completely missed its target on the 10x frame. Mask-RCNN can accurately detect and segment GTV across all acceleration levels on Re-Con-GAN and CS reconstructed images, while Re-Con-GAN achieved a moderately higher confidence score (98%) than that of CS (90%) at 10x acceleration.

*Table 4-2 Results from Re-Con-GAN under 3x, 6x and 10x acceleration rate and baselines.*

| Model | Generator | Acceleration | Loss | PSNR↑ | 1-SSIM↓ | RMSE↓ | Time (s)↓ |
|---|---|---|---|---|---|---|---|
| Re-Con-GAN | ResNet9 | 3x | $L_1 + L_2$ | 25.65 ± 2.89 | 0.06 ± 0.02 | 0.09 ± 0.03 | 0.15 |
| | | | $L_1 + L_2$ +MS-SSIM | **26.13 ± 3.02** | **0.05 ± 0.02** | **0.08 ± 0.03** | |
| | | 6x | $L_1 + L_2$ | 21.68 ± 2.88 | 0.10 ± 0.03 | 0.13 ± 0.03 | |
| | | | $L_1 + L_2$ +MS-SSIM | **23.97 ± 3.84** | **0.07 ± 0.03** | **0.11 ± 0.04** | |
| | | 10x | $L_1 + L_2$ | 20.01 ± 2.81 | 0.11 ± 0.03 | 0.15 ± 0.05 | |
| | | | $L_1 + L_2$ +MS-SSIM | **21.61 ± 2.93** | **0.09 ± 0.03** | **0.13 ± 0.04** | |
| | U256 | 3x | $L_1 + L_2$ | 25.09 ± 2.74 | 0.10 ± 0.03 | 0.10 ± 0.03 | 0.16 |
| | | | $L_1 + L_2$ +MS-SSIM | 25.41 ± 2.70 | 0.08 ± 0.03 | 0.09 ± 0.02 | |
| | | 6x | $L_1 + L_2$ | 21.82 ± 3.12 | 0.10 ± 0.04 | 0.13 ± 0.04 | |
| | | | $L_1 + L_2$ +MS-SSIM | 22.01 ± 3.13 | 0.08 ± 0.03 | 0.12 ± 0.04 | |
| | | 10x | $L_1 + L_2$ | 19.95 ± 3.01 | 0.12 ± 0.03 | 0.14 ± 0.05 | |
| | | | $L_1 + L_2$ +MS-SSIM | 20.08 ± 2.99 | 0.11 ± 0.03 | 0.12 ± 0.05 | |
| | SWT-T | 3x | $L_1 + L_2$ | 19.22 ± 2.64 | 0.18 ± 0.07 | 0.18 ± 0.11 | 0.73 |
| | | | $L_1 + L_2$ +MS-SSIM | 20.21 ± 2.76 | 0.16 ± 0.07 | 0.15 ± 0.09 | |
| | | 6x | $L_1 + L_2$ | 18.05 ± 3.11 | 0.21 ± 0.11 | 0.20 ± 0.13 | |
| | | | $L_1 + L_2$ +MS-SSIM | 18.85 ± 3.14 | 0.21 ± 0.11 | 0.19 ± 0.12 | |
| | | 10x | $L_1 + L_2$ | 15.78 ± 3.09 | 0.28 ± 0.10 | 0.24 ± 0.14 | |
| | | | $L_1 + L_2$ +MS-SSIM | 15.97 ± 3.01 | 0.27 ± 0.09 | 0.23 ± 0.13 | |
| - | U256 | 3x | $L_1 + L_2$ +MS-SSIM | 22.23 ± 2.95 | 0.12 ± 0.04 | 0.13 ± 0.05 | 0.16 |
| | | 6x | | 19.02 ± 3.12 | 0.13 ± 0.05 | 0.16 ± 0.06 | |
| | | 10x | | 17.34 ± 2.95 | 0.15 ± 0.05 | 0.19 ± 0.06 | |
| | SWT-T | 3x | | 18.91 ± 2.81 | 0.21 ± 0.09 | 0.20 ± 0.09 | 0.73 |
| | | 6x | | 18.08 ± 2.95 | 0.24 ± 0.08 | 0.22 ± 0.12 | |
| | | 10x | | 14.52 ± 3.11 | 0.29 ± 0.12 | 0.27 ± 0.17 | |
| | ResNet9 | 3x | | 22.45 ± 3.01 | 0.11 ± 0.03 | 0.12 ± 0.04 | 0.15 |
| | | 6x | | 20.08 ± 3.12 | 0.12 ± 0.04 | 0.14 ± 0.05 | |
| | | 10x | | 18.25 ± 3.10 | 0.14 ± 0.06 | 0.17 ± 0.06 | |
| CS | | 3x | - | 25.31 ± 2.56 | 0.08 ± 0.02 | 0.19 ± 0.05 | 120 |
| | | 6x | | 20.73 ± 2.95 | 0.12 ± 0.03 | 0.16 ± 0.05 | |
| | | 10x | | 19.29 ± 2.99 | 0.13 ± 0.05 | 0.21 ± 0.08 | |

*Table 4-3 : Statistical results from Mask-RCNN detection and segmentation from our proposed Re-Con-GAN under 3x, 6x and 10x acceleration rate and their corresponding baselines.*

| Image Modality | Generator | Acceleration | Detection | | | Segmentation | | | |
|---|---|---|---|---|---|---|---|---|---|
| | | | Precision↑ | Recall↑ | *Dice* ↑ | Precision↑ | Recall↑ | *Dice* ↑ | HD 95↓ (mm) |
| FS nuFFT | - | - | **94.40** | **72.91** | **82.27** | **92.52** | **72.55** | **81.33** | **8.87** |
| US nuFFT | - | 3x | 80.27 | 61.45 | 69.61 | 76.35 | 57.23 | 65.42 | 13.29 |
| | - | 6x | 55.34 | 45.27 | 49.80 | 49.37 | 40.56 | 44.53 | 18.79 |
| | - | 10x | 25.34 | 17.21 | 20.50 | 21.56 | 13.75 | 16.79 | 19.31 |
| Re-Con-GAN | ResNet9 | 3x | **93.57** | **71.38** | **80.98** | **91.26** | **70.07** | **79.27** | **8.95** |
| | | 6x | **91.05** | **70.32** | **79.35** | **89.77** | **68.38** | **77.63** | **9.13** |
| | | 10x | **85.45** | **67.46** | **75.40** | **82.06** | **62.57** | **71.00** | **9.27** |
| | U256 | 3x | 92.47 | 70.35 | 79.91 | 89.46 | 68.32 | 77.47 | 9.07 |
| | | 6x | 87.78 | 68.25 | 76.79 | 86.87 | 65.34 | 74.58 | 9.18 |
| | | 10x | 81.46 | 64.57 | 72.04 | 79.34 | 60.54 | 68.68 | 9.47 |
| | SWT-T | 3x | 82.74 | 61.36 | 70.46 | 78.35 | 59.37 | 67.55 | 12.89 |
| | | 6x | 78.35 | 57.23 | 66.15 | 75.47 | 53.27 | 62.46 | 14.73 |
| | | 10x | 60.45 | 49.46 | 54.41 | 59.48 | 48.57 | 53.47 | 16.81 |
| U256 | - | 3x | 92.35 | 70.37 | 79.88 | 89.02 | 67.99 | 77.10 | 9.10 |
| | - | 6x | 86.89 | 68.12 | 76.37 | 86.08 | 65.24 | 74.22 | 9.25 |
| | - | 10x | 81.01 | 63.75 | 71.35 | 78.99 | 60.12 | 68.28 | 9.90 |
| SWT-T | - | 3x | 81.35 | 60.12 | 69.14 | 77.24 | 59.01 | 66.91 | 13.52 |
| | - | 6x | 77.45 | 56.34 | 65.23 | 75.06 | 53.17 | 62.25 | 15.04 |
| | - | 10x | 68.72 | 54.36 | 60.70 | 58.77 | 47.62 | 52.61 | 17.08 |
| ResNet9 | - | 3x | 93.06 | 71.12 | 80.62 | 91.05 | 69.57 | 78.87 | 9.02 |
| | - | 6x | 90.05 | 69.23 | 78.28 | 88.01 | 67.03 | 76.10 | 9.12 |
| | - | 10x | 83.53 | 66.89 | 74.29 | 80.27 | 61.05 | 69.35 | 9.37 |
| CS | - | 3x | 93.46 | 71.06 | 80.74 | 91.11 | 70.02 | 79.18 | 8.99 |
| | - | 6x | 90.89 | 70.01 | 79.10 | 88.72 | 67.33 | 76.56 | 9.12 |
| | - | 10x | 84.56 | 67.02 | 74.78 | 81.99 | 62.04 | 70.63 | 9.35 |

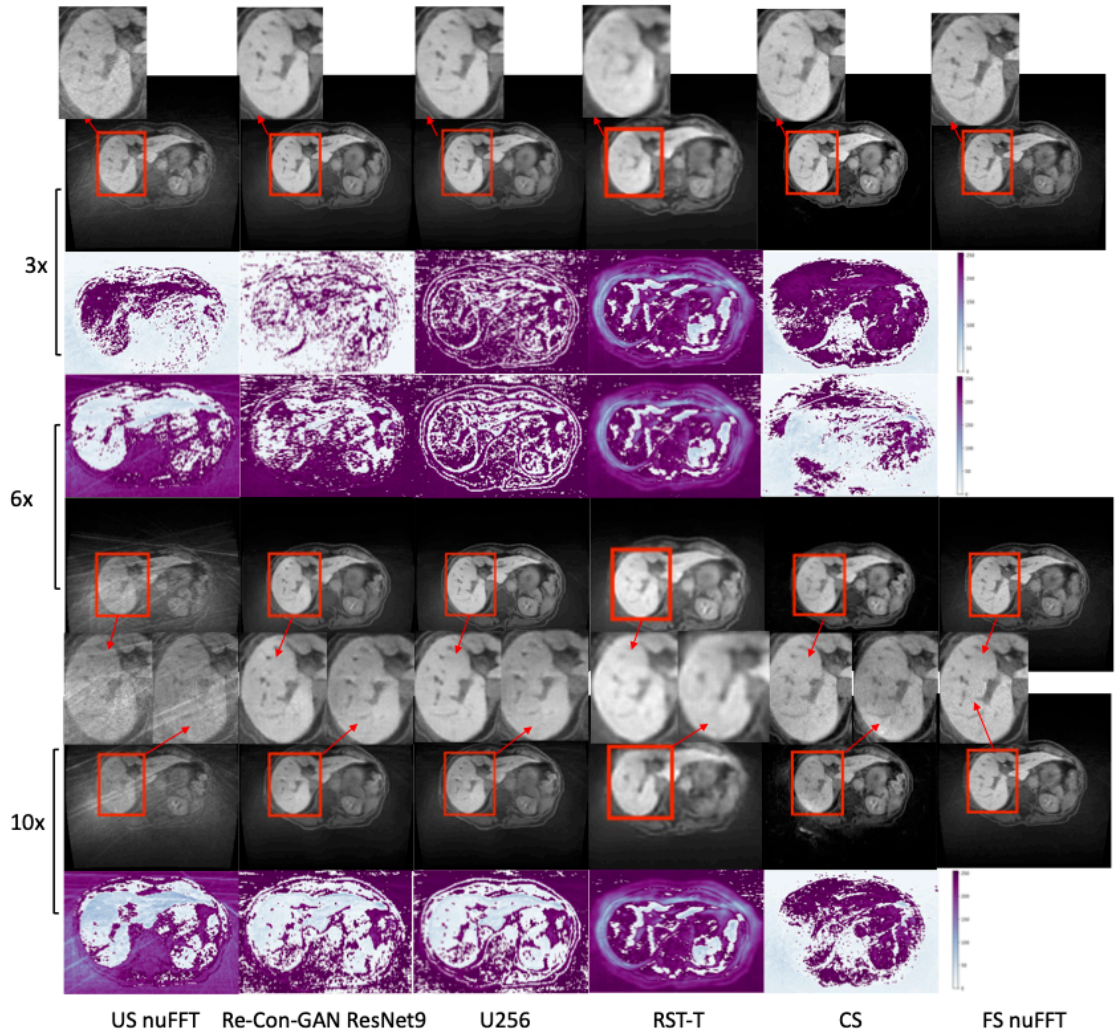

*Figure 4-9 Visualization of 3x, 6x and 10x reconstruction results of an axial view slice from a patient in validation set.*

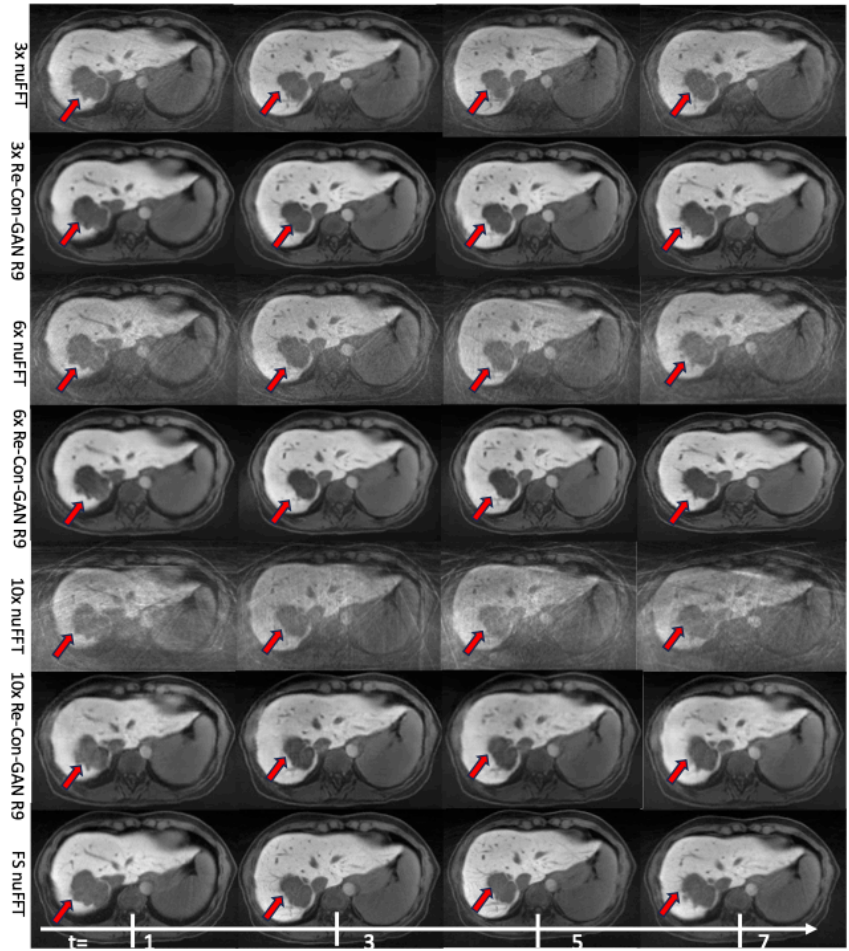

*Figure 4-10 Visualization of a selected temporal profile (motion binning = 1, 3, 5, 7) from a patient in the validation set. 3x, 6x, and 10x reconstruction results from input, GT and our proposed method.*

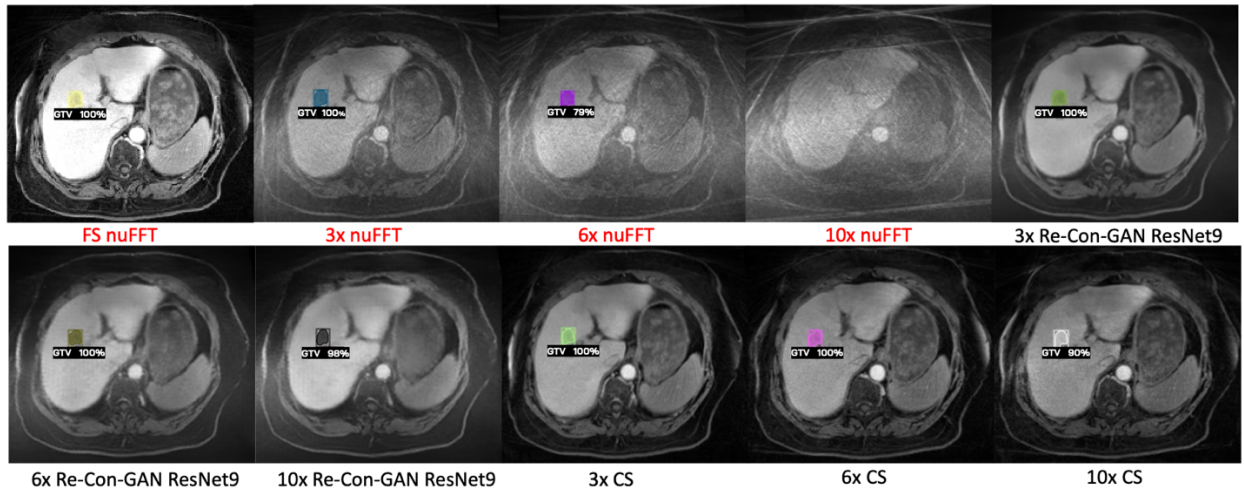

*Figure 4-11 Visualization of Mask-RCNN detection and segmentation results on a validation set patient from selected Re-Con-GAN and baseline models.*

### *4.2.4 Discussion*

The current project focuses on liver 4D MRI, which is particularly relevant to image-guided liver cancer radiotherapy. Hepatocellular carcinoma (HCC) is the fifth most common cancer worldwide in men and the seventh in women. HCC represents the third most frequent and fast-rising cause of cancer deaths[223,224]. The past few decades have witnessed a continuous decrease in the average age at HCC diagnosis, with most HCC patients now diagnosed between 45 and 60[225]. Additionally, the liver is one of the most common metastatic sites of several cancer types, including colorectal, pancreatic, melanoma, lung, and breast cancer[226].

Surgical resection remains the standard of care for hepatic primary and metastatic tumors and continues to demonstrate persistent positive prognosis outcomes in surgical-qualified candidates[227]. For non-surgical patients, orthotopic liver transplants, ablative procedures, chemotherapy, and radiation therapy (RT) are considered effective alternatives[228]. SBRT, delivering intense and highly conformal radiation doses, has shown promising results in hepatic malignancy and metastasis management[228–231]. The success of liver SBRT, however, depends on the ability to focus the high radiation dose on the tumor while minimizing the dose to the normal liver tissue, which is sensitive to radiation[232]. A prerequisite for successful liver SBRT is accurate liver tumor imaging and motion management of the highly mobile organ.

Unlike lung tumors, which are often clearly visualized in CT and 4D CT, liver tumors have low soft tissue X-ray contrast but high MR contrast, making MRI and 4D MRI an ideal pre and during-treatment liver imaging. 4D MRI requires densely sampled k-space data for spatiotemporal reconstruction. Fully sampling the required k-space data results in lengthy MR sequences that are challenging for MR simulation due to limited patient tolerance and available scanner time and impractical for online MR-guided RT[233]. Acceleration of MR acquisition via down-sampling the k-space and rapid image reconstruction without compromising the usability of the image quality is thus highly desired. Non-cartesian k-space sampling and compressed sensing have achieved remarkable success in the former goal but struggled with the latter due to the slow iterative algorithms. Though some previous works attempted to utilize DL methods, such as 3D UNet, RNN, and Transformers variants[139,143,144], to tackle the problem, these methods are heterogeneous in reconstructing dynamic liver images, as shown in the non-adversarial trained DL benchmarking results in Figure 4-9 and Table 4-2. We postulate that the difficulty of defining a loss function suitable for simultaneous detail retention and artifacts suppression is a contributing factor.

Therefore, we developed Re-Con-GAN in this work. Re-Con-GAN is structured with pair-trained conditional GAN architecture constraint with loss objective fused from $L_2$, $L_1$ and MS-SSIM. Three types of generators, including 3D ResNet9, 3D UNet and 3D RST-T, are demonstrated. $70 \times 70$ PatchGAN is utilized as the discriminator. Re-Con-GAN is validated on an in-house dynamic liver MRI dataset with 48 patients having a total of 12332 2D+t image series. Further downstream validation tasks of GTV detection and segmentation were also conducted on Re-Con-GAN reconstructed images. Re-Con-GAN showed competitive performance to CS in image quality and is significantly faster. The real-time inference speed and sharp liver GTV morphology visualization of Re-Con-GAN are conducive to image-guided liver radiotherapy. Our proposed methods achieve 1-SSIM of $0.05 \pm 0.02$ at 3x acceleration, which outperforms previous GAN-based 3D stack-of-radial liver MRI reconstruction studies conducted by Gao et al. reporting 1-SSIM of $0.16 \pm 0.01$ at 3x acceleration[211]. The study based on raw k-space data and undersampling radial spokes of the stack of stars can be readily deployed.

There are several theoretical and practical advantages to using GAN for 4D MRI. Standard NNs, such as U256, ResNet9, and RST-T, fully parameterize their loss function and use the fixed loss function to conduct representation learning from training information. In GANs, the penalty imposed by the discriminator is a nonparametric loss function, mitigating the inflexibility of an explicitly defined loss function and tradeoffs in noise, uniformity, detail retention, and computational tractability. As shown here, Re-Con-GAN reconstructs sharper and more consistent images than the compared DL benchmarks (U256, ResNet9, and RST-T). The improvement is more evident in quantitative image quality assessment using SSIM, PSNR, and RMSE than in the automated liver segmentation task. Liver segmentation using deep learning is less sensitive to image quality but more dependent on the training data size, which is the common bottleneck of the current study. This is evidenced by a larger improvement in the segmentation accuracy with a higher acceleration ratio, where the image quality degradation is evident. We also note that the All generators significantly outperformed Transformers (RST-T). Among all the compared generators (U256, ResNet9, and RST-T), CNN architectures achieve similar performance, with U256 slightly inferior to ResNet9. We attribute the result to the current limited size of training samples. Evidence has shown that Vision Transformers architecture performance declines when trained on small datasets due to the lack of locality, inductive biases, and hierarchical structure of the representation commonly observed in CNNs. Therefore, Vision Transformers architectures, including RST-T, require large-scale training data or domain-relevant pre-training + fine-tuning to learn such properties from the data distribution[234].

The current work can be improved or extended in several areas. First, our implementation is restricted to learning 2D+t image series. 3D+t training would allow more effective learning of the inter-slice anatomy but requires an exceedingly large GPU memory footprint. Second, the current validation is conducted on a dataset collected from a single institute. Although our pipeline is shown robust to the single institutional held-out test, its performance in the external data needs further testing. Despite the recent rapid increase of medical images in the public domain, raw k-space data of 4D MRI essential for realistic undersampling are rarely stored and shared. Third, as more aggressive acceleration ratios (6x and 10x) are pursued, tiling artifacts was suppressed but still noticeable. Model structures more robust to tiling artifacts, such as diffusion-based frameworks[235] or post-processing techniques, are worth exploring to combat such artifacts. Fourth, despite the real-time image reconstruction speed, acquiring the highly under-sampled stack of star k-space data is not real-time. As a result, 4D MRI using Re-Con-GAN does not reflect real-time anatomy. Sparser sampling in combination with prior retrospective 4D MRI may be necessary for real-time 3D MR reconstruction. Lastly, the current method requires transformation from k-space to images as input. The additional step leads to information loss and adds latency. Future work will explore networks using k-space or coil data as the input.

### 4.2.5 Summary

An efficient yet robust liver 4D MRI reconstruction framework, Re-Con-GAN, is proposed. Re-Con-GAN uses a flexible framework with 3D ResNet9, 3D UNet and 3D RST demonstrated as generator, PatchGAN as discriminator, and $L_1$, $L_2$ and MS-SSIM fused measurements as loss objectives. Validation from the in-house liver 4D MRI dataset substantiates the superior inference speed of Re-Con-GAN to its CS benchmark as well as higher predicted image quality to the compared 3D UNet, 3D ResNet and 3D RST-T DL Benchmarks.

## 4.3 Rapid Reconstruction of Extremely Accelerated Liver 4D MRI via Chained Iterative Refinement

### *4.3.1 Introduction*

As previously discussed, fruitful studies have demonstrated that well-trained DL models can match/exceed CS with significantly less inference time[91,139,140]. Existing works have explored using CNNs[139], RNNs[140], Transformers[147], or GAN assisted networks for 4D MR reconstruction[148]. Yet, neither of them can go beyond 10-times acceleration (10x). Since those frameworks require carefully designed regularization and optimization tricks, which are often hard to seek, in challenging tasks to tame optimization instability[236,237] and mode collapse[238,239]. Recently, stochastic diffusion probabilistic models, such as Denoising Diffusion Probabilistic Model (DDPM)[240,241] and Super Resolution via Repeated Refinement (SR3)[235], have been demonstrated to be more powerful in natural image super-resolution tasks than simple regression based CNNs/RNNs/Transformers and GAN enhanced frameworks. Such architectures are superior in two major perspectives. 1) Modelling imaging noise instead of the mean to form a more well-defined optimization target. 2) Incrementally approaching the optimization target for modelling stability. Therefore, we proposed chained iterative refinement network (CIRNet), a framework rooted on stochastic diffusion, to achieve the reconstruction in ultra-sparse undersampled 4D MRI sequences.

### *4.3.2 Methods*

#### 4.3.2.1 Data Cohort

The same data cohort as Section 4.2 was utilized in the current project.

### 4.3.2.2 CIRNet Framework

The source-target image sequence pairs, $D = \{x_i, y_i\}_{i=1}^{N}$, are samples drawn from unknown distribution $p(y|x)$. Our goal is to learn a parametric approximation of $p(y|x)$ that can map $x$ to $y$. We approach this problem through the forward and reverse stochastic iterative diffusion refinement process. As seen in Figure 3-12, the forward diffusion process $q$ incrementally adds Gaussian noise to the fully sampled image sequence $y_0$ over $T$ timesteps through a fixed Markov chain $p_\theta(y_t|y_{t-1})$. The reverse diffusion process $p_\theta$ is parameterized with CIRNet which aims to iteratively recover signals from noise $y_T$ conditioned on the information in source $x$.

The CIRNet architecture is a variant of U-Net, with modification adapts from SR3[235]. We replace the original residual blocks with that from BigGAN[242], rescale the skip connections by $\frac{1}{\sqrt{2}}$, replace the input/output number of channels ($C = 3$) in natural images with the number of motion bins ($C = 8$) in our image sequence. The forward training noise schedule uses a piece-wise distribution $\gamma$ that conducts uniformly sampling through the iteration timesteps $T$, where $T = 800$ across our experiments.

Assuming the forward diffusion process can be viewed as a fixed approximate posterior to the inference process, we can derive the variational lower bound on the marginal log-likelihood as Equation (4-18). Given a particular parameterization $p_\theta$, we can derive that the negative variational lower bound as the simplified $L_2$ loss for $p_\theta$ optimization defined in Equation (4-19), up to a constant weighting of each term for each time step[240].

$$E_{(x,y_0)} \log p_\theta(y_0|x) \geq E_{x,y_0} E_{q(y_{1:T}|y_0)} \left[ \log p_\theta(y_T) + \sum_{t \geq 1} \log \frac{p_\theta(y_{t-1}|y_t, x)}{q(y_t|y_{t-1})} \right] \quad (4\text{-}18)$$

$$E_{x,y_0,\epsilon} \sum_{t=1}^{T} \frac{1}{T} ||\epsilon - \epsilon_\theta(x, \sqrt{\gamma_t}y_0 + \sqrt{1-\gamma_t}\epsilon, \gamma_t)||_2^2 \qquad (4\text{-}19)$$

Where $\epsilon \sim \mathcal{N}(0, I)$.

CIRNet is trained for 1,000,000 iterations with a batch size of $4 \times 1$. Adam optimizer with a linear warmup schedule over 10,000 training iterations, followed by a fixed learning rate of 0.0001 for the rest iterations. All the experiments are carried out on a $4\times RTXA6000$ GPU cluster.

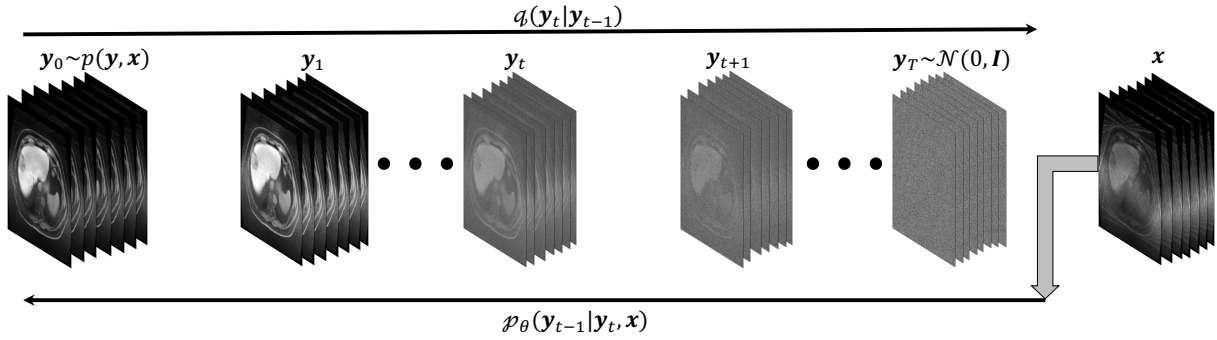

*Figure 4-12 CIRNet structure.*

### 4.3.3 Results

The statistical results and selected visualization of the test set are reported in Table 4-4, Figure 4-14 and Figure 4-15. CIRNet outperforms CS and Re-Con-GAN consistently across all acceleration levels. From Figure 4-13, we can also observe that the reconstruction from CIRNet is much sharper and detailed in comparison to Re-Con-GAN. From Figure 4-15, we can see that CIRNet is robust across all acceleration level and can maintain stable contrast over the entire motion bins.

*Table 4-4 Statistical results from our proposed CIRNet under 3x, 6x, 10x, 20x and 30x acceleration rate and their corresponding baselines.*

| Model | Acceleration | PSNR (dB) | 1-SSIM | RMSE | Inference Time (s) |
|---|---|---|---|---|---|
| CS | 3x | 25.31±2.56 | 0.08±0.02 | 0.19±0.05 | 120 |

| Model | Acceleration | PSNR (dB) | 1-SSIM | RMSE | Inference Time (s) |
|---|---|---|---|---|---|
| Re-Con-GAN | | 26.13±3.02 | 0.05±0.02 | 0.08±0.03 | **0.15** |
| CIRNet | | **29.43±2.67** | **0.04±0.02** | **0.06±0.02** | 11 |
| CS | | 20.73±2.95 | 0.08±0.02 | 0.19±0.05 | |
| Re-Con-GAN | 6x | 23.97 ± 3.84 | 0.12 ± 0.03 | 0.16 ± 0.05 | |
| CIRNet | | **28.36±2.78** | **0.05±0.02** | **0.07±0.03** | |
| CS | | 19.29±2.99 | 0.13±0.05 | 0.21±0.08 | |
| Re-Con-GAN | 10x | 21.61±2.93 | 0.09±0.03 | 0.13±0.04 | |
| CIRNet | | **26.36±2.89** | **0.06±0.03** | **0.09±0.03** | - |
| CS | | 15.35±3.67 | 0.18±0.09 | 0.26±0.15 | |
| Re-Con-GAN | 20x | 17.02±3.45 | 0.15±0.07 | 0.22±0.12 | |
| CIRNet | | **25.03±2.92** | **0.08±0.05** | **0.12±0.05** | |
| CS | | 13.27±3.89 | 0.19±0.12 | 0.29±0.19 | |
| Re-Con-GAN | 30x | 15.89±3.65 | 0.17±0.09 | 0.25±0.15 | |
| CIRNet | | **22.35±2.94** | **0.11±0.07** | **0.15±0.07** | |

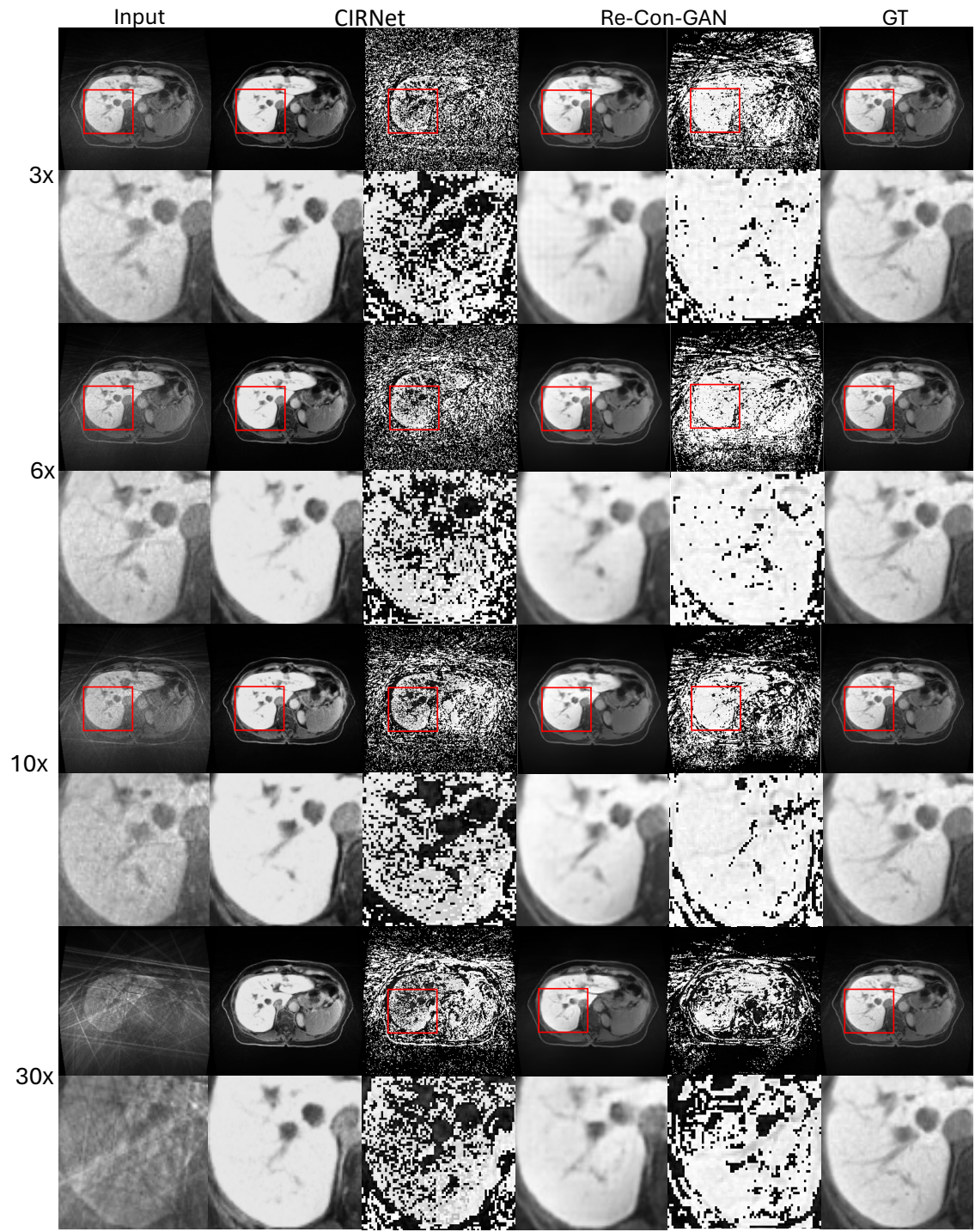

Figure 4-13 Visualization of 3x, 6x, 10x and 30x reconstruction results from CIRNet and Re-Con-GAN of an axial view slice from a patient in the test set.

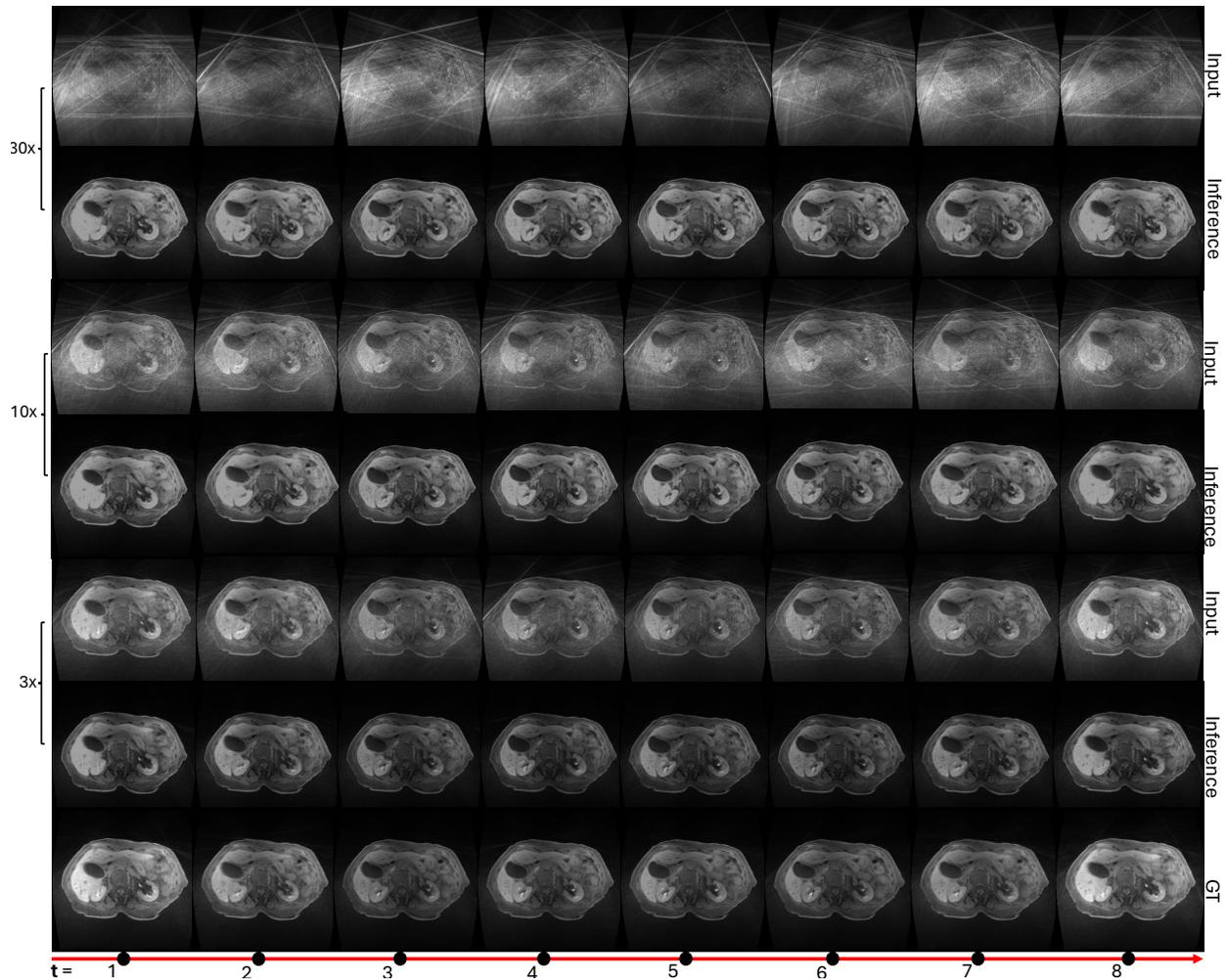

*Figure 4-14 Visualization of the temporal profile of a patient in the test set. 3x, 10x, and 30x reconstruction results from input, GT and our proposed CIRNet.*

### 4.3.4 Disscussion

Hepatocellular carcinoma (HCC) ranks at the top ten most prevalent cancers globally, and it has become one of the leading and fastest-growing causes of cancer-related death[223,224]. Moreover, the liver is a frequent site for metastases from various cancer types, including colorectal, pancreatic, melanoma, lung, and breast cancers[226]. Although surgical resection remains the standard of care for HCC[227], RT is an effective alternative for unresectable patients[228]. Owing to its superior soft-tissue contrast, 4D MRI is a crucial imaging tool for image-guided RT (IGRT) in liver cancer.

Though existing DL and analytical methods reported promising results for accelerated 4D MRI reconstruction, none maintain usable image quality with >10x acceleration. The current paper shows that CIRNet can maintain its reconstructed image quality up to 30x acceleration, leading to around 20-second acquisition, significantly reducing the patient burden. There are several theoretical and practical advantages to using CIRNet for 4D MRI reconstruction. First, CIRNet demonstrates superior retention of subtle tissue textures, which is essential for the accurate delineation of liver tumors. In comparison, Re-Con-GAN suffers from detail loss due to the inherent smoothing effects of CNNs, while CS methods tend to lose the structures and leave more pronounced artifacts that worsen steeply with high acceleration rates. Second, CIRNet offers significantly faster reconstruction (11s) than the CS method (120s), making them practical for time-sensitive adaptive IGRT. Though CIRNet is slower than Re-Con-GAN (<1s) inference-wise, CIRNet offsets longer reconstruction time with shorter acquisition. The acquisition time using CIRNet approaches the duration of a breathing cycle, indicating the potential of capturing real-time motion with minimal binning/averaging.

Nevertheless, the current work is not without room for improvement. First, our implementation is limited to learning 2D+t image series. Training with 3D+t data would enable more effective learning of inter-slice anatomy but would demand an impractically large GPU memory capacity. Second, the current method functions in the image domain, which leads to potential information loss and added latency due to preliminary k-space to image transformation. Future work will explore frameworks working in k-space or using multi-coil data as the model input. Lastly, CIRNet can be complex and less interpretable, especially when trained on 4D data, which could pose potential challenges in clinical validation, where model interpretability is essential. In the future, explainable artificial intelligence techniques, such as gradient-weighted class activation mapping[243], saliency maps[244], and feature attribution analysis[245], can be incorporated to improve the transparency and trustworthiness of the current framework.

### *4.3.5 Conclusion*

An ultra-sparse liver 4D MRI reconstruction framework, CIRNet, is proposed in the current work. CIRNet employs a stochastic diffusion process to iteratively model the source noise for robust image reconstruction. The evaluation conducted on an in-house data cohort demonstrates promising imaging quality at ultra-high acceleration ratios.

**Publishing Agreement**

It is the policy of the University to encourage open access and broad distribution of all theses, dissertations, and manuscripts. The Graduate Division will facilitate the distribution of UCSF theses, dissertations, and manuscripts to the UCSF Library for open access and distribution. UCSF will make such theses, dissertations, and manuscripts accessible to the public and will take reasonable steps to preserve these works in perpetuity.

I hereby grant the non-exclusive, perpetual right to The Regents of the University of California to reproduce, publicly display, distribute, preserve, and publish copies of my thesis, dissertation, or manuscript in any form or media, now existing or later derived, including access online for teaching, research, and public service purposes.

_Di Xu_  4/7/2025

Author Signature                                                                                        Date